\documentclass[aps,prb,twocolumn,longbibliography,superscriptaddress]{revtex4-2}

\usepackage{soul}
\usepackage[english]{babel}
\babelprovide[main]{english}
\babelprovide{en}
\usepackage{graphicx}	
\usepackage{bm}			
\usepackage{soul}
\usepackage{mathrsfs} 
\usepackage{amssymb}
\usepackage{multirow}
\usepackage{color}
\usepackage[normalem]{ulem}
\usepackage{mathtools}
\usepackage{tikz}
\usepackage{cancel}
\usetikzlibrary{arrows.meta}
\usepackage{braket}
\usepackage{booktabs}

\definecolor{LinkColor}{rgb}{0.223,0.223,1}
\usepackage{hyperref}
\hypersetup{colorlinks=true,citecolor=LinkColor,linkcolor=LinkColor,urlcolor=LinkColor}

\newcommand{\avg}[1]{\left\langle #1 \right\rangle}

\newcommand{\comments}[1]{}   

\newcommand{\autoeqref}[1]{Eq.~(\ref{#1})}

\AtBeginDocument{}
\AtBeginDocument{}
\AtBeginDocument{}
\AtBeginDocument{}
\AtBeginDocument{}
\newcommand{\appref}[1]{App.~\ref{#1}}

\begin{document}

\title{Large deviations in the many-body localization transition: \\ The case of the random-field XXZ chain}

\date{\today}

\author{Greivin~Alfaro\hphantom{-}Miranda}
\affiliation{Sorbonne Universit\'e, Laboratoire de Physique Th\'eorique et Hautes Energies, CNRS UMR 7589,
    4 Place Jussieu, 75252 Paris Cedex 05, France}
\author{Fabien~Alet}
\affiliation{Univ. Toulouse, CNRS, Laboratoire de Physique Th\'eorique, Toulouse, France}

\author{Giulio~Biroli}
\affiliation{Laboratoire de Physique Statistique, Ecole Normale Supérieure,
PSL Research University, 24 rue Lhomond, 75005 Paris, France}

\author{Leticia~F.~Cugliandolo}
\affiliation{Sorbonne Universit\'e, Laboratoire de Physique Th\'eorique et Hautes Energies, CNRS UMR 7589,
    4 Place Jussieu, 75252 Paris Cedex 05, France}

\author{Nicolas~Laflorencie}
\affiliation{Univ. Toulouse, CNRS, Laboratoire de Physique Th\'eorique, Toulouse, France}

\author{Marco~Tarzia}
\affiliation{Sorbonne Universit\'e, Laboratoire de Physique Th\'eorique de la  Mati\`ere Condens\'ee, CNRS UMR 7600,
    4 Place Jussieu, 75252 Paris Cedex 05, France}

\begin{abstract} 
The effect of rare system-wide resonances in the many-body localization (MBL) transition has recently attracted significant attention. They are expected to play a prominent role in the stability of the MBL phase, prompting the development of new theoretical frameworks to properly account for their statistical weight. We employ a method based on an analogy with mean-field disordered glassy systems to characterize the statistics of transmission amplitudes between distant many-body configurations in Hilbert space, and apply it to the random-field XXZ spin chain. By introducing a Lagrange multiplier, which formally plays the role of an effective-temperature controlling the influence of extreme outliers in the heavy-tailed distribution of propagators, we identify three distinct regimes: (i) an ergodic phase with uniform spreading in Hilbert space, (ii) an intermediate regime where delocalization is driven by rare, disorder-dependent long-range resonances, and (iii) a robust MBL phase where such resonances cannot destabilize localization. We derive a finite-size phase diagram in the disorder–interaction plane both in the spin and in the Anderson basis that quantitatively agrees with recent numerical results based on real-space spin-spin correlation functions. We further demonstrate that even infinitesimal interactions can destroy the Anderson insulator at finite disorder, with the critical disorder remaining finite down to small interaction strengths. By visualizing resonant transmission pathways on the Hilbert space graph, we provide a complementary perspective to real-space and spectral probes, revealing how the destabilization of the MBL phase at finite sizes stems from the emergence of resonant paths that become progressively rarer and shorter-ranged deep in the localized phase.

\vspace{0.5cm}

\noindent

\end{abstract}
{
\maketitle
\newpage
\tableofcontents
}

\section{Introduction}
Many-body localization (MBL) is a dynamical phase of matter in which an isolated, interacting quantum system with sufficient disorder fails to thermalize. Although the possibility of localization in interacting systems was first suggested by Anderson in his pioneering work~\cite{anderson_absence_1958} (for a review, see~\cite{Evers2008}), it was only about two decades ago that the perturbative stability of the Anderson insulator in the presence of weak interactions was explored~\cite{Gornyi2005, Basko2006}. For recent reviews, see Refs.~\cite{Nandkishore2015, abanin_colloquium_2019, alet_many-body_2018, sierant_many-body_2025}. This failure is attributed to the emergence of an extensive set of quasi-local integrals of motion (LIOMs or `$\ell$-bits') that inhibit thermalization, allowing the system to retain local memory of its initial conditions indefinitely \cite{Serbyn2013, Huse2014, Ros2015, Imbrie2017, Nandkishore2015, abanin_colloquium_2019, alet_many-body_2018, sierant_many-body_2025}. 

In recent years, the stability (particularly with respect to {\it non-perturbative} events) of the MBL phase has been put into question \cite{de_roeck_stability_2017, suntajs_quantum_2020, sierant_polynomially_2020, suntajs_ergodicity_2020, sierant_thouless_2020, kiefer-emmanouilidis_evidence_2020, kiefer-emmanouilidis_slow_2021, sierant_challenges_2022, sels_bath-induced_2022, sels_thermalization_2023,Beraetal2017, WeinerEversBera2019,EversModakBera2023}. This skepticism stems from the exponential increase in Hilbert space volume with the number of degrees of freedom, which severely limits exact diagonalization studies to relatively small system sizes, potentially missing the asymptotic behavior characteristic of large systems. In these studies, commonly used spectral observables---such as average spectral statistics, eigenstate participation entropies, imbalance decay, and entanglement entropy~\cite{ZnidaricProsenPrelovsek2008, luitz_many-body_2015, mace_multifractal_2019, sierant_challenges_2022}---fail to yield a consistent estimate of the critical disorder strength $W_{\rm MBL}$. As the system size $L$ is increased, this putative critical point drifts toward higher disorder strengths indicating that larger systems require stronger randomness to localize. This raised concerns that the true transition in the thermodynamic limit might occur only at infinitely strong disorder. In other words, the MBL regime seen in finite chains might be a finite-size effect rather than an asymptotically stable phase~\cite{suntajs_quantum_2020, suntajs_ergodicity_2020,
kiefer-emmanouilidis_evidence_2020, 
sierant_polynomially_2020,sels_dynamical_2021, sels_bath-induced_2022, kiefer-emmanouilidis_slow_2021,  sierant_challenges_2022, sels_thermalization_2023,Beraetal2017, WeinerEversBera2019, EversModakBera2023}. 

This possibility has been further supported by a simple theoretical argument, which suggests that the MBL phase may be unstable with respect to a runaway avalanche thermalization mechanism. This instability would 
be  triggered by rare regions in the system where the disorder is anomalously weak~\cite{de_roeck_stability_2017, luitz_how_2017, thiery_many-body_2018, goihl_exploration_2019, Crowley2020, sels_bath-induced_2022, leonard_probing_2023, peacock_many-body_2023, szoldra_catching_2024}. The basic idea is that rare regions with atypically weak disorder inevitably occur in sufficiently large systems, albeit with small but finite probability. In an otherwise localized system, these rare regions behave like small thermal bubbles that can resonantly couple to nearby degrees of freedom. When the coupling strength between a thermal bubble and its surroundings exceeds the bubble’s internal level spacing, the neighboring degrees of freedom become thermalized and are effectively absorbed into the bubble. As the bubble grows, its internal level spacing decreases, which in turn makes it easier to thermalize additional nearby regions. This process can initiate a self-sustained avalanche, wherein the thermal region expands through the system, eventually destroying localization and restoring ergodic behavior. 

Yet, the system sizes accessible through numerical simulations are too small to accommodate these statistically rare, locally thermal regions, making hard to probe the consequences of their existence. To address this limitation, several studies have investigated how disordered quantum chains in the MBL phase respond when coupled to an artificially introduced thermal region at one end~\cite{luitz_how_2017, Suntajs2022, peacock_many-body_2023, leonard_probing_2023, morningstar_avalanches_2022, Ha2023, szoldra_catching_2024, Pawlik2024, Colmenarez2024, pawlik2025, Sajid2025}.
The core idea is that if the equilibration time of the localized region---quantified by its slowest decay rate---is long enough, a thermal bubble cannot trigger the thermal avalanche. This approach sets a lower bound on the disorder strength required for the MBL phase to remain stable, and the critical disorder thus estimated turns out to be much stronger than those previously obtained using standard observables~\cite{luitz_many-body_2015, Pietracaprina2018, mace_multifractal_2019, Laflorencie2020, sierant_polynomially_2020}.

In parallel, many recent numerical studies on finite-size systems have revealed the presence and significance of rare many-body resonances, even deep within regions of the phase diagram that were previously considered to be in the MBL phase. These hybridizations take place among `localized' many-body eigenstates that differ by a number of spin flips within a finite spatial range in real space~\cite{imbrie_diagonalization_2016, VillalongaArxiv, Garratt2021, crowley_constructive_2022, Gopalakrishnan2015,HerbrychMierzejewskiPrelovsek2022, HerbrychPrelovsek2025}. Importantly, these resonant eigenstates are nearly degenerate, with energy differences that decrease exponentially with the spatial range of the spin reconfigurations involved~\cite{Gopalakrishnan2015}. The hybridization of eigenstates differing by an extensive number of spin flips---known as long-range resonances---is central to formal proofs concerning the stability of the MBL phase~\cite{Imbrie2016, imbrie_diagonalization_2016, DeRoeck2024} and also believed to play a key role in destabilizing the MBL phase. They have been indirectly observed in microscopic models~\cite{Gopalakrishnan2015, Geraedts2016, Khemani2017, Colmenarez2019, villalonga_characterizing_2020} and, more recently, identified more directly through numerical probes~\cite{morningstar_avalanches_2022, VillalongaArxiv, Garratt2021, Garratt2022, Ha2023, Kjll2018,HerbrychMierzejewskiPrelovsek2022, PrelovsekHerbrychMierzejewski2023, HerbrychPrelovsek2025}. 

It is important to emphasize that these two destabilization mechanisms---many-body resonances and avalanches---are not necessarily independent. In particular, Ref.~\cite{Ha2023} argues that avalanches primarily propagate through a certain type of strong, rare (near-)resonances, suggesting a close connection between the two.  However, the precise relationship between avalanches and long-range resonances remains elusive. Besides, the growing body of evidence for many-body resonances places them at the center of the discussion, as they seem to play a crucial role in determining the physical properties of the MBL transition, its stability, and its associated crossover regimes~\cite{Khemani2017, Crowley2020, villalonga_characterizing_2020, De_Tomasi2021-ua, Garratt2021, morningstar_avalanches_2022, long_phenomenology_2023, Ha2023}. Despite this, a comprehensive microscopic understanding of the origin, the statistics, and the effect of long-range resonances is still lacking. 

As a result, recent studies have shifted focus to the properties of many-body resonances in strongly disordered regimes---regions previously identified as many-body localized using standard observables. Analyses of the scaling of extreme values of spectral diagnostics~\cite{morningstar_avalanches_2022, Ha2023}, show that system-wide resonances, initially rare deep in the localized phase, become prevalent across all length scales in the ergodic phase, given large enough system sizes. Under strong disorder, longitudinal spin-spin correlations exhibit broad distributions~\cite{colbois_interaction-driven_2024, colbois_statistics_2024}; rare events with unusually large correlations lead to an algebraic decay of the average with real space distance, even though the typical decay remains exponential. In Hilbert space, new observables have been introduced as proxies for the probability of decorrelation from a randomly initialized configuration. This probability---and its proxy---are dominated by rare disorder realizations that produce anomalously large matrix elements connecting distant configurations in the Hilbert space graph~\cite{biroli_large-deviation_2024}.

\subsection{Summary of the main results}

In the present work, we build upon the approach introduced in Ref.~\cite{biroli_large-deviation_2024}, previously applied to the out-of-equilibrium phase diagram of the random-field Ising model in a transverse field (also known as the Imbrie model), where the absence of diffusion at strong disorder has been rigorously established under minimal assumptions~\cite{imbrie_diagonalization_2016, Imbrie2016, DeRoeck2024}. We extend this method to the more debated case of the U(1) symmetric random-field Heisenberg chain, a model that has been central to most numerical investigations of the MBL transition~\cite{pal_many-body_2010, DeLuca2013, luitz_many-body_2015, Gray2018, doggen_many-body_2018, mace_multifractal_2019, sierant_polynomially_2020, Laflorencie2020, sierant_thouless_2020, abanin_distinguishing_2021, morningstar_avalanches_2022, EversBera2025}. Specifically, we study the general random-field XXZ chain and present an updated phase diagram at high energy (in the middle of the many-body spectrum) in the disorder–interaction $(W, \Delta)$ plane, accounting for the role of long-range resonances. 

Leveraging an analogy with a class of mean-field disordered glassy systems, our method aims to evaluate the statistical weight of rare events—specifically, those that span a large portion of the Hilbert space. The central quantity we focus on is the probability that a system initialized in a random configuration $\ket{0}$ at time $t = 0$ is found in a configuration $\ket{f}$, located far from $\ket{0}$ in Hilbert space (with a measure given by the number of 'hops' present in the Hamiltonian), at infinite time. These probabilities are estimated via the amplitudes of the propagators $|\mathcal{G}_{0f}|^2$, which are significantly easier to compute numerically. Rare resonances are identified as outliers in the probability distributions of these propagators, corresponding to pairs of resonant configurations separated by large distances on the Hilbert space graph.

Exploiting the analogy with classical disordered models, we introduce a Lagrange multiplier, which formally plays the role of an effective-temperature, 
which controls the influence of extreme outliers and enables us to isolate their contribution to transport. This reveals three distinct regimes, illustrated in the phase diagram of Fig.~\ref{fig:phase_diagram}: (i) an ergodic phase in which a wavepacket initialized in $\ket{0}$ spreads uniformly over configurations at large Hilbert space distances; (ii) an intermediate regime where delocalization is driven by rare, disorder-dependent long-range resonances that appear only in atypical disorder realizations; and (iii) a robust many-body localized phase, where such resonances are neither strong nor frequent enough to destabilize localization. Importantly, numerically accessible typical samples in the intermediate regime do not exhibit the system-wide resonances responsible for asymptotic delocalization. Nevertheless, our approach, guided by the analogy with mean-field glassy systems, captures the effect of these rare contributions in the large $L$ limit.

We also investigate the spatial structure of these rare events within Hilbert space. In the MBL regime, rare resonances become increasingly short-ranged: the portion of Hilbert space accessible from the initial configuration progressively shrinks with disorder, and transmission remains confined near the initial state. In contrast, in the ergodic regime, uniform delocalization is recovered only at large distances in Hilbert space. Our results thus provide a Hilbert space-based complement to real space and spectral probes of MBL, highlighting the crucial role of rare, system-wide resonances in driving finite-size delocalization.

The emergence of heterogeneous resonant pathways facilitating delocalization is further confirmed by examining the structure of many-body eigenstates, which exhibit pronounced amplitude fluctuations in the intermediate phase, both between rare and typical disorder realizations and between distinct branches of the Hilbert space graph.

We further explore these differences by visualizing the transmission pathways on the Hilbert space graph using techniques originally developed for quantum transport in mesoscopic systems~\cite{Pichard1991, Marko2010, lemarie_glassy_2019}. This graphical perspective offers new insight into the destabilization of the MBL phase at finite sizes, interpreting it as the emergence of resonant transmission paths that are abundant in the ergodic regime but become increasingly rare and short-ranged deep in the localized phase.

Our results provide new numerical evidence that the introduction of interactions induces delocalization of the Anderson insulator through genuinely non-perturbative mechanisms. Notably, even at very small interaction strengths $\Delta$, the critical disorder strength separating ergodic from non-ergodic phases remains finite. This implies the existence of a broad region of the phase diagram at small but finite disorder where the addition of an infinitesimal interaction is sufficient to destroy the Anderson insulator at $\Delta=0$, and possibly even restore a full ergodic behavior. This observation is consistent with the behavior of longitudinal correlation functions and with the recently updated phase diagram of the XXZ model at mid-spectrum energies reported in Refs.~\cite{colbois_interaction-driven_2024, colbois_statistics_2024}.

\subsection{Outline}

The paper is organized as follows.
The model and the main representations of the basis states considered are introduced in \autoref{sec:model}. The method we developed to characterize rare resonances is described in detail in \autoref{sec:method}. Our main results are presented in \autoref{sec:results}; they include the behavior of key observables, an updated finite-size phase diagram of the model obtained using our approach, and a range of additional analyses, across the distinct regimes identified. Technical details and supplementary information that support the main discussion are provided in four appendices.

\section{The Model}
\label{sec:model}

We consider the XXZ spin-$\tfrac{1}{2}$ chain with interaction anisotropy parameter $\Delta$ ($0 \leq \Delta \leq 1$~\footnote{The sign of the interaction is not relevant at high energy, see for instance Ref.~\cite{Lin2018}}), and a random magnetic field $h_i$ uniformly drawn from $[-W, W]$. The Hamiltonian, for a chain of $L$ sites, is given by 
\begin{equation}
    \hat{\mathcal{H}} = \sum^L_{i=1} \left(\hat S_i^x \hat S_{i+1}^x+\hat S_i^y \hat S^y_{i+1}+\Delta \hat S_i^z \hat S^z_{i+1} + h_i \hat S_i^z \right) \;.
    \label{eq:XXZ_SpinsHamiltonian}
\end{equation} 
For $\Delta=1$, we recover the strongly interacting random-field Heisenberg chain, which serves as the paradigmatic model of the MBL transition.
This model is equivalent, through a Jordan-Wigner transformation, to that of interacting spinless fermions hopping on a chain of $L$ sites, written as 
\begin{equation}
\hat{\mathcal{H}} = \frac{1}{2}\sum^L_{i=1} \left(\hat c^\dagger_i \hat c_{i+1}^{\vphantom{\dagger}}+ {\rm{h.c.}} +2 \Delta \hat n_i \hat n_{i+1} + 2 h_i \hat n_i \right)\;,
\label{eq:XXZ_FermionsHamiltonian} 
\end{equation} 
with $\hat c_i^\dagger (\hat c_{i}^{\vphantom{\dagger}})$ the creation (annihilation) fermionic operators and $\hat{n}_i = \hat{c}_i^\dagger \hat{c}_{i}^{\vphantom{\dagger}}$ the local occupation number operator. 

In both cases, setting $\Delta = 0$ corresponds to the non-interacting limit, which---when described in the language of spinless fermions---is known to exhibit Anderson localization for any nonzero value of the disorder strength $W$. In this regime, the localization length of the single-particle Anderson-localized orbitals diverges as $1/W^2$ in the limit $W \to 0$. The Hamiltonian preserves the $U(1)$ symmetry thus conserving the magnetization in the $z$ direction---or total particle number---of the system. In the present work, we restrict to studying the sector of zero magnetization, or half-filling, with periodic boundary conditions.

\subsection{The Hilbert space picture}
\label{sec:HilbertSpace}

The Hilbert space structure of  
the model described in the previous section has been extensively studied~\cite{Altshuler1997, Gornyi2005, DeLuca2013, Biroli2017, Logan2019, Tikhonov2021, roy_fock-space_2024}
Any quantum state $\ket \Psi$ can be expressed as superpositions of the many-body basis states in the form
\begin{equation}
\ket{\Psi} = \sum_{I=1}^\mathcal{N} \psi_I \ket{I} \;,
\label{eq:FS-state}
\end{equation} 
where $\{ \ket{I}\}$ is the appropriate set of many-body basis states. We will focus on two bases: spin configurations $\{ \ket I\}_S$
and the Anderson basis $\{ \ket I\}_A$. These are the most relevant for the model in~\autoeqref{eq:XXZ_SpinsHamiltonian}, as each diagonalizes the Hamiltonian in specific limits where localization is well-understood, the infinite-disorder limit $W \to \infty$ for $\{ \ket I\}_S$, and the non-interacting limit $\Delta = 0$ for $\{ \ket I\}_A$.

Once the computational basis is chosen,  the Hamiltonian (\ref{eq:XXZ_SpinsHamiltonian})-(\ref{eq:XXZ_FermionsHamiltonian}) can be recast as
\begin{equation} 
    \hat{\mathcal H} = \sum_{I} E_I | I \rangle \langle I | + \sum_{\langle IJ \rangle} T_{I J} | I \rangle  \langle J | \;.
    \label{eq:TBH}
\end{equation} 
This is now a tight-binding model of a fictitious single-particle hopping on a graph. The number of vertices $\mathcal{N}$ in this high-dimensional graph represents the possible basis states associated to the problem (i.e., the Hilbert space volume). In a system with $L$ spins or fermions, the dimension of Hilbert space grows exponentially with $L$. In the zero magnetization (or half-filling) sector it is given by $\mathcal{N} = \binom{L}{L/2}\approx 2^L\sqrt{\frac{2}{\pi L}}$. 
Each of these states have an associated `on-site' energy $E_I$, given by the expectation value of the diagonal part of the Hamiltonian in the chosen basis, which is random through the fields $h_i$. The edges of the Hilbert space graph joining different vertices are given by the non-zero off-diagonal elements $T_{IJ}$, which gives the `hopping' amplitudes of the fictitious particle between two neighboring vertices.

Each vertex---basis state---is connected to a large number of other vertices via local interactions, such as spin flips or particle hoppings. Even though the underlying physical system is one-dimensional, the connectivity in Hilbert space is governed by the action of local operators on basis states. This gives rise to a complex, high-dimensional network in which each vertex (or basis state) is connected to a number of others that scales with $L$ or a power of $L$, depending on the chosen computational basis (see below).

This reformulation provides a framework in which spectral and dynamical properties of the many-body quantum disordered system can be understood in terms of the effective single-particle problem on an underlying complex graph structure. The formulation of the many-body localization problem in this Hilbert space graph has been exploited in several previous works (see for instance Refs.~\cite{Roy2020,Roy2021,sutradhar_2022}, and the review Ref.~\cite{roy_fock-space_2024}). Generically, this Hilbert space graph exhibits an effective infinite dimensionality: the number of vertices at a given distance from a reference vertex grows exponentially with that distance.

The main differences compared to the Anderson localization problem on high-dimensional graphs~\cite{Tikhonov2016, Tikhonov2019-ku, Tikhonov2019-tb, Tikhonov2021} lie in the scaling and structure of the problem. In the many-body case, the corresponding Hilbert space (or Fock-space) graph is deterministic and contains loops of various lengths. Moreover, the many-body problem features strongly correlated on-site energies ${ E_I }$ and matrix elements ${ T_{IJ} }$, further complicating the analysis compared to standard single-particle Anderson localization on tree-like structures such as the Bethe lattice~\cite{roy_fock-space_2024,Roy2020,Roy2020b}.

In the following sections we introduce and explain in detail the two computational bases used to investigate the random-field XXZ model.

\subsection{Spin basis}
\label{sec:spinbasis}
The first computational basis we will be working with is the simultaneous eigenstates of the $S_i^z$ operators: $\{ \ket{I}\}_S$ are just classical Ising spin configurations on the chain with zero global magnetization in the $z$. For example, for $L=6$ one of its basis states is $\ket{\uparrow \downarrow \downarrow \uparrow \uparrow \downarrow}$. The states in the set $\{\ket{I}\}_{S}$ are the eigenstates of $\mathcal{\hat{H}}_0= \sum^L_{i=1} (\Delta \hat S_i^z \hat S^z_{i+1} + h_i \hat S_i^z)$ with eigenenergy $E_I=\langle I|{\mathcal{\hat{H}}_0}|I\rangle$. In this basis the off-diagonal elements, $T_{I J}$, are not random and are in fact all set to $T_{I J} = 1/2$ for those basis states---$I$ and $J$---that differ by the exchange of two nearest neighbor spins forming domain walls in the chain (we will often speak in terms of `spin flips', keeping in mind that they occur in pairs: a single nearest-neighbor exchange corresponds to two simultaneous spin flips due to the Hamiltonian’s $U(1)$ symmetry).
It is clear that the connectivity of the vertices is not constant and depends on the number of domain walls in each basis state. For example, the N\'eel state $\ket{\uparrow \downarrow \uparrow \downarrow ... \uparrow \downarrow}$, and its time-reversed symmetric, 
are maximally connected as they both have $L$ domain walls, but these two N\'eel states are distant from each other by $L/2$ spin flips. Instead, the state $\ket{\uparrow \uparrow ...\uparrow \downarrow \downarrow ...\downarrow}$ has only two domain walls (with periodic boundary conditions) thus having just two neighboring vertices in the Hilbert space graph. The average degree of the Hilbert space graph is $L/2$.

Through a Jordan-Wigner transformation, these basis states can be mapped to Fock states of spinless fermions hopping on a lattice at half-filling. For example, the N\'eel state introduced above becomes $\ket{1\,0\,1\,0\,\dots\,1\,0}$ in this representation. Throughout the text, we will refer to both representations as the \textit{spin basis}, and we will switch between the conventional spin configuration notation (e.g., $\ket{\uparrow\downarrow\uparrow\downarrow\dots\uparrow\downarrow}$) and the corresponding \textit{Fock state} notation (e.g., $\ket{1\,0\,1\,0\,\dots\,1\,0}$) whenever it is convenient for clarity.
\subsection{Anderson basis}
\label{sec:AndersonBasis}
Another natural choice that, by construction, captures the fact that the system remains localized at any finite disorder in the absence of interactions is the Anderson basis of single particle localized orbitals, described below (see also Refs.~\cite{Prelovsek2018,Thomson2018,Detomasi2019,Laflorencie2020,Colbois2023}). This second computational basis corresponds to the eigenstates of the non-interacting part of the Hamiltonian in Eq. (\ref{eq:XXZ_FermionsHamiltonian}), defined as $\hat{\mathcal{H}}_{\rm NI} = \frac{1}{2}\sum^L_{i=1} (\hat c^\dagger_i \hat c_{i+1}^{\vphantom{\dagger}}+ {\rm{h.c.}} + 2 h_i \hat n_i )$. 
This $\hat{\mathcal{H}}_{\rm NI}$ defines the Anderson model of a spinless fermion hopping on a chain, in a random potential. Eigenpairs of this single-particle Hamiltonian stem from $\hat{\mathcal{H}}_{\rm NI} \phi_\alpha = \epsilon_\alpha \phi_\alpha$, and are known to be Anderson localized. 
The single-particle orbitals $\phi_\alpha$ are used to construct the unitary transformation that diagonalizes the non-interacting Hamiltonian $\hat{\mathcal{H}}_{\rm NI}$, 
\begin{equation}
    \hat{\mathcal{H}}_{\rm NI} = \sum_{\alpha} \epsilon_\alpha \hat b^\dagger_{\alpha}\hat  b_{\alpha}^{\vphantom{\dagger}}  \;, \qquad \hat b_\alpha = \sum^L_{i=1} \phi_\alpha (i) \hat c_i 
    \;.
    \label{eq:unitary-transformation}
\end{equation} 
The basis states are thus defined as the occupation numbers of the $L$ single particle orbitals, i.e.,  the tensor product of the simultaneous eigensates of the number operators $\hat b_\alpha^\dagger \hat b_\alpha^{\vphantom{\dagger}} $. The Anderson basis $\{ \ket{I} \}_A$, is thus built as
\begin{equation}
    \ket{I} = \prod_{\alpha=1}^L (\hat b_\alpha^\dagger)^{n_\alpha} \ket{\varnothing}\;, \qquad E^{\rm NI}_I = \sum_\alpha n_\alpha \epsilon_\alpha \;,
    \label{eq:AndersonNI-basis-states}
\end{equation} 
where $\ket{\varnothing}$ is the vacuum state, and $n_\alpha$ is the fermion occupation number for the $\alpha$-th orbital, that in our case for half-filling fulfills $\sum_\alpha  n_\alpha = L/2$. $E^{\rm NI}_I$ is the non-interacting energy associated to the basis state $\ket{I}$. 
The interacting part, $\sum_{i=1}^L \hat{n}_i \hat{n}_{i+1}$, transforms under the unitary transformation defined in Eq.~(\ref{eq:unitary-transformation}) into
\begin{eqnarray}
    \hat{V}  &= &\sum_{\alpha \beta \gamma \delta} V_{\alpha \beta \gamma \delta}(\Delta)~ \hat{b}^\dagger_\alpha \hat{b}^\dagger_\beta \hat{b}_\gamma^{\vphantom{\dagger}}  \hat{b}_\delta^{\vphantom{\dagger}}  
    \end{eqnarray}
    with  
\begin{eqnarray}   
    V_{\alpha \beta \gamma \delta}(\Delta) &=& \Delta  \sum_{i=1}^L \phi^*_\alpha(i) \phi^*_\beta(i+1) \phi_\gamma(i+1) \phi_\delta(i) \;.
    \label{eq:fock_nondiag}
\end{eqnarray} 
We can unfold Eq.~(\ref{eq:fock_nondiag}) to consider explicitly the three non-vanishing contributions to $V_{IJ} = \bra{J} \hat{V} \ket{I}$, according to the groupings of indices $\alpha, \beta, \gamma$ and $\delta$: 
\textbf{(i)}
    $\alpha = \delta$ and $\beta = \gamma$, gives a diagonal contribution in the form
\begin{equation}
    \hat V_d = 2 \sum_{\alpha > \beta} \left(V_{\alpha\beta\beta\alpha} - V_{\beta \alpha \alpha \beta}\right) \hat{n}_\beta \hat n_\alpha \;, 
\end{equation} 
making the on-site energy of the associated basis state $\ket{I}$ to be $E_I = E^{\rm NI}_I + \bra{I} \hat{V}_d \ket{I}$. 
The other two non-zero contributions to $V_{IJ}$ come from the off-diagonal entries $\ket{I} \neq \ket{J}$, that will construct the hopping terms $T_{IJ}$.  The second contribution comes from \textbf{(ii)} $\alpha = \delta, \beta \neq \gamma $:
\begin{equation}
    \hat V_1 = 2 \sum_{\alpha \beta \gamma} (V_{\alpha \beta \gamma \alpha} - V_{\beta \alpha \alpha \gamma}) \hat n_\alpha \hat{b}^\dagger_\beta \hat{b}_\gamma^{\vphantom{\dagger}}  \;, 
    \label{eq:V1}
\end{equation} 
where the occupation of the $\gamma$-th orbital has been replaced by a new occupation in the orbital labeled by $\beta$, given that $n_\alpha = n_\gamma = 1$ and $n_\beta = 0$. This `assisted' hopping connects each vertex with $L^2/4$ nearest-neighbors. The third and final contribution corresponds to \textbf{(iii)} $\alpha \neq \beta \neq \gamma \neq \delta$:
\begin{equation}
    \hat{V}_2 = \sum_{\substack{\alpha > \beta \\ \gamma > \delta}} (V_{\alpha \beta \gamma \delta} + V_{\beta \alpha \delta \gamma} - V_{\alpha \beta \delta \gamma} - V_{\beta \alpha \gamma \delta}) \hat{b}^\dagger_\alpha  \hat{b}^\dagger_\beta \hat{b}_\gamma^{\vphantom{\dagger}} \hat{b}_\delta^{\vphantom{\dagger}} \;.
    \label{eq:V2} 
\end{equation} 
This contribution is non-zero provided that $n_\delta = n_\gamma = 1$ and $n_\alpha = n_\beta = 0$. This contribution adds $L^{4}/64 - L^3/16 +L^2/16$ neighboring vertices to each single vertex. Thus, the Fock-space graph has constant connectivity $z = L^2/4 \; [  \left(L/2-1 \right)^2/4 + 1 ]$, for each basis state. Note that unlike the spin basis, here the $T_{IJ}$ are all random, and broadly distributed~\cite{Prelovsek2018}.

\section{The method}
\label{sec:method}
\subsection{The delocalization probability}\label{sec:deloc_prob}
A suitable order parameter for detecting delocalization is the probability that a system, initially prepared in the basis state $\ket{0}$, is found in distant basis states $\ket{f}$ after infinite-time.

We select the initial basis state $\ket{0}$ such that the expectation value of the energy lies in the middle of the many-body spectrum, in order to probe localization in this highly-excited regime. This is done by selecting the diagonal elements of the Hamiltonian, that correspond to the random part $\hat {\mathcal{H}}_0$ in the spin basis. These energies  $E_0 = \bra 0 \hat{\mathcal{H}}_0 \ket 0$ are extensive and normal distributed with zero mean and a variance proportional to $L$. As a result, $E_0/L \sim 1/\sqrt{L}$.

Basically, we want to study the long-time spreading of the wave-packet starting from $\ket{0}$, and examine whether it can reach distant configurations on the Hilbert space graph. For most of the work reported here, and unless otherwise stated, we will consider the distant basis states $\ket{f}$ to be states maximally uncorrelated with the initial one $\ket{0}$. The criteria of selection for these `distant' or maximally uncorrelated configurations will be based on the overlap (as defined below) between 
the basis states $\ket{0}$ and $\ket{f}$, which depends on the choice of basis. In the spin basis the overlap is defined as
\begin{equation}
    q^S_{0f} = \frac{4}{L} \sum_{i=1}^L S_i^z(0) S^z_i(f) 
    \;, 
    \label{eq:spin_overlap}
\end{equation} 
where $S_i^z \in \left\{ \pm \frac{1}{2} \right\}$ is the eigenvalue of the spin operator in the $z$ direction $\hat{S}_i^z$ acting on the $i$-th site of the chain. 
Whereas, in the Anderson basis the basis states are occupation numbers, i.e. bit-strings made of 1s and 0s and the overlap reads
\begin{equation}
    q^A_{0f} = \frac{1}{L} \sum_{\alpha=1}^L (2 n_\alpha(0) -1) (2 n_\alpha(f) - 1)
    \;,
    \label{eq:anderson_overlap}
\end{equation} 
where $n_\alpha$ is the occupation number of the $\alpha$-th single-particle orbital, i. e. the eigenvalues of the $\hat{b}^\dagger_\alpha \hat{b}_\gamma \hat{b}_\alpha$ operators. 

In both cases, $-1 \le q^{S,A}_{0f} \le 1$, and basis states that are completely uncorrelated with the initial state are characterized by zero overlap. In the spin basis this corresponds to flipping half of the spins between $\ket{0}$ and $\ket{f}$. In the Anderson basis this implies that half of the $L/2$ spinless fermions have transitioned to different single-particle orbitals.
Such overlaps are closely related to the concept of \emph{imbalance}, 
a measurement frequently used in the study of many-body quantum systems to quantify the degree to which memory of an initial basis state is retained over time~\cite{sierant_challenges_2022, richter_pal_2022}.
Based on the overlap value $q$, with respect to the initially prepared basis state $\ket{0}$, we define two sets of basis states:
\begin{equation}
\begin{aligned}
    \mathcal{S}_0(q) &= \left\{ \ket{f} \; : \; q^S_{0f} = q \right\} \;, \\
    \mathcal{A}_0(q) &= \left\{ \ket{f} \; : \; q^A_{0f} = q \right\}\;,
\end{aligned}
\label{eq:set_targets}
\end{equation}
for the spin and Anderson bases, respectively. In the following, we will primarily focus on $\mathcal{S}_0(q = 0)$ and $\mathcal{A}_0(q = 0)$, which we will collectively denote by the symbol $\mathcal{E}$ and refer to as the `equator' of the Hilbert space graph. It will be clear from the context whether $\mathcal{E}$ refers to $\mathcal{S}_0(q = 0)$ or $\mathcal{A}_0(q = 0)$. Whenever ambiguity arises, we will explicitly add the superscript $S$ or $A$ to indicate that the quantity is measured in the spin or Anderson basis, respectively. In the zero-magnetization (or half-filling) sector, the equator possesses $\mathcal{N}_\mathcal{E} = \binom{L/2}{L/4}^2 \sim 2^L/ (\pi L /4) $ target basis states, for any initial condition $\ket{0}$ chosen. 
The fact that the volume of the equator of the graph scales asymptotically with $L$ in the same way as the volume of the entire graph, $\mathcal{N}_\mathcal{E}/\mathcal{N} \sim \sqrt{8/(\pi L)}$, is a hallmark of infinite-dimensional geometries. This justifies our primary choice of $q^{S,A}_{0f}$ for the analysis of localization:
If the wave packet can reach $\mathcal E$ in the infinite-time limit, it can essentially reach any configuration on the Hilbert space graph that is compatible with energy conservation.
\begin{figure*}[t]
\vspace{0.5cm}
\centering
  \centering
  \includegraphics[width=.32\linewidth]{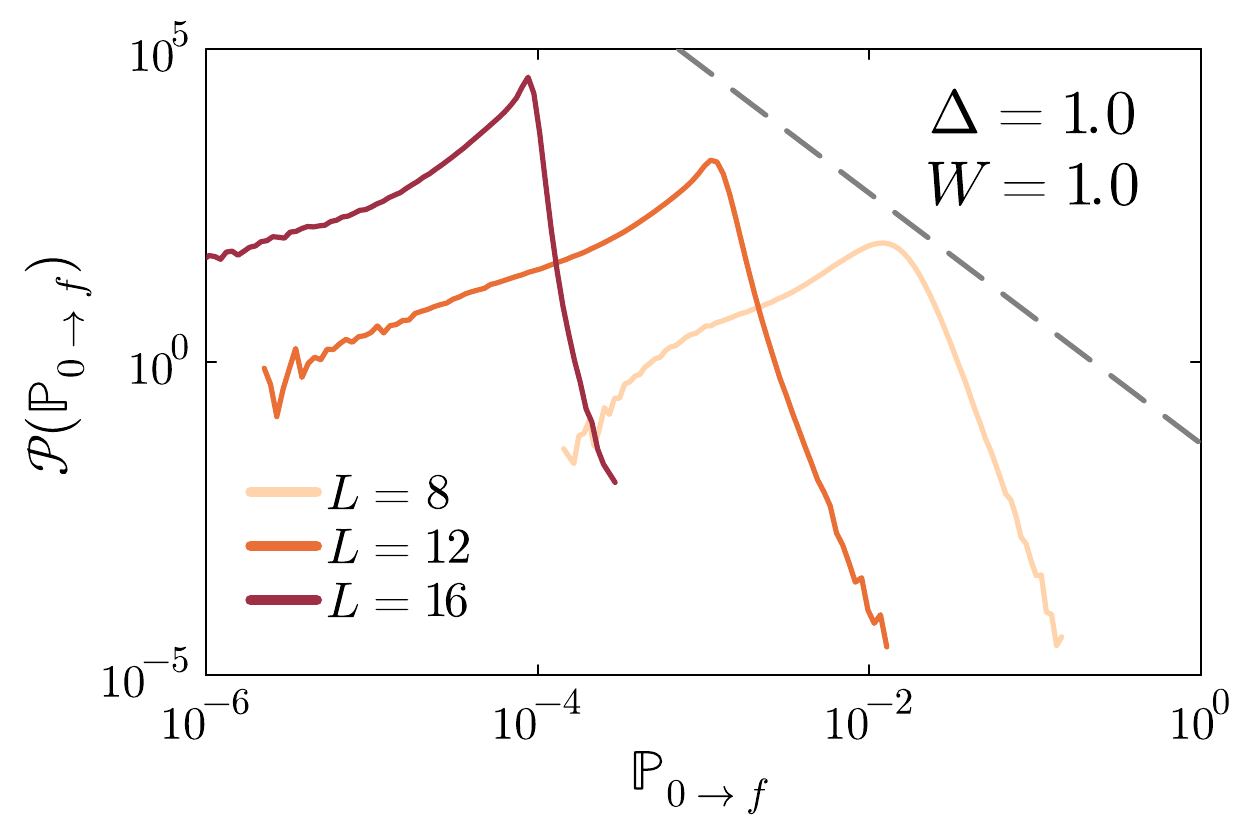}
  \includegraphics[width=.32\linewidth]{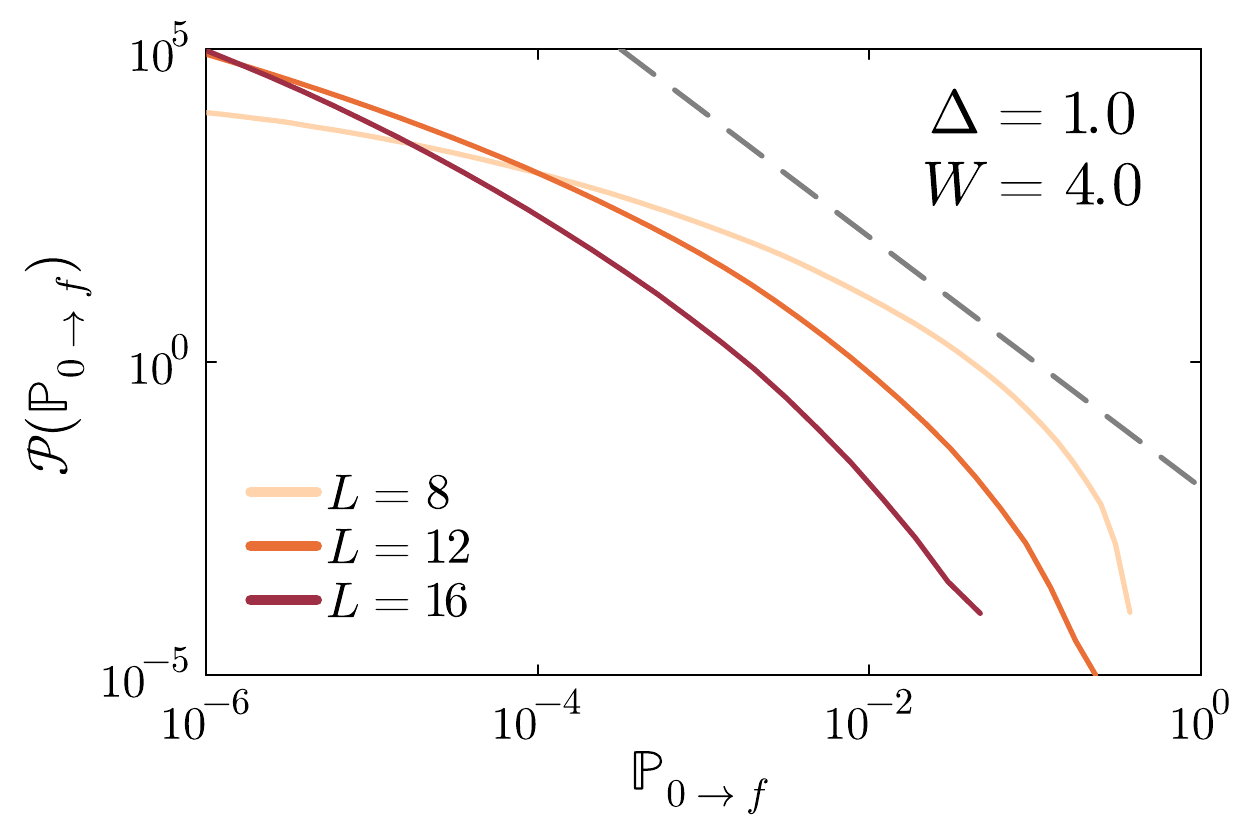}
  \includegraphics[width=.32\linewidth]{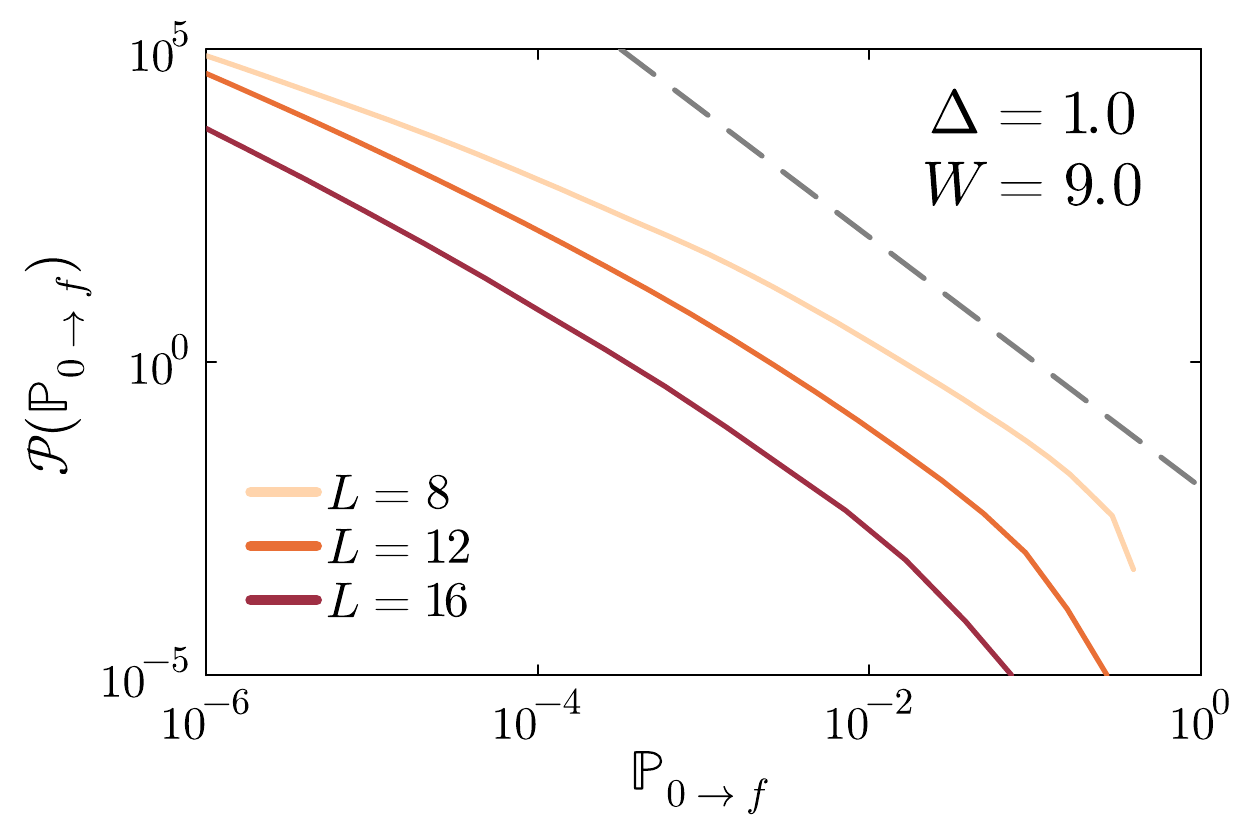}
\caption{Probability distribution function for the infinite time 
probabilities, Eq.~(\ref{eq:prob}), of finding a system, 
initially prepared in the basis state $\ket{0}$, 
in a basis state $\ket{f} \in \mathcal{E}$. 
The gray dashed line indicates a reference power-law decay with exponent 2. For weak disorder ($W = 1$, left panel), the distribution is relatively narrow. As the disorder strength increases ($W = 4$ and $W = 9$, center and right panels), the distribution broadens significantly. The total number of samples used to compute these distributions is $N_{\rm tot} = N_0 \times N_S \times \mathcal{N}_\mathcal{E}$, where $N_0 = 2^{L/2-2}$, and $N_S = 500,\, 5\times10^3,\, 5\times10^4$ for $L = 8, 12, 16$, respectively.}
\label{fig:pdf_ProbToEq}
\end{figure*} 

The asymptotic probability that a system prepared
in $\ket{0}$ reaches a basis state in the equator, $\ket{f} \in \mathcal{E}$, 
can be calculated exactly in terms of the eigenstates $\ket{n}$ of the Hamiltonian $\hat{\mathcal{H}}$ as
\begin{equation}
\mathbb{P}_{0 \to f} = \lim_{t \to \infty} |\langle f| e^{-i\mathcal{\hat{H}} t} | 0 \rangle |^2 =  \sum_{n} |\langle f | n \rangle  \langle n | 0\rangle|^2 
\;.
\label{eq:prob}    
\end{equation}
Our key observable of interest is the overall \emph{delocalization probability} that quantifies the likelihood that a system, initially prepared in the state $\ket{0}$, evolves into \emph{any} state belonging to the equator:
\begin{equation}
    \mathbb{P}_{\mathcal{E}} = \sum_{f \in \mathcal{E}} \mathbb{P}_{0 \to f} \;.
    \label{eq:delocProb}
\end{equation}
The sum runs over an exponentially large number of terms in $L$. This object is a random variables that depends on the disorder realization and on the choice of the initial state. In a localized phase, we expect the typical value of the delocalization probability to vanish with increasing system size, and to go to 1 in the extended phase. 

Using exact diagonalization of the full spectrum, we calculated the probability density function (PDF) of $\mathbb{P}_{0 \to f}$ for finite-sized systems for $\Delta = 1$. The results are shown in Fig.~\ref{fig:pdf_ProbToEq}. 
To construct this PDF, we selected 
$N_0$ initial states--with an expectation value of energy located near the center of the energy spectrum--for each of the $N_S$ disorder realizations of the random fields $\{ h_i \} $. For each initial state, we computed $\mathbb{P}_{0 \to f}$ for all ${\cal{N}}_{\mathcal{E}}$ basis states with zero overlap from the initial one (i.e., at the equator of the Hilbert space graph). Consequently, each pdf is built from a total of $N_S \times N_0 \times \mathcal{N}_\mathcal{E}$ data points. The specific values of $N_S $, $N_0 $, and $\mathcal{N}_\mathcal{E}$ used for each system size are provided in the caption of Fig.~\ref{fig:pdf_ProbToEq}.

For weak disorder ($W = 1$), the distribution is sharply peaked. 
Consequently, the sum Eq.~\eqref{eq:delocProb} is primarily governed by the bulk of the distribution 
$\mathbb{P}_{0 \to f}$, with dominant contributions coming from its peak. This makes $\mathbb{P}_{\mathcal{E}}$ a self-averaging quantity. The peak of the distribution of $\mathbb{P}_{0 \to f}$ shifts  toward smaller values as the system size $L$ increases. This indicates that as $L$ is increased an exponentially increasingly large number of terms must contribute to the sum Eq.~\eqref{eq:delocProb} in order to 
have that $\mathbb{P}_\mathcal{E}$  is of order 1.

As disorder is increased, instead, the tails of the distribution of $\mathbb{P}_{0 \to f}$ decay increasingly slowly. 
At strong enough disorder this decay becomes slower than a square power-law, shown as a reference with a dashed-gray line.
 As a result, the sum in Eq. (\ref{eq:delocProb})  becomes dominated by contributions from the tails of the distribution—that is, by rare events. These rare outliers from the tails 
correspond precisely to the system-wide strong resonances between states $\ket{0}$ and $\ket{f}$, which have been extensively  discussed in the recent literature~\cite{Khemani2017, villalonga_characterizing_2020,
Garratt2021, De_Tomasi2021-ua, crowley_constructive_2022, long_phenomenology_2023,  morningstar_avalanches_2022, Ha2023}. 

\begin{figure}[b!]
\centering
  \includegraphics[width=\linewidth]{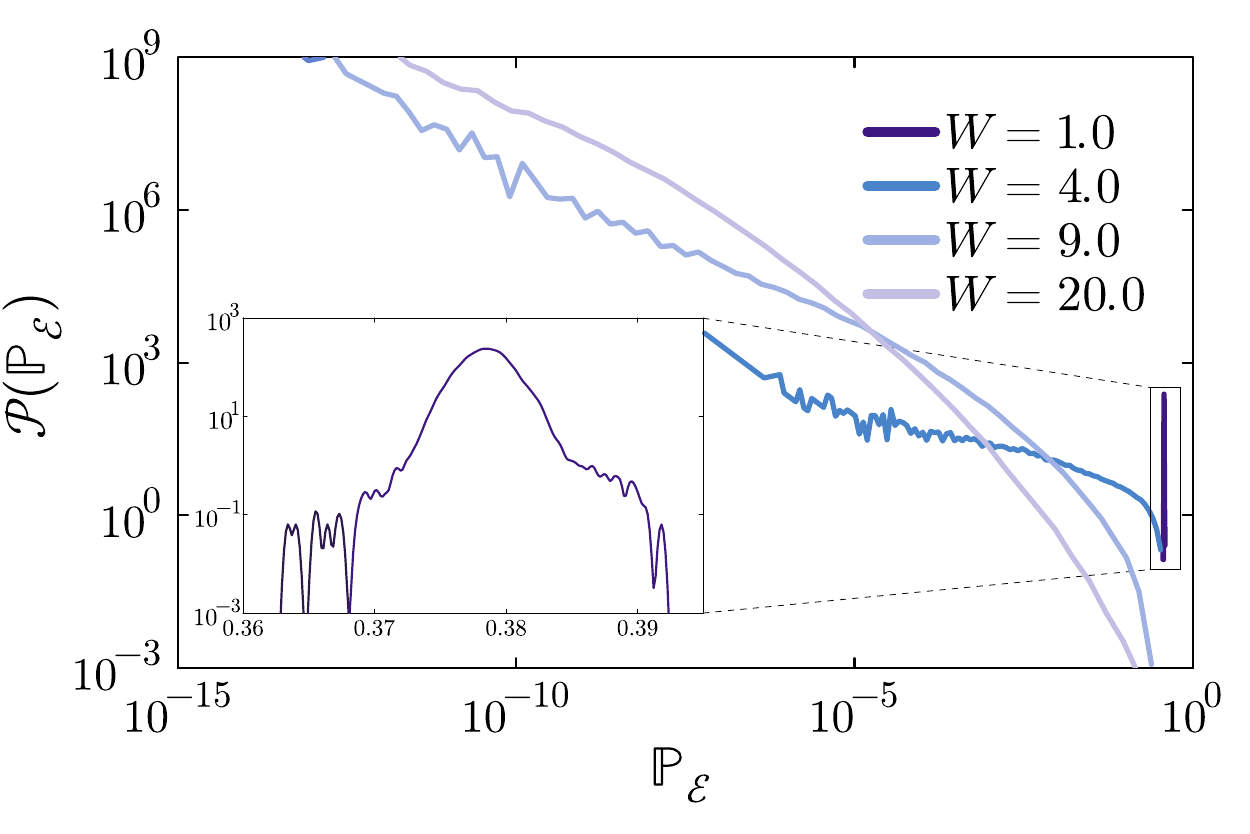}
\caption{Probability distributions of the delocalization probability $\mathbb{P}_\mathcal{E}$ in log-log scale, for the disorder strengths shown in the legend, with $L=16$ and $\Delta = 1$. The inset is a zoom-in for the $W = 1$ distribution, shown in linear-log scale instead. This latter distribution is heavily peaked at finite values of $\mathbb{P}_\mathcal{E}$ with fast decaying tails.}
\label{fig:violins}
\end{figure} 

The properties of $\mathcal{P} ( \mathbb{P}_{0 \to f})$ directly reflect on those of  $\mathcal{P}(\mathbb{P}_{\mathcal{E}})$. In Fig.~\ref{fig:violins}, we present these PDFs for four different disorder strengths: $W = 1$, $4$, $9$, and $20$, with system size $L = 16$ and interaction strength $\Delta = 1$.
For $W = 1$, $\mathbb{P}_\mathcal{E}$ is strongly peaked, with rapidly decaying tails.
This is better appreciated in the inset, where we show a zoom-in with respect to the $\mathbb{P}_{\mathcal{E}}$ axis. While most delocalization probabilities concentrate near very small values, the distributions develop increasingly heavy-tailed behavior. These fat tails signal the presence of rare disorder realizations for which the delocalization probability remains non-negligible---of order $O(1)$---even within a strong disorder regime that was previously associated with the MBL phase~\cite{Oganesyan2007, luitz_many-body_2015}.

\subsection{Mapping to classical disordered systems}\label{sec:disordered}
In this work, we argue that these rare events are responsible for the presence of ergodic instabilities inside the MBL at finite sizes, within a broad intermediate region of the phase diagram. The central goal of the method proposed here is to put forward a computational scheme that allows one to account correctly for the statistical weight of the fat-tailed distributions $\mathcal{P} ( \mathbb{P}_{0 \to f})$.  
Such a computational scheme is based on an analogy with classical mean-field disordered systems that undergo phase transitions exhibiting phenomenological features similar to those described above for the probability of delocalization from a random initial state. Specifically, the partition function of a classical disordered system with \( N \) degrees of freedom is given by
\begin{equation} \label{eq:Zcds}
\mathcal Z_N= \sum_\mu e^{-\beta E_\mu} \,,
\end{equation}
where the \( \mu \)'s label microscopic configurations of the system, whose number grows exponentially with \( N \). In general, the energies \( E_\mu \) are random (and  correlated). The inverse temperature \( \beta \) controls the spread of the Boltzmann weights. In this analogy, the partition function \( {\cal Z}_N \) corresponds to the delocalization probability \( \mathbb{P}_{\mathcal{E}} \), while the Boltzmann factors \( e^{-\beta E_\mu} \) play the role of the transition probabilities \( \mathbb{P}_{0 \to f} \). 

A broad class of classical mean-field disordered systems exhibit a sharp phase transition in the thermodynamic limit (\( N \to \infty \)), from a high-temperature phase---where the partition function receives contributions from an exponential number of configurations---to a low-temperature phase, where the Boltzmann measure freezes onto a few rare configurations with anomalously large weights in the tails of the Boltzmann factor distribution~\cite{derrida_random-energy_1981, derrida_generalization_1985, mezard_spin_1987, b_derrida_solution_1986, carpentier_glass_2001, Charbonneau2023}. 

The key result for such systems is that, in the frozen phase, in the thermodynamic limit the \emph{typical} value of the partition function (which corresponds to the free-energy in classical disordered systems terminology) is dominated by rare configurations in the far tails of the Boltzmann weight distribution. In the MBL context, this implies that the \emph{typical} value of the delocalization probability \( \mathbb{P}_{\mathcal{E}} \) is asymptotically controlled by a few anomalously strong, system-wide resonances.

Yet, accurately estimating the asymptotic typical value of the partition function (or equivalently \( \mathbb{P}_{\mathcal{E}} \)) from finite-size numerical simulations is extremely challenging, since accessible system sizes typically do not include the rare events that dominate the measure in the thermodynamic limit, resulting in strong finite-size corrections. 

Our approach therefore consists of adapting the set of tools and methods developed in the study of mean-field classical disordered systems to properly account for these rare-event effects, allowing us to evaluate the correct asymptotic behavior of the delocalization probability in a disorder regime where it is given by a sum over an exponentially large number of correlated and broadly distributed random variables.

However, the computational cost of evaluating Eq.~(\ref{eq:delocProb}) is substantial, specifically, it requires computing all eigenstates $\ket{n}$ of the Hamiltonian $\hat{\mathcal{H}}$. 
Carrying out exact diagonalization of the full spectrum over a sufficiently large number of disorder realizations to obtain reliable statistics is computationally intensive and can only be done for relatively small sizes, $L \le 16$ (corresponding to a maximal Hilbert space dimension of $12\,870$). 
To address this limitation, we introduce a proxy quantity for $\mathbb{P}_{0 \to f}$ that is easier to access computationally, yet retains the same physical information. This allows us to capture the essential features of the transport properties under investigation. An additional advantage of this approach is that it enables the exploration of larger system sizes, up to $L = 22$, which corresponds to a Hilbert space dimension of $705\,432$.

\subsection{A proxy for transport in Hilbert space}
\label{sec:proxy}

Let us now go back to the set-up defined in \autoref{sec:deloc_prob}, in which we probe how a system, initially prepared in the random basis state $\ket{0}$, spreads to an exponentially large subset of basis states located at the equator $\mathcal{E}$. In this setting, Eqs.~\eqref{eq:prob} and~\eqref{eq:delocProb} give the definition of the probability to decorrelate from a random initial state after infinite time. In a many-body system, the projections of the eigenstates $\vert n \rangle$ with energy densities different from the initial energy $\langle 0 \vert {\cal H} \vert 0 \rangle$ are expected to vanish exponentially with the system size. As a result, the sum over $\vert n \rangle$ is expected to be dominated by the eigenstates having the same expectation value of energy as the initial state. Hence, the probability that the system is initialized in $\vert 0 \rangle$ and is found in $\vert f \rangle$ after an infinite time can be expressed in terms of the squared modulus of the infinite-time propagator  between the nodes of the Hilbert space corresponding to the basis states $\vert 0 \rangle $ and $\vert f \rangle$:
\begin{equation} \label{eq:PdelocG}
    \mathbb{P}_{\mathcal{E}} \approx \lim_{\eta \to 0} \sum_{f \in \mathcal{E}} \eta |{\cal G}_{0f} (E - {\rm i} \eta) \vert^2 \;,
\end{equation}
where ${\cal G}_{0f}$ are the off-diagonal elements of the resolvent $\hat{\mathcal{G}} (E - {\rm i} \eta) = (E - {\rm i} \eta - \hat{\mathcal{ H}})^{-1}$ computed
on the nodes $\vert 0 \rangle $ and $\vert f \rangle$ of the Hilbert space, and $\eta$ is a small imaginary regulator.
    
The interpretation of Eq.~\eqref{eq:PdelocG} is quite intuitive: The spreading of the many-body states is driven by energy-resonant hybridization among basis states with energies close to the selected energy $E$. In closed systems, such resonances are captured by the off-diagonal elements of the resolvent operator, whose matrix elements, $\mathcal{G}_{0f} = \bra{f} \hat{\mathcal{G}} \ket{0}$, quantify the effective hopping amplitude for an energy-resolved transition between the initial state $\ket{0}$ and the target state $\ket{f}$.

As detailed in Appendix~\ref{sec:SimulationDetails}, computing the matrix elements \(  \mathcal{G}_{0f}  \) is computationally more efficient than evaluating the probabilities \( \mathbb{P}_{0 \to f} \), as it only requires solving a sparse linear system (see Eq.~\eqref{eq:linear_system}), rather than obtaining the full energy spectrum. Nevertheless, \( \mathcal{G}_{0f} \) encodes the same physical information as \( \mathbb{P}_{0 \to f }\).
	
\begin{figure}[h!]
		\centering
		\includegraphics[width=\linewidth]{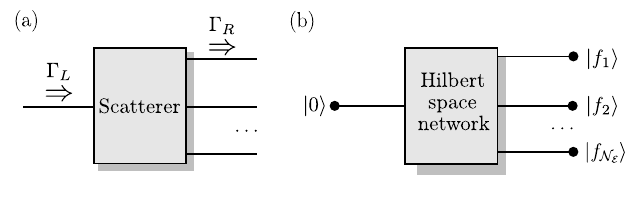}
		\caption{(a) Quantum transport on a network in a scattering geometry, receiving particles from a semi-infinite lead on the left and transmits them through several semi-infinite leads connected to its right-hand side.
			(b) Schematic of the transport of the 'fictitious particle'--initially prepared in the basis state $\ket 0$--on the Hilbert space network.}
		\label{fig:schematic_proxy}
	\end{figure}

Specifically, Eq.~\eqref{eq:PdelocG} can be interpreted in terms of an analogy from the study of quantum mesoscopic systems in real space, where transport properties at fixed energy $E$ are typically characterized by the Landauer transmission~\cite{Datta2013-vq}. These are commonly computed using the Fisher-Lee formula~\cite{Fisher1981, lemarie_glassy_2019}, which relates the dimensionless conductance to the Green's function of the scatterer dressed by the incoming and outgoing leads---creating channels of transport. Explicitly, the formula reads
	\begin{equation} \label{eq:LandauerFL}
		\mathcal{T}_{\rm FL} = \rm{Tr}\{ \Gamma_L \mathcal G^r \Gamma_R \mathcal G^a \} \;,
	\end{equation}
	where the energy dependence $E$ is implicit in all quantities. The superscripts $r$ and $a$ denote the retarded and advanced Green’s functions, respectively, which are related through $\mathcal{G}^a = \left( \mathcal{G}^r \right)^\dagger$. The quantities $\Gamma_L$ and $\Gamma_R$ represent the level broadening due to the coupling with the left (L) and right (R) leads, respectively, and are given by
	\begin{equation}
		\Gamma_{L,R} = -2 \, \rm{Im} \Sigma_{L, R}
		\;,
	\end{equation} with $\Sigma_{L, R}$ being the self-energies of the leads. A schematic representation of this construction is shown in Fig.~\ref{fig:schematic_proxy}(a). In this respect, $\mathbb{P}_{\mathcal{E}}$ is the analogue of the conductance of a complex network (i.e., the Hilbert space graph) in a scattering geometry in which a semi-infinite lead through which `particles' are  injected is connected to the node $\vert 0 \rangle$, and ${{L/2}\choose{L/4}}^2$ semi-infinite leads through which `particles' are extracted are connected to the nodes \( \ket{f} \in \mathcal{S}_0(0) \) (or \( \mathcal{A}_0(0) \), depending on the chosen basis). This construction is illustrated in \autoref{fig:schematic_proxy}(b), where the incoming and outgoing semi-infinite leads are replaced by the initial state $\ket{0}$ and the set of target states $\mathcal{E}$, respectively.  It is clear that in this analogy the broadening of the energy levels provided by the imaginary part of the self-energy of the leads plays the role of the small imaginary regulator $\eta$ in Eq.~\eqref{eq:PdelocG}. In the ergodic phase---where all channels contribute comparably to transport---$\mathcal{T}_{\mathrm{FL}}$ is of order $1$. By contrast, in the MBL regime transport is suppressed and the typical value of $\mathcal{T}_{\rm FL}$ decays exponentially with $L$.

In order to further simplify the numerical computations, in most of the following we do not evaluate the true delocalization probability Eq.~\eqref{eq:delocProb} (or, equivalently, the Landauer transmission Eq.~\eqref{eq:LandauerFL}), but rather a simplified proxy in which the imaginary parts are neglected: 
    \begin{equation}
	   \mathcal{T}_0 = \sum_{f \in \mathcal{E}} \left| \mathcal{G}_{0f} \right|^2\;,   
	   \label{eq:FSLandauerTrans}
	\end{equation}
    where $\mathcal{G}_{0f} = \bra{f} (E - \hat{\mathcal{ H}})^{-1}  \ket{0}$ denotes the real-part off-diagonal propagator between $\vert 0 \rangle$ and $\vert f \rangle$. Restricting the computation to real parts only reduces the numerical cost---both in terms of computational time and, more importantly, memory usage---by more than a factor of two.
    
    Yet, in the absence of the imaginary regulator, the amplitudes $|\mathcal{G}_{0f}|^2$ are no longer bounded and can take arbitrarily large values. This occurs because the poles of $\mathcal{G}$ associated with many-body eigenstates whose energies lie very close to $E$ are no longer regularized, leading to spuriously large contributions. As a result, the $|\mathcal{G}_{0f}|^2$'s, and hence $\mathcal{T}_0$, can no longer be strictly interpreted as the probabilities to delocalize from $\vert 0 \rangle$ to $\vert f \rangle$ after infinite time.

The absence of this regularization has, in particular, important consequences for the asymptotic scaling of $\mathcal{T}_0$ in the delocalized phase. While the properly regularized transmission $\mathcal{T}_{\rm FL}$ and the probability to reach the equator $\mathbb{P}_{\mathcal{E}}$ saturate to a value of order one on the ergodic side of the transition, their unregularized counterpart $\mathcal{T}_0$ grows exponentially with the system size $L$, proportionally to the number of outgoing channels. In contrast, in the MBL phase both $\mathcal{T}_{\rm FL}$ and $\mathcal{T}_0$ exhibit the same exponential decay with $L$. Therefore, although $\mathcal{T}_0$ formally overestimates the true Landauer transmission, it reproduces its correct large-$L$ scaling in the localized phase and can thus be reliably (and more efficiently) used to identify the localization transition, as discussed below.

Here, we present several arguments to support this claim and justify our choice.  The first argument stems from an analogy with the Anderson model on the Bethe lattice. Indeed, the Hilbert-space graph associated with a quantum many-body problem---which lies at the core of our discussion---is, in general, a sparse and high-dimensional network, as discussed in~\autoref{sec:HilbertSpace}. Although this graph exhibits strong correlations and complex loop structures that are absent in simpler tree-like models such as the Bethe lattice, the latter nonetheless serves as a valuable toy model that captures the essential features of the Hilbert-space network~\cite{Altshuler1997}. This analogy provides a powerful framework for gaining qualitative insight into the behavior of many-body systems~\cite{DeLuca2013, de_luca_anderson_2014, Biroli2017, Logan2019, Tikhonov2021, Herre2023, Scoquart2024, lemarie_glassy_2019}.

\begin{figure}
    \centering
    \includegraphics[width=0.8\linewidth]{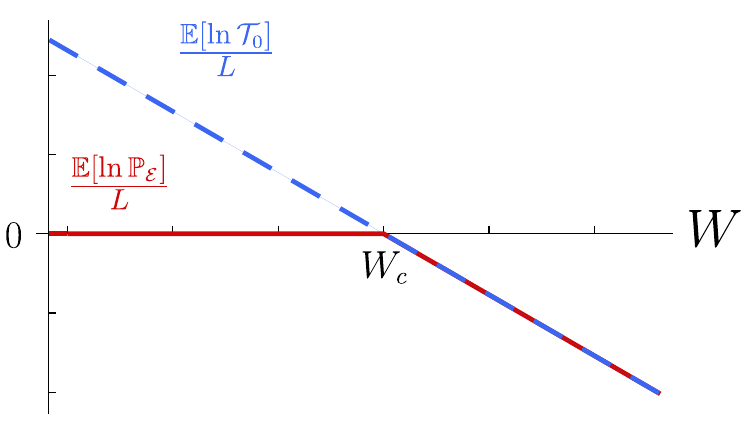}
    \caption{Sketch of the different scaling behavior with $L$ of the typical value of the Landauer transmissions with and without imaginary parts for the Anderson model on the Bethe lattice.
    }
    \label{fig:sketch}
\end{figure}

The order parameter for the localization transition in the single-particle Anderson model is the typical value of the imaginary part of the local Green's function, $\exp \{ \mathbb{E}\![ \ln \mathrm{Im} \, \mathcal{G}_{ii}] \}$, which is finite in the delocalized phase and vanishes in the localized one. For locally tree-like graphs, asymptotically exact recursive relations can be derived that express these local Green’s functions in terms of those on neighboring nodes~\cite{RAbouChacra1973}. Considering a specific node $0$ of the tree, the recursive equation for the imaginary part (in the $\eta \to 0$ limit) can be telescoped as:
\begin{equation}
    {\rm Im} \, {\cal G}_{00} = \left| {\cal G}_{00} \right|^2 \sum_{i \in \partial 0} {\rm Im} \, {\cal G}_{ii} \, ,
\end{equation}
where the sum runs over all neighbors $i$ of node $0$, and $\left| {\cal G}_{00} \right|^2$ is the squared modulus of the full local Green's function, including both its real and imaginary parts.  Starting from node $0$, we iteratively apply the recurrence to unfold the expression, rewriting the 
imaginary parts ${\rm Im}\,{\cal G}_{ii}$ on the right-hand side in terms of those from successive generations of the tree, i.e., of nodes at increasing distance from $0$. Repeating this procedure over $L$ generations, and writing the two-points propagators as product of the local Green's functions yields (see Refs.~\cite{biroli2020anomalous,biroli_large-deviation_2024} for more details):
\begin{equation} \label{eq:ImG00}
    {\rm Im} \, {\cal G}_{00} = \sum_{f=1}^{k^L} \left| {\cal G}_{0f} \right|^2 {\rm Im} \, {\cal G}_{ff} \, ,
\end{equation}
where $f$ labels the $k^L$ nodes at distance $L$ from $0$ (where $k$ is the branching ratio of the tree).  This equation is the analogue of Eqs.~\eqref{eq:PdelocG} and~\eqref{eq:LandauerFL}, and its physical interpretation is intuitively clear: ${\rm Im}\, {\cal G}_{00}$ represents the inverse lifetime of a particle created at node $0$, while the propagators $\left| {\cal G}_{0f} \right|^2$ correspond to the probabilities that the particle escapes from $0$ to $f$ after an infinite time.

We now set the imaginary part of the Green's functions on all these distant nodes $f$ to a small value $\eta$, and take $L$ to be large. In the delocalized phase, the imaginary parts grow under iteration and eventually reach a stationary value for large $L$. In the localized phase, by contrast, they decay exponentially under iteration. In this regime, the recursion relations can be linearized with respect to the imaginary parts, and to leading order Eq.~\eqref{eq:ImG00} becomes:
\begin{equation} \label{eq:ImG00real}
    {\rm Im}\, {\cal G}_{00} = \eta \sum_{f=1}^{k^L} \left| {\rm Re}\, {\cal G}_{0f} \right|^2 = \eta\, {\cal T}_0 \, .
\end{equation}
The exponential decay of the typical value of ${\rm Im}\, {\cal G}_{00}$ under iteration in this linearized regime is governed by the largest eigenvalue of an integral operator that encodes the critical properties of Anderson localization~\cite{RAbouChacra1973,Tikhonov2019-ku}. One finds
$\exp \{\mathbb{E}[\ln {\rm Im}\, {\cal G}_{00} ] \} \simeq \eta\, \lambda_{\rm max}^L$,  
with $\lambda_{\rm max} \simeq 1 - c(W - W_c)$ near the localization transition, hence
\begin{equation} \label{eq:T0lambda}
    \frac{1}{L}\, \mathbb{E} \! \left[ \ln {\cal T}_0 \right] = \ln \lambda_{\rm max} \simeq -\, c (W - W_c) \, .
\end{equation}
Consequently, the scaling of the typical Landauer transmission with $L$ (evaluated from the real propagators) reflects the $W$-dependence of the Lyapunov exponent that controls the response of the typical ${\rm Im}\,{\cal G}$ to perturbations, thereby indicating whether the system is in a localized or a delocalized phase. Note that in the context of the Anderson model on tree-like graphs, this is an exact result.

Of course, in the delocalized phase the recursion equations can no longer be linearized, and Eq.~\eqref{eq:ImG00real} no longer describes the probability of delocalization from a random initial state: instead of saturating to a finite value, as its counterpart~\eqref{eq:ImG00} which contains also the imaginary parts, it diverges exponentially. Nevertheless, as discussed below and explicitly demonstrated in~\cite{biroli_large-deviation_2024}, the scaling of its typical value with $L$ can still be used effectively and reliably to locate the localization transition in the Anderson model on the Bethe lattice. This argument is schematically depicted in Fig.~\ref{fig:sketch} which illustrates the different behavior of the scaling with $L$ of the typical value of the Landauer transmissions---or equivalently, the delocalization probability---with and without imaginary parts for the Anderson model on the Bethe lattice.

The second argument supporting the choice of the typical value of ${\cal T}_0$ as an order parameter for MBL comes from the benchmark analysis presented in Ref.~\cite{biroli_large-deviation_2024}. In that work, some of us employed this quantity not only to locate the localization transition in the Anderson model on the Bethe lattice (for which, as discussed above, Eq.~\eqref{eq:T0lambda} can be explicitly and rigorously justified), but also to study abstract random matrix ensembles that display three distinct regimes: a fully delocalized phase, a localized phase, and an intermediate delocalized yet fractal phase.  Ref.~\cite{biroli_large-deviation_2024} demonstrated that the numerical scaling analysis of the typical value of ${\cal T}_0$ successfully reproduces the analytically known phase diagram of these models, accurately identifying the transitions between the different phases.

A final argument in favor of our approximation is provided by direct numerical tests~\cite{Tarzia2026} on the random-field transverse-field Ising model. We explicitly examined the scenario schematically illustrated in Fig.~\ref{fig:sketch}.  In particular, we computed the probability to delocalize from a random initial state, using Eqs.~\eqref{eq:prob} and~\eqref{eq:delocProb} (via exact diagonalization), and compared it to the Landauer transmission ${\cal T}_0$ for the same disorder realizations and initial states. At sufficiently strong disorder---where $\mathbb{P}_{\mathcal{E}}$ decays exponentially with the system size---we find that the typical value of ${\cal T}_0$ is proportional to that of $\mathbb{P}_{\mathcal{E}}$. Moreover, their covariance increases with system size and approaches unity in the strong-disorder regime for the system sizes accessible to our numerics~\cite{Tarzia2026}. This implies that samples and initial configurations with an anomalously large probability to decorrelate after infinite time also exhibit anomalously large values of ${\cal T}_0$.

To summarize, even though neglecting the imaginary regulator in Eqs.~\eqref{eq:PdelocG} and~\eqref{eq:LandauerFL} leads to a distinct scaling behavior in the delocalized phase (due to the proliferation of spurious poles in the denominator of ${\cal G}$), the typical value of the Landauer transmission ${\cal T}_0$ nonetheless provides an efficient and reliable order parameter for MBL.  It offers a much simpler computational route while faithfully reflecting the same physical information: instead of directly evaluating the probability for a random initial state to delocalize, we compute a proxy quantity that  signals whether this probability decays exponentially or not.

Yet, we are still left with the problem of how to perform a correct statistical analysis of ${\cal T}_0$, which is a sum of an exponential number of correlated random variables, whose distribution becomes broader as $W$ is increased. 

As explained above, Eq.~\eqref{eq:FSLandauerTrans}  is formally equivalent to the partition function of a classical disordered system~\eqref{eq:Zcds}. Specifically, the factors $| \mathcal{G}_{0f} |^2$ play the role of the Boltzmann weights associated to each one of the exponentially numerous target states $\ket{f}$. 
As mentioned above, for a broad class of mean-field classical disordered systems, such as directed polymers in disordered media on high-dimensional graphs~\cite{derrida_polymers_1988, Derrida1989, Derrida1991} and related models~\cite{derrida_random-energy_1981, derrida_generalization_1985, Mezard1986, carpentier_glass_2001}, it is well known that the partition function~\eqref{eq:Zcds} can undergo a phase transition: if the probability distribution of the Boltzmann weights becomes too broad (which in classical systems is induced by reducing the temperature), then the measure is dominated by few $O(1)$ outliers of the distribution, corresponding to a few configurations of the system with particularly low energy---the so-called frozen phase. By contrast, at high temperature the partition function receives substantial contributions from exponentially many terms.

In this context, our ultimate goal is to estimate the typical value of \( \mathcal{T}_0 \) in the large-\( L \) limit (see Eq.~\eqref{eq:T0lambda}), which is analogous to estimating the free-energy density in classical disordered systems at a given temperature. However, in the many-body problem, the broadening of the distribution of the \( | \mathcal{G}_{0f} |^2 \)'s is indirectly controlled by the disorder strength \( W \). For a given \( W \), it is not obvious a priori whether \( \mathcal{T}_0 \) corresponds to the `high-temperature' or `low-temperature' phase in this analogy. This distinction is crucial, as estimating the asymptotic behavior of the typical value in the frozen phase is particularly subtle. As mentioned earlier, one must properly account for the contribution of rare outliers—namely, atypical disorder realizations that lead to anomalously large values of \( \mathcal{T}_0 \) and are unlikely to appear in small or moderate system sizes.

In the following, we begin by explaining how this type of analysis is carried out in the standard setting of classical disordered systems, where the distribution of Boltzmann weights is known and analytical calculations are possible. As a guiding example, we will focus on the directed polymer in disordered media~\cite{comets2017directed}, which has already been connected to various tightly-related quantum systems—most notably to single-particle Anderson localization~\cite{shiferaw_goldschmidt_2001, monthus_anderson_2011, lemarie_glassy_2019}, two-interacting-particle models in one-dimensional disordered systems~\cite{Mu2024-aj}, and random-field quantum Ising models~\cite{dimitrova_mezard_2011, monthus_garel_2012}. Following this discussion, we will outline how we adapt this theoretical framework to compute the typical value of $\mathcal{T}_0$ in our quantum many-body context.
\subsection{The transition indicators}
\subsubsection{Freezing transition 
in mean-field classical disordered systems}
\label{sec:DPRM}

The discrete model of a directed polymer in random media (DPRM) consists of a self-avoiding directed random walk on a $d$-dimensional lattice. Each edge of the lattice, $(ij)$, has an associated energy $\epsilon_{ij}$ which is a quenched random variable. The set of edges that the self-avoiding random walk follows defines a path $\mathcal P$, and 
its energy is given by $E_\mathcal{P} = \sum_{(ij) \in \mathcal{P}} \epsilon_{ij}$.  The partition function is 

\begin{equation}
    \mathcal Z_{N}(\beta) = \sum_{ \mathcal{P} \in \mathcal{P}_N} e^{-\beta E_{\mathcal{P}}} \;,
    \label{eq:Zdprm}
\end{equation} 
where $E_\mathcal{P}$ is the total random energy collected along the path 
$\mathcal P$ of length $N$. The sum runs over all possible directed paths of length $N$, here denoted as the set $\mathcal{P}_N$, and $\beta$ is the inverse temperature. The energies of different paths, say $E_\mathcal{P}$ and $E_{\mathcal{P}'}$, are correlated through their common edges. A formal relationship between this model and Anderson localization has been exploited in the past~\cite{monthus_anderson_2011, somoza_unbinding_2015, lemarie_glassy_2019, pietracaprina_forward_2016, tarquini_level_2016}. DPRM exhibits a well-known freezing transition in the infinite-dimensional limit~\cite{derrida_polymers_1988, Derrida1991}, when the problem is studied on an infinite tree. In this class of hierarchical lattices, the number of directed paths for polymers of length $N$ is $(k+1)k^{N-1}$, $k$ being the branching ratio of the tree. 

The transition occurs at a critical inverse temperature $\beta = \beta_\star$.
At high temperature $\beta < \beta_\star$, the partition function receives contributions from an exponential number of directed paths. 
Instead, in the low temperature phase $\beta > \beta_\star$, the polymer freezes in a few $O(1)$ specific disorder-dependent paths. This corresponds to a condensation of the Boltzmann measure on a few paths with particularly low energy, which directly reflects in the  non-analytic behavior of the {\it quenched} free-energy density, defined as 
\begin{equation}
    f_q(\beta) = -\lim_{N \to \infty} \frac{1}{\beta N} \langle \ln \mathcal{Z}_{N}(\beta) \rangle \;,
    \label{eq:quenched-free-energy}
\end{equation} where $\langle \cdots \rangle$, denotes the average over the disorder realizations. For the specific problem of directed polymers on a tree, the exact solution of Refs.~\cite{derrida_polymers_1988, Derrida1991} yields:
\begin{equation} \label{eq:fqDPRM}
f_q (\beta) =  
\left \{
\begin{array}{ll}
- \ln \left ( k \avg{e^{-\beta \epsilon}} \right) / \beta  \,\, & \textrm{for~} \beta< \beta_\star \, , \\
- \ln \left ( k  \avg{e^{-\beta_\star \epsilon}} \right ) / \beta_\star & \textrm{for~} \beta \ge \beta_\star \, .
\end{array}
\right .
\end{equation}
In the thermodynamic limit, the quenched free-energy density develops a plateau for \( \beta > \beta_\star \). For finite system sizes, however, the quenched free-energy remains a concave function of \( \beta \) for any arbitrarily large but finite \( N \). The convergence of the finite-\( N \) quenched free-energy to its asymptotic plateau value for \( \beta > \beta_\star \) is slow, with finite-size corrections scaling as \( \log N / N \)~\cite{Derrida1990, Evans1992}. This slow convergence arises because typical finite-size samples do not contain the rare configurations that dominate the behavior in the thermodynamic limit.

In this situation, it is particularly insightful to study also the behavior of the finite-size {\it annealed} free-energy density, defined as
\begin{equation}
    f_a(\beta, N) = - \frac{1}{\beta N}  \ln \langle \mathcal{Z}_{N}(\beta) \rangle \;,
    \label{eq:annealed-free-energy}
\end{equation} 
and compare it with the finite-size quenched free-energy density~\eqref{eq:quenched-free-energy}.

The basic idea is the following: In the high-temperature phase (\( \beta < \beta_\star \)), the sum~\eqref{eq:Zdprm} is dominated by the bulk of the probability distribution of the Boltzmann weights. As a result, the average and typical values of \( \mathcal{Z}_N \) exhibit the same asymptotic scaling with \( N \), i.e. the partition function is self-averaging. Consequently, the annealed and quenched free-energy densities converge to the same value upon increasing the system size $N$. By contrast, in the low-temperature phase (\( \beta > \beta_\star \)), only a few disorder-dependent paths with particularly low energies \( E_\mathcal{P} \) dominate the sum in~\autoeqref{eq:Zdprm}. These rare, large Boltzmann weights induce strong sample-to-sample fluctuations, leading to a broad distribution \( P(\mathcal{Z}_N) \) characterized by power-law tails~\cite{derrida_polymers_1988}. When these tails are sufficiently heavy, the typical and average values of \( \mathcal{Z}_N \) exhibit different scaling with \( N \): the former is governed by the bulk of the distribution, while the latter is dominated by rare configurations with anomalously low energies. Consequently, the self-averaging property of the partition function is lost. In particular, the extreme outliers skew the average $\langle \mathcal{Z}_N \rangle$, causing the finite-$N$ annealed free-energy density to develop a maximum close to $\beta_\star$. Such maximum of the annealed free-energy curve is completely unphysical and purely a consequence of the biased finite sampling of anonymously large outliers of $\mathcal{Z}_N$. 

Yet, in models for which the analytical solution in the thermodynamic limit is unavailable (as is the case for the asymptotic typical value of \( \mathcal{T}_0 \) in the MBL problem), the behavior of the annealed free-energy provides a valuable practical tool to estimate the asymptotic critical behavior of the quenched free-energy in the low-temperature phase using finite-\( N \) numerical results.

The basic idea is to first identify the position of the maximum of the annealed free-energy curve, which gives an estimate of \( \beta_\star \). Then, the unphysical portion of the annealed free-energy for \( \beta > \beta_\star \) is replaced by a flat segment at height \( f_a(\beta_\star, N) \). For models in which the large-\( N \) solution is known analytically, this construction has been shown to provide a more accurate approximation of the asymptotic value of the quenched free-energy density in the frozen phase \cite{Evans1992, biroli_large-deviation_2024}

Formally, this construction is as follows:
\begin{equation}
\begin{aligned}  
\tilde{f}_{a}(\beta, N) = \left \{
\begin{array}{ll}
f_a(\beta, N)  \,\, & \textrm{for~} \beta< \beta_\star \, , \\
[5pt]
f_a(\beta_\star, N) & \textrm{for~} \beta \ge \beta_\star \, ,
\end{array}
\right .   
\end{aligned}
\label{eq:approx_free-energy}
\end{equation} 
where $f_a(\beta, N)$ is the finite-size annealed free-energy density given in~\autoeqref{eq:annealed-free-energy}. With this modification one obtains two equivalent quantities for the entire temperature range in the thermodynamic limit:
\begin{equation}
    f(\beta) = \lim_{N \to \infty} f_q(\beta, N) = \lim_{N \to \infty} \tilde{f}_a(\beta, N) \;.
\end{equation}
At finite $N$, $f_q(\beta, N)$ and $\tilde{f}_a(\beta, N)$ are complementary, and allow us to assess the role that finite
size effects play in the problem at hand. The former describes the behavior of typical samples at the chosen value of $N$; while the latter provides a more accurate estimation of the transition point at large $N$~\cite{biroli_large-deviation_2024}.

\subsubsection{Probing the rare events in the quantum many-body problem}

In the following, we apply the construction~\eqref{eq:approx_free-energy} to estimate the asymptotic behavior of the typical value of \( \mathcal{T}_0 \) in the large-\( N \) limit, while properly accounting for the statistical contribution of strong, system-wide resonances which, for the system sizes accessible numerically, typically form only between a few specific distant configurations, and only in rare disorder realizations.

The key assumption underlying our approach is that Eq.~\eqref{eq:FSLandauerTrans} belongs to the same universality class as the classical mean-field disordered models discussed above. This connection can be rigorously established in the case of single-particle Anderson localization on the Bethe lattice~\cite{biroli_large-deviation_2024}. More generally, this universality holds whenever the elements of the sum exhibit ultrametric correlations, as in the DPRM case~\cite{derrida_generalization_1985, b_derrida_solution_1986}. In Appendix~\ref{sec:ultrametric}, we provide numerical evidence that the correlations $\langle | \mathcal{G}_{0f} | |\mathcal{G}_{0f^\prime} | \rangle_c$ are consistent with this assumption for the system sizes we are able to study.

To apply the machinery described in the previous section to the MBL problem, we enlarge the parameter space by formally introducing an auxiliary parameter $\beta$, which plays the role of the inverse temperature in the classical problem.  We thus define the $\beta$-dressed version of the transmission, $\mathcal{T}_0(\beta)$: 
\begin{equation}
    \mathcal{T}_0(\beta) = \sum_{f \in \mathcal{E}} \left| \mathcal{G}_{0f} \right|^\beta 
    \;.
    \label{eq:FGLT}
\end{equation}
This auxiliary parameter has the role of tuning the strength of the tails of the probability distribution of the $| \mathcal{G}_{0f} | $'s for a fixed value of the disorder $W$ of the original MBL problem, thereby allowing us to identify the threshold value $\beta_\star$ at which the distribution of $\mathcal{T}_0(\beta)$ develops sufficiently broad tails. At this point, rare outliers begin to dominate its typical value, signaling the onset of a freezing transition analogous to that in the DPRM.
Note that $\mathcal{T}_0(\beta)$  has been already introduced and studied in the mathematical literature on single-particle Anderson localization in hierarchical lattices~\cite{Aizenman1993-mp, Aizenman2013-yd}, in the context of the so-called fractional moment method.

Our primary task in this work is to compute the asymptotic scaling behavior of the typical value of $ \mathcal{T}_0(\beta=2)$, which is a proxy for the probability that a randomly chosen initial state reaches arbitrarily distant configurations on the Hilbert space graph after infinite time. In order to do this, we have to determine the position of $\beta = 2$ relative to the threshold $\beta_\star$, which determine whether $\mathcal{T}_0(\beta = 2)$ is dominated by the broad tails of its distribution (i.e., rare events) or whether the tails decay rapidly enough for it to be dominated by the bulk of the distribution (i.e., typical realizations of the disorder). 

Proceeding as in the theoretical framework outlined in the previous section, we define two quantities analogous to the annealed and quenched free-energy densities: 
\begin{equation}
\label{eq:phia}
\phi_a(\beta, L) = \frac{\ln \mathbb{E}[\mathcal T_0(\beta, L)]}{\beta \ln \mathcal{N}_\mathcal{E}} \;,
\end{equation} 
\begin{equation}
\begin{aligned}  
\tilde{\phi}_{a}(\beta, L) = \left \{
\begin{array}{ll}
\phi_a(\beta, L)  \,\, & \textrm{for~} \beta< \beta_\star \, , \\
[5pt]
\phi_a(\beta_\star, L) & \textrm{for~} \beta \ge \beta_\star \, ,
\end{array}
\right .   
\end{aligned}
\label{eq:approx_phia}
\end{equation}

\begin{equation}
\label{eq:phiq}
\phi_q(\beta, L) = \frac{\mathbb{E}[\ln \mathcal T_0(\beta, L)]}{\beta \ln \mathcal{N}_\mathcal{E}} \;,
\end{equation} corresponding to the annealed, annealed with the plateau replacement---similar to~\autoeqref{eq:approx_free-energy}---and the quenched free-densities, respectively. Here $\mathbb{E}[\cdots]$ denotes the average with respect to initial conditions $\ket{0}$ and disorder realizations of the random fields. 

The partition function of the classical DPRM $\mathcal{Z}_N(\beta)$, has been replaced by the biased Hilbert space Landauer transmission $\mathcal{T}_0(\beta)$. In the classical DPRM model, the total number of configurations---that is, the number of terms contributing to the partition function---scales exponentially with the length of the polymer as $k^N$, previously denoted by $N$. In this case, the number of terms contributing to the sum is $\mathcal{N}_\mathcal{E}$, which represents the number of target states. Accordingly, we adapt the definition of the free-energy functions by normalizing with $\ln \mathcal{N}_\mathcal{E}$. Additionally, we have omitted the overall minus sign in the definition, so that the free-energies become negative when the typical value of $\mathcal{T}_0$ vanishes---signaling localization---while a positive values of the free-energies indicates delocalization. 
Due to this modification the maximum in the  annealed free-energy density of the original DPRM at $\beta_\star$ now appears as a minimum of $\phi_a(\beta)$.
Beyond the dependence on $\beta$ and $L$, $\mathcal{T}_0(\beta, L)$ also depends on $W$ and $\Delta$. From this point onward, we will omit the explicit dependence on $L$, $W$ and $\Delta$ in the functions $\mathcal{T}_0$, $\phi_a$, $\tilde{\phi}_a$, and $\phi_q$, unless otherwise stated. 

Exploiting the analogy with DPRM and  similar mean-field disordered models~\cite{derrida_polymers_1988, Evans1992, derrida_generalization_1985, Derrida1991, fyodorov_freezing_2008}, the large-$L$ behavior of the probability of delocorrelate from a random initial state, $\mathcal{T}_0(\beta = 2)$, can be estimated indirectly in two steps, as described below:
\begin{itemize}
    \item {\bf Ergodic Phase.} We first determine the position of the minimum $\beta_\star$ of the finite-$L$ annealed free-energy density. If $\beta_\star > 2$ the Hilbert space Landauer transmission defined in~\autoeqref{eq:FSLandauerTrans} lies within the ergodic (or high-temperature) phase, and its typical value can be directly obtained from $\phi_q(\beta=2)$. In this regime the system is delocalized since $\phi_q(\beta=2) > 0$.
\item {\bf Non-Ergodic Phases.} Conversely, when $\beta_\star < 2$, physical transport occurs within the freezing region, where the Hilbert space Landauer transmissions are dominated by the tails of their distribution. According to our analogy with classical disordered systems and DPRM, in this regime, the finite-$L$ value of $\phi_q(\beta = 2)$ provides a poor estimate of the typical value of $\mathcal{T}_0$ in the large-$L$ limit. This is because typical samples at small $L$ lack the rare events that will dominate the statistics at large $L$. A more accurate estimation is instead provided by the value of $\tilde{\phi}_a(\beta=2) = \phi_a(\beta_\star)$. Then:
\begin{itemize}
    \item {\bf Non-Ergodic Delocalized Phase.} If $\phi_a(\beta_\star) > 0$, the elements in the tails of the distribution of $| \mathcal{G}_{0f} |^2$ give such a large contribution that the typical value of  $\mathcal{T}_0$ remains positive upon increasing system size. This corresponds to a regime in which the system eventually delocalizes for large enough $L$ through a small number of long-range resonances that destabilize the MBL phase.

    \item {\bf Non-Ergodic Many-Body  Localized Phase.} If instead $\phi_a(\beta_\star) < 0$, even with the inclusion of large rare-events of $\mathcal{T}_0$, the typical value of $\mathcal{T}_0$ vanishes with increasing system size. This corresponds to a genuine MBL phase, where rare system-wide resonances are unable to make a random initial state completely decorrelate.
\end{itemize}
\end{itemize}
These three distinct criteria provide us with independent estimates of finite-sized critical disorder strengths, separating the different regimes observed in the model:
\begin{enumerate}
    \item $W_{\rm ergo}(L)$ is defined by the condition $\beta_\star(W_{\rm ergo}) = 2$. It provides an estimate for the disorder strength at which the system transitions from an ergodic regime---where physical transport at $\beta = 2$ receives contributions from an exponential number of terms $|\mathcal{G}_{0f}|^2$ and its behavior is governed by typical instances---to a regime where $\mathcal{T}_0$ is dominated by few anomalously large outliers from the distribution’s tails of $|\mathcal{G}_{0f}|^2$, and the distribution of $\mathcal{T}_0$ becomes broad.  
    \item $W_{\rm MBL}(L)$ is defined by the condition $\phi_a(\beta_\star, W_{\rm MBL}) = 0$. It provides an estimate of the critical disorder strength at which the system enters a genuinely localized regime. In this phase, even rare resonant inclusions between basis states are insufficient to induce delocalization.
    \item $W^{\rm typ}_{\rm MBL}$ is defined by the condition $\phi_{q}(\beta = 2, W^{\rm typ}_{\rm MBL}) = 0$. As previously discussed, the quenched free-energy suppresses the contribution of rare, large outliers, making it a good indicator of the typical behavior at a given system size $L$. Consequently, we expect this estimator to capture the localization transition of typical disorder realizations within the accessible sizes. It should therefore align with the critical disorder strength estimated through standard spectral observables and conventional approaches to the MBL transition. As will be discussed below, this may also apply to $W_{\rm ergo}$. For instance, applying this analysis to the single-particle Anderson localization (AL) on random regular graphs (RRG), one finds that $W^{\rm typ}_{\rm AL}$ coincides with the disorder strength at which the average gap ratio shows an apparent transition from RMT to Poisson behavior, with $W_{\rm ergo} \lesssim W_{\rm AL}^{\rm typ}$.
\end{enumerate}

It is important to note that the phase diagram derived from these estimators is a finite-$L$ phase diagram. We observe significant finite-size effects and drifts in the positions of the transitions between different regimes (see Figs.~\ref{fig:phase_diagram}). Concerning the fate of this phase diagram in the $L \to \infty$ limit, two main scenarios are possible. In the first scenario, the intermediate phase $W_{\rm ergo} < W < W_{\rm MBL}$---where delocalization occurs through rare resonances---is merely a finite-size crossover region that progressively shrinks and eventually disappears as $L \to \infty$, yielding a direct transition from the fully chaotic phase to the MBL phase (as for the Anderson model on the RRG). In the second scenario, this intermediate phase persists in the thermodynamic limit as a genuine new type of delocalized phase. Our numerical results for the accessible system sizes seem to favor the first scenario: the intermediate region shrinks progressively as $L$ increases (see Sec.~\ref{sec:results}).

In the first scenario, the apparent non-ergodic behavior is a statistical artifact: typical finite-$L$ samples are not representative of the thermodynamic limit. As established earlier, for $W > W_{\rm MBL}^{\rm typ}$, typical instances are localized with exponentially decaying long-range resonances, yet rare disorder realizations---identified by $W_{\rm MBL}(L)$---can still drive delocalization. Although such realizations are uncommon at currently accessible sizes, they become typical at larger $L$. Therefore, in this scenario, it is natural to identify $W_{\rm MBL}^{\rm typ}$ as the disorder strength at which apparent ergodicity is effectively broken for finite systems. We thus expect $W_{\rm MBL}^{\rm typ}$ and $W_{\rm ergo}(L)$ to exhibit the same trend, both drifting toward $W_{\rm MBL}$ as $L \to \infty$. In the second (less likely) scenario, $W_{\rm MBL}^{\rm typ}$ would still drift toward $W_{\rm MBL}$ in the thermodynamic limit, while $W_{\rm ergo}$ would converge to a finite value strictly smaller than both $W_{\rm MBL}$ and $W_{\rm MBL}^{\rm typ}$.

In the rest of the article we systematically track the behavior of $\beta_\star$, $\phi_a(\beta_\star)$, and $\phi_q(\beta = 2)$, varying the parameters of the model. We present results in both the spin and Anderson bases, identified by different color scales in the following figures.

\section{Results}
\label{sec:results}
\subsection{The free-energies and the relevant indicators}

\begin{figure*}[t]
\centering
  \centering
  \includegraphics[width=\linewidth]{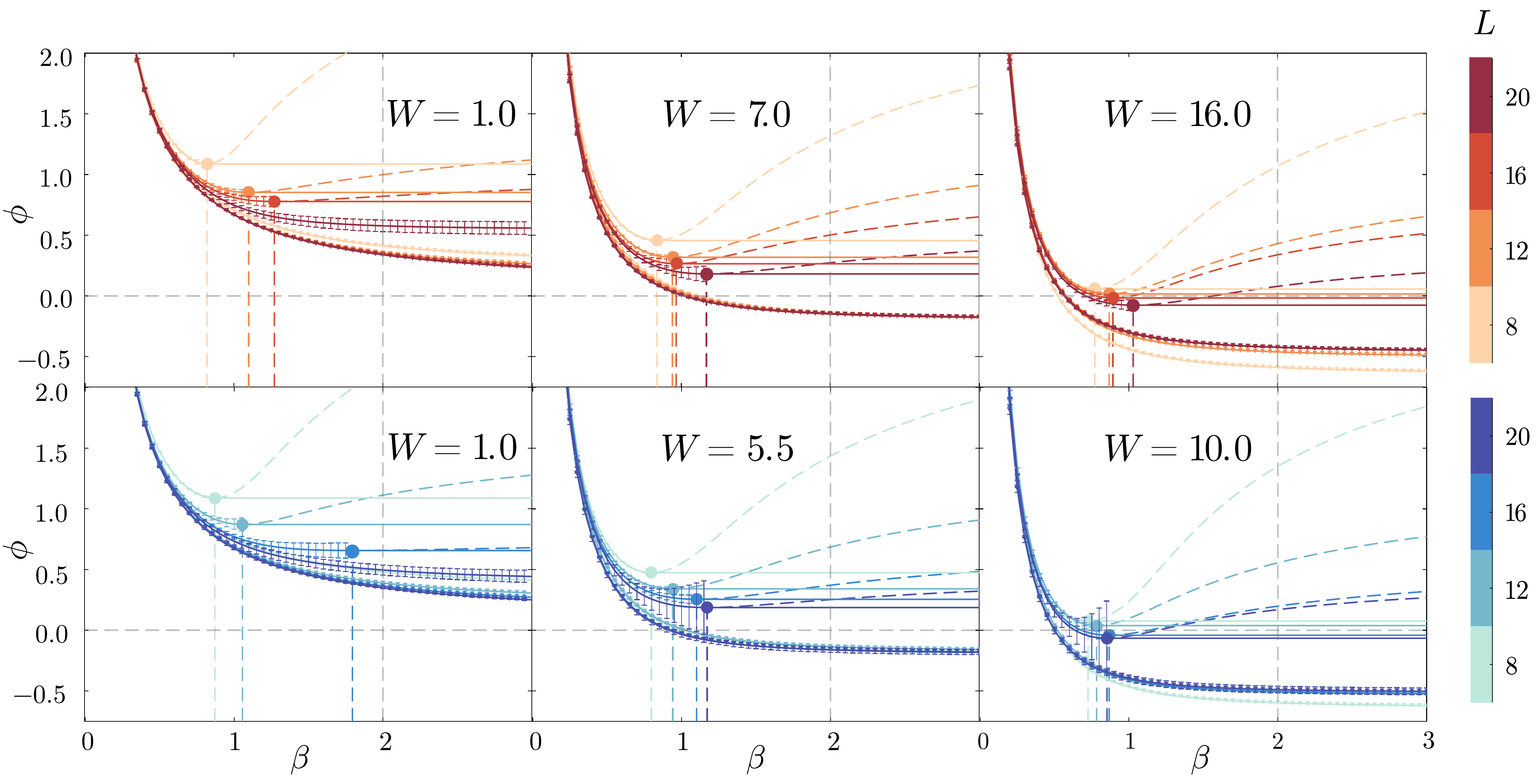}
\caption{The annealed free-energy $\phi_a$ (dashed), the modified annealed free-energy $\tilde{\phi}_a$ (solid) and the quenched free-energy $\phi_q$ (solid with triangular markers) in the spin (warm colors) and Anderson (cold colors) bases. Low (left panel), intermediate (middle panel) and large (right panel) disorder strengths. The sizes $L$ are distinguished by the colors of the scale. The dashed gray lines show the relevant values at $\beta = 2$ (physical transport) and $\phi(\beta) = 0$ (delocalization/localization). The dots and the vertical colored dashed lines mark explicitly  the position of the minima of the annealed free-energy $\beta_\star$ for each curve. Note that to improve the image readability, the $x$ axis is restricted to the interval $\beta \in [0,3]$. However, for $W=1$ (left panels) and larger system sizes ($L \ge 20$), the position of $\beta_\star$ shifts outside the plotted $x$-axis range.}
\label{fig:free_energies}
\end{figure*} 

In Fig.~\ref{fig:free_energies} we show examples of the free-energy densities as a function of the auxiliary parameter---or `inverse temperature'---$\beta$. In the Figure we show the case $\Delta = 1$ for several disorder strengths across the phase diagram, and for system sizes, shown in the color scale on the right of the figures. Results are shown for both spin basis (red, top row) and Anderson basis (blue, bottom row). The corresponding biased Hilbert space Landauer transmission, $\mathcal{T}_0(\beta)$, is computed using initial states $\ket{0}$ with expectation value of energy in the middle of the many-body spectrum. This choice, along with other simulation details, are described in~\appref{sec:SimulationDetails}.

The solid lines denote the free-energy functions $\tilde{\phi}_a$, i.e. the annealed free-energy with the replacement defined in Eq.~(\ref{eq:approx_phia}). The increasing part of the annealed free-energy functions $\phi_a$ for $\beta>\beta_\star$ are shown as dashed lines, with their respective minima marked by filled circles and vertical dashed lines of the same color. The error bars associated with these annealed free-energy functions are omitted for $\beta > \beta_\star$, as the average value of $\mathcal{T}_0$ becomes ill-defined in this regime, and $\phi_a$ depends strongly on the number of samples, being dominated by extreme value statistics.

The quenched free-energy functions $\phi_q(\beta)$ are also shown in solid lines with triangular markers.
The two relevant values $\beta = 2$ and $\tilde{\phi}_a = \phi_q = 0$ are also shown as vertical and horizontal dashed-gray lines, respectively. 

For $\beta < \beta_\star$, the annealed free-energy $\phi_a(\beta)$ (and accordingly its modified counterpart $\tilde{\phi}_a(\beta)$) closely follows the quenched free-energy $\phi_q(\beta)$, with the gap between them narrowing as the system size $L$ increases. In the vicinity of $\beta = \beta_\star$ the annealed free-energy $\phi_a(\beta)$ begins to deviate, developing into a convex function with a minimum. 
Both $\beta_\star$ and the corresponding value of the annealed free-energy $\phi_a(\beta_\star)$ exhibit a systematic drift with system size, while the value of the quenched free-energy $\phi_q(\beta = 2)$ shows a much weaker finite-size effect.

For small disorder ($W = 1$, left panels) the position of $\beta_\star$ systematically shifts toward larger values of $\beta$ with increasing $L$, until it crosses $\beta = 2$ for the largest system size considered ($L = 20$), indicating that $\mathcal{T}_0(\beta=2)$ receives contributions from an exponential number of terms. To resolve the relevant structure of the curves, Fig.~\ref{fig:free_energies} displays only the interval $\beta \in [0,3]$; in the excluded region no additional features appear apart from the minimum at larger $\beta$.

This behavior indicates that at weak enough disorder strengths, the non-concavity of the annealed free-energy is progressively lost and $\phi_a$ eventually approaches the quenched free-energy density. As a consequence, the minimum drifts toward larger $\beta$ and eventually moves outside the boundary of the sampled interval, as in this case $W = 1$. This reflects the absence of strong fluctuations dominating the sum $\mathcal{T}_0(\beta=2)$. Upon increasing the disorder strength, fluctuations become more pronounced, and the minimum at $\beta_\star$ becomes clearly visible again in the considered interval while remaining at $\beta_\star > 2$ within the ergodic phase even for the largest accessible size. This behavior is observed, for example, at $W = 2.5$ in the spin basis (see Fig.~\ref{fig:minima_Delta1} below).

In contrast, for both intermediate ($W = 5.5$ and $W = 7$, middle panels) and strong disorders ($W = 10$ and $W = 16$, right panels), we observe $\beta_\star < 2$, signaling that, at least for the accessible system sizes, the sum~\eqref{eq:FSLandauerTrans} is dominated by a few outliers in the tails of the distributions of the propagators. 
 
For the intermediate disorder regime $\phi_a(\beta_\star) > 0$, implying that the typical value of $\mathcal{T}_0$ in the large $L$ limit should remain finite due to the inclusion of rare, large outliers of $\mathcal{T}_0$. Upon increasing even more the disorder, however, $\phi_a(\beta_\star) < 0$, indicating that the typical value of $\mathcal{T}_0$ vanishes exponentially upon increasing the size of the system. 

This analysis relies on the determination of the minimum $\beta_\star$. This value is obtained by minimizing the  function $\phi_a(\beta)$, which is constructed from discrete data points, using a quasi-Newton (BFGS) method with automatically evaluated gradients. While such a procedure may, in principle, introduce some numerical noise due to the gradient evaluation, the robustness of the result can be independently assessed by determining $\beta_\star$ directly from the stationarity condition $\partial \phi_a / \partial \beta  |_{\beta = \beta_\star} = 0$. Using Eq.~\eqref{eq:phia}, this yields the following self-consistent equation for $\beta_\star$:
\begin{equation}
    \beta_\star = \frac{\ln \mathbb E \left[\sum_{f \in \cal E} |\mathcal{G}_{0f}|^{\beta_\star} \right] \mathbb E \left[\sum_{f \in \cal E} |\mathcal{G}_{0f}|^{\beta_\star} \ln |\mathcal{G}_{0f}| \right] }{\mathbb E \left[\sum_{f \in  \cal E} |\mathcal{G}_{0f}|^{\beta_\star} \right]} \;.
\end{equation}
This equation yields an independent and more stable determination of $\beta_\star$ (particularly in cases where the annealed free-energy is very flat) directly from the numerical data for $|\mathcal{G}_{0f}|$, that avoids gradient evaluations. Comparing it with the BFGS result provides a direct consistency check of the procedure.

The results in both bases exhibit qualitatively similar behavior and allow us to distinguish the three relevant regimes. 
Yet, the crossover between these regimes occur at different values of disorder strength $W$, indicating a quantitative difference between the two bases---a point to which we will return in later sections.

\subsection{The finite size phase diagrams} 
\label{sec:phasediagrams}

 \begin{figure*}[!ht]
	\centering
	\includegraphics[width=\linewidth]{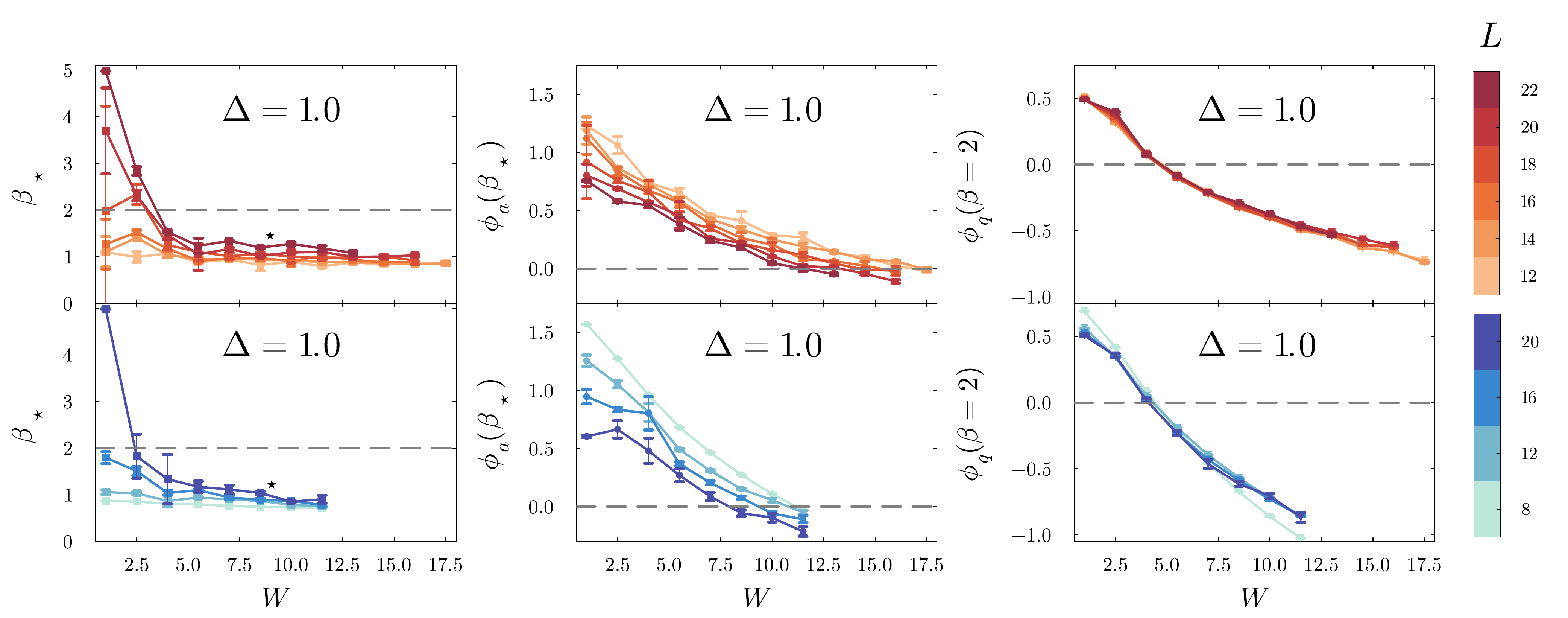}
	\caption{(From top to bottom) Disorder strength $W$ dependence of 
	$\beta_\star$, $\phi_a(\beta_\star)$, and $\phi_q(\beta = 2)$ for different system sizes,  
	in the  spin basis (left panels) and the Anderson basis (right panels). 
	The relevant values $\beta_\star = 2$ and $\phi = 0$ are correspondingly indicated with an horizontal dashed-gray line. The crossing of the curves with these lines identify the position of $W_{\rm ergo}(L)$, $W_{\rm MBL}(L)$ and $W^{\rm typ}_{\rm MBL}(L)$, accordingly. The star symbol at disorder $W = 9$ in the first row is obtained by computing the same quantity, $\beta_\star$, through another method, in order to verify the consistency of the results (see  \autoref{sec:DistributionOfLandauer}).}
	\vspace{0.5cm}
	\label{fig:minima_Delta1}
\end{figure*}

In what follows, we show the behavior of $\beta_\star$, $\phi_a(\beta_\star)$ and $\phi_q(\beta = 2)$ as a function of $W$ with increasing $L$. From these curves we extract the relevant finite-size critical disorder strengths $W_{\rm ergo}(L)$, $W_{\rm MBL}(L)$ and $W^{\rm typ}_{\rm MBL}(L)$, respectively, as it is shown in the example of Fig.~\ref{fig:minima_Delta1} for $\Delta = 1.0$. From top to bottom the panels are: the value $\beta_\star$ for which the minimum is reached, the height of the plateau $\phi_a(\beta_\star)$, and the value of the quenched free-energy for the physical transport $\phi_q(\beta = 2)$. The critical disorder strengths are computed through a linear or cubic spline interpolation using the points closest 
to the relevant crossings.  

Note however that an accurate numerical determination of the  values of $\beta_\star$, the corresponding minimum $\phi_a(\beta_\star)$, and their associated statistical uncertainties is rather delicate. In fact, the behavior of $\phi_a(\beta)$ for $\beta > \beta_\star$ is intrinsically affected by very large and uncontrolled statistical fluctuations. This originates from the very mechanism that generates the minimum: for $\beta > \beta_\star$ (i.e., at lower effective temperatures), the annealed free energy becomes dominated by rare, anomalously large contributions to the partition sum. As a consequence, the variance of $\phi_a(\beta)$ in this regime becomes ill-defined (or extremely large and dominated by extreme statistics at finite sampling), making conventional empirical error estimates unreliable.

For this reason, although the region $\beta < \beta_\star$ exhibits well-controlled and clearly bounded error bars (as shown in Fig.~\ref{fig:free_energies}), the statistical uncertainty in the vicinity of and beyond $\beta_\star$ is necessarily much larger.

To estimate the uncertainty on $\beta_\star$ and $\phi_a(\beta_\star)$, we therefore adopted a heuristic strategy based on sampling stability. Specifically, we monitor the evolution of these quantities as a function of the number of disorder realizations, computing cumulative averages and tracking their convergence upon increasing the sample size. The resulting fluctuations provide an empirical estimate of the associated error bars. The procedure is described in detail in~\appref{sec:ErrorBars}.

We emphasize that, since this approach relies on cumulative averaging, successive estimates are statistically correlated. As a consequence, the resulting error bars are inevitably underestimating the true statistical uncertainty. To the best of our knowledge, however, there is no straightforward alternative method to reliably quantify the uncertainty in a regime dominated by rare-event statistics of this type.

Finally, the error bars associated with the extracted critical disorder strengths are obtained either through standard error propagation—when the corresponding covariance matrix is well-conditioned (see~\appref{sec:ErrorBars})—or via a Monte Carlo bootstrap resampling procedure.

\begin{figure*}[!ht]
    \centering
        \includegraphics[width=.485\linewidth]{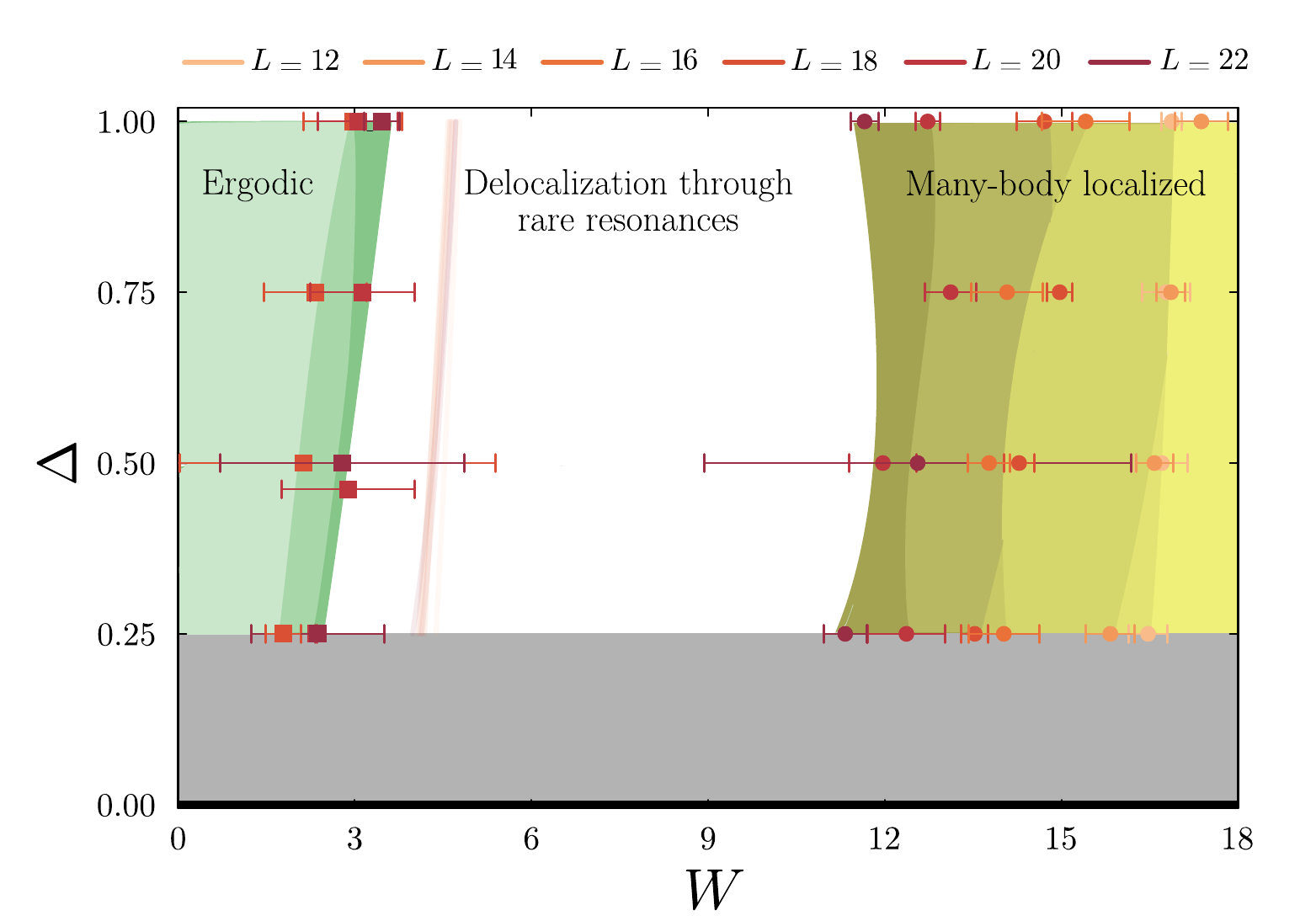}
        \includegraphics[width=.485\linewidth]{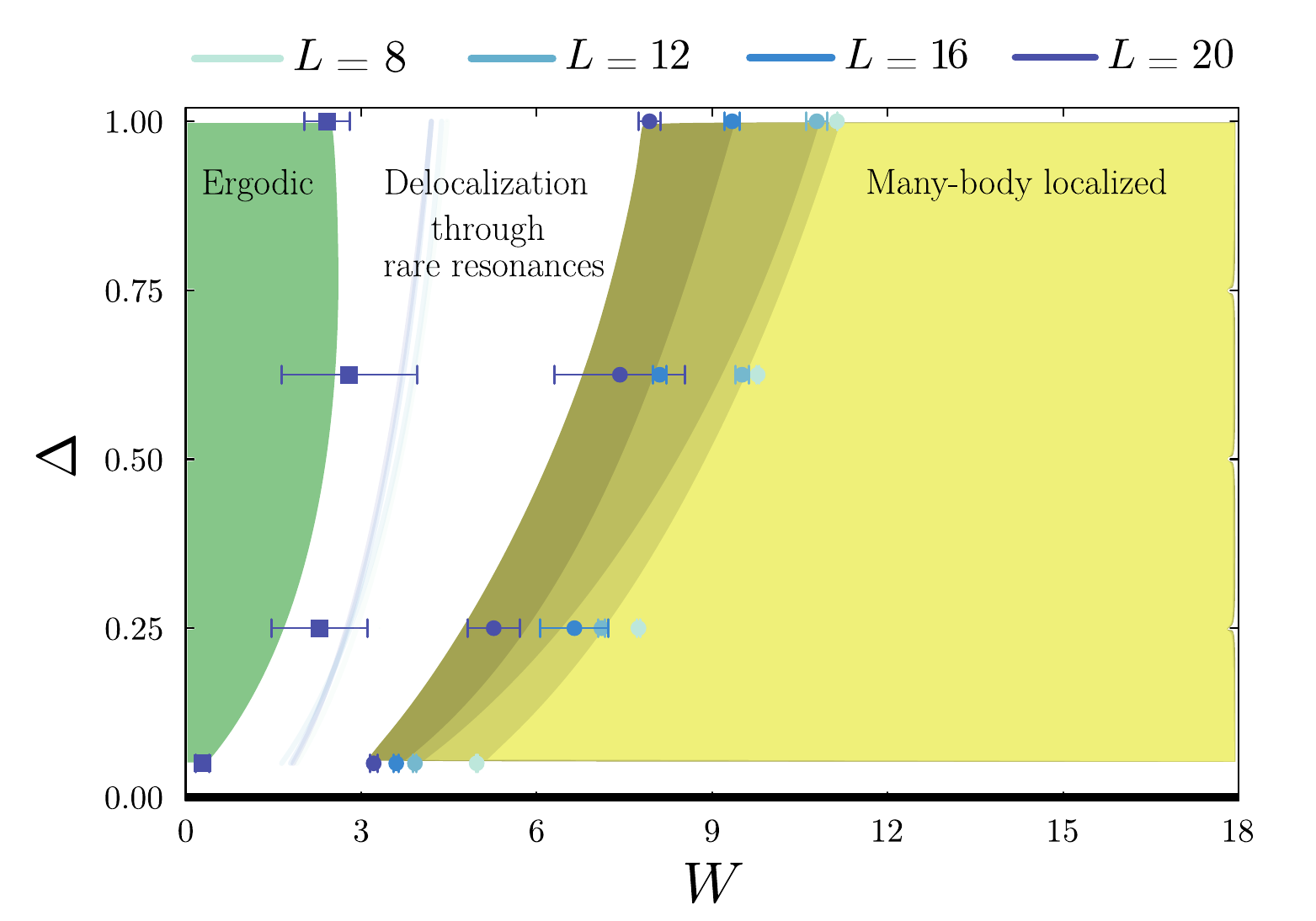}
    \caption{Phase diagram of the system at the center of the energy spectrum, shown in the $\Delta$--$W$ plane. Crossover lines between the three regimes are indicated as follows: $W_{\rm ergo}$ (square markers), $W_{\rm MBL}$ (circular markers), and $W^{\rm typ}_{\rm MBL}$ (solid transparent lines), each plotted for the system sizes indicated in the labels (top of the frame). The three identified regimes---ergodic, delocalization via rare resonances, and many-body localization---are colored in green, white, and yellow, respectively. Gradients within each colored region serve as a visual guide to distinguish crossover lines estimated for different system sizes: regions identified from smaller sizes are lighter and get darker upon increasing $L$. The limits of each region are intended as guides to the eye and are obtained through  a B\'ezier interpolation between the corresponding data points at each size.
    }
    \label{fig:phase_diagram}
\end{figure*}

We repeat this procedure for several values of $\Delta$ to construct a finite size phase-diagram in the $W-\Delta$ plane. The results of this procedure are 
shown in Fig.~\ref{fig:phase_diagram} for both the spin and Anderson bases.

We identify three distinct finite-size regimes. At low disorder strengths, where $\beta_\star > 2$ and $\phi_{q,a}(\beta = 2) > 0$, the system is in an \emph{ergodic regime}, in which $\mathcal{T}_0 (\beta=2)$ receives contributions from an exponentially large number of terms in the sum. As a result, it is self-averaging, its distribution has rapidly decaying tails, and sample-to-sample fluctuations are small. The typical and average values of $\mathcal{T}_0(2)$ both increase with the number of target states connected to the initial state $\ket{0}$, which grows exponentially with $L$. This behavior corresponds to a standard metallic regime for the conductance on the Hilbert space graph, meaning that the probability of delocalization from a random initial state $\ket{0}$ approaches 1 as $L$ increases. This regime is depicted in green in the leftmost part of the phase diagrams \autoref{fig:phase_diagram}.

A second regime emerges at intermediate disorder, $\beta_\star<2$ and $\phi_a(\beta_\star) > 0$. This implies that $\mathcal{T}_0 (\beta=2)$ is dominated by rare outliers in the tails of the distribution of the propagators, corresponding to atypical delocalization events involving basis states that differ from the initial configuration by an extensive number of spin flips.
In this regime the distribution of $\mathcal{T}_0 (2)$ develops heavy tails. According to our analogy with mean-field classical disordered systems, its typical value at large $L$ will be ultimately dominated by samples that are rare at the accessible system sizes, and that feature anomalously large transmission events, leading to delocalization through rare resonances. As a result, the typical value of the probability that the system reaches one of the configurations at zero overlap from a random initial state will eventually approach 1 in the thermodynamic limit. However, only a few specific disorder-dependent configurations will be reached under the unitary evolution, corresponding to an extremely heterogeneous spreading of the wave-packet on the Hilbert space graph. This regime is depicted in green in \autoref{fig:phase_diagram}, and is separated from the weak-disorder ergodic regime by the crossover line $W_{\text{ergo}}(L, \Delta)$, indicated by square markers (darker colors correspond to larger $L$). 

Note that the line where $\phi_q(\beta = 2) = 0$, which defines the typical disorder strength $W_{\rm typ}$, lies within the intermediate region of the phase diagram. This line separates a regime at $W<W_{\rm typ}$, where typical samples are delocalized for the system sizes accessible numerically, from a regime of stronger disorder, where typical samples are localized. The position of this line agrees reasonably well with previous estimates of the MBL transition based on standard observables and conventional approaches~\cite{luitz_many-body_2015, mace_multifractal_2019}.
Importantly, most of the intermediate region corresponds to parameters where typical samples appear localized. This implies that rare resonances, ultimately responsible for  delocalization to distant configurations over very long times in the asymptotic regime, are typically absent in the disorder realizations we can currently probe numerically. Nevertheless, our approach, inspired by mean-field theories of disordered glassy systems, provides a way to capture the asymptotic effects of such atypical disorder realizations.

At stronger disorder, the system enters a third regime, characterized by $\beta_\star<2$ and $\phi_a(\beta_\star) < 0$. This corresponds to a genuine localized behaviour, as the typical value of the transmission $\mathcal{T}_0(\beta=2)$ vanishes (exponentially) in the large $L$ limit even with the contributions of anomalously large outliers coming from the right tails of $\mathcal{G}_{0f}$. This phase is shown in shades of orange, with the corresponding crossover lines $W_{\text{MBL}}(L, \Delta)$ marked by circular symbols. The graded coloring within each region serves as a visual guide to the system sizes $L$ used in the analysis: darker tones indicate larger system sizes.  

A clear finite-size trend is observed: as $L$ increases, $W_{\text{ergo}}(L)$ shifts toward higher disorder (rightward), while $W_{\text{MBL}}(L)$ shifts toward lower disorder (leftward), reflecting a systematic drift of the phase boundaries with increasing system size. This implies that the broad regime where the system delocalizes through rare events shrinks upon increasing $L$. This leaves open the possibility that the entire intermediate region corresponds to a finite-size crossover~\cite{morningstar_avalanches_2022, Ha2023}, and that in the thermodynamic limit there is a direct transition from the ergodic phase to the MBL phase, i.e. that the two transition lines $W_{\rm ergo}(L, \Delta)$ and $W_{\rm MBL}(L, \Delta)$ may converge as $L \to \infty$.

This systematic drift of the transition point differs significantly from what was observed when applying the same method to the random field Ising model in a transverse field (also known as the Imbrie model), studied in Ref.~\cite{biroli_large-deviation_2024}. The precise origin of the discrepancy between the two models remains unclear at present, though it certainly calls for further investigation.

Furthermore, the crossover lines found in the spin and in the Anderson basis are quantitatively different, and the discrepancy is most pronounced at small $\Delta$. In particular, the spin basis seems to perform poorly in detecting localization at small values of the interaction. This is clearest at $\Delta = 0$ (horizontal black line), where the eigenstates of the Hamiltonian are tensor products (Slater determinants) of the single-particle eigenstates of the $1d$ Anderson model. The latter are exponentially localized over a disorder- and energy-dependent characteristic length $\xi_{\rm loc}$ (larger near the band center and smaller near the edges~\cite{Colbois2023}). Because these localized orbitals are not aligned with the spin basis states, a random spin-basis product state has support on many such single-particle orbitals (see App.~\ref{app:IPR}). As a result, even at $\Delta = 0$ the state exhibits apparent spreading in the spin basis: the wavepacket spreads dynamically until it 'accommodates' into a superposition of eigenstates where the initial condition has a strong support. In other words, the `delocalization' seen in the spin basis at $\Delta = 0$ is largely basis-rotation-induced rather than a true transport-type delocalization between Anderson-basis states. As explained in App.~\ref{app:IPR} with an simple heuristically argument, a many-body eigenstate constructed as a tensor product of single-particle localized orbitals necessarily occupies an exponentially large volume of the Hilbert space in the spin or particle basis, while, on the contrary, the same eigenstate is localized on a single node of the Hilbert-space graph when expressed in the Anderson basis by construction. These eigenstates appear thus as multifractal in the spin basis (although for a rather trivial reason), with a fractal dimension roughly given by Eq.~\eqref{eq:D2}. We expect this separation between basis-rotation–induced spreading and genuine delocalization to persist with the addition of interactions ($\Delta \neq 0$), where the LIOM picture applies. 


To correctly recognize that this initial partial spreading does not correspond to proper delocalization, it is necessary for the system size to satisfy $L \gg \xi_{\rm loc}$. For the system sizes accessible in numerical simulations, this condition is only met at sufficiently strong disorder, where the localization length is small enough compared to the system size. At smaller disorder, the localization length $\xi_{\rm loc}$ becomes large, introducing strong finite-size effects that hinder the observation of the localization behavior.
This implies that at small $\Delta$ and moderate $W$, the spin basis is not well suited for the method, as it differs too much from the basis formed by the many-body localized eigenstates of the Hamiltonian. This is why, in our analysis using the spin basis, the transition lines appear essentially vertical as $\Delta$ decreases. In particular, we can explicitly verify that our method in the spin basis fails to correctly capture localization at small $W$ when $\Delta = 0$. For these reasons, we have restricted our analysis to the spin basis to the regime $\Delta \geq 0.25$, below which it becomes unreliable.

\begin{figure*}[!ht]
\centering
  \includegraphics[width=.45\linewidth]{Figures/XXZFSgraph_Spins.pdf}
  \includegraphics[width=.45\linewidth]{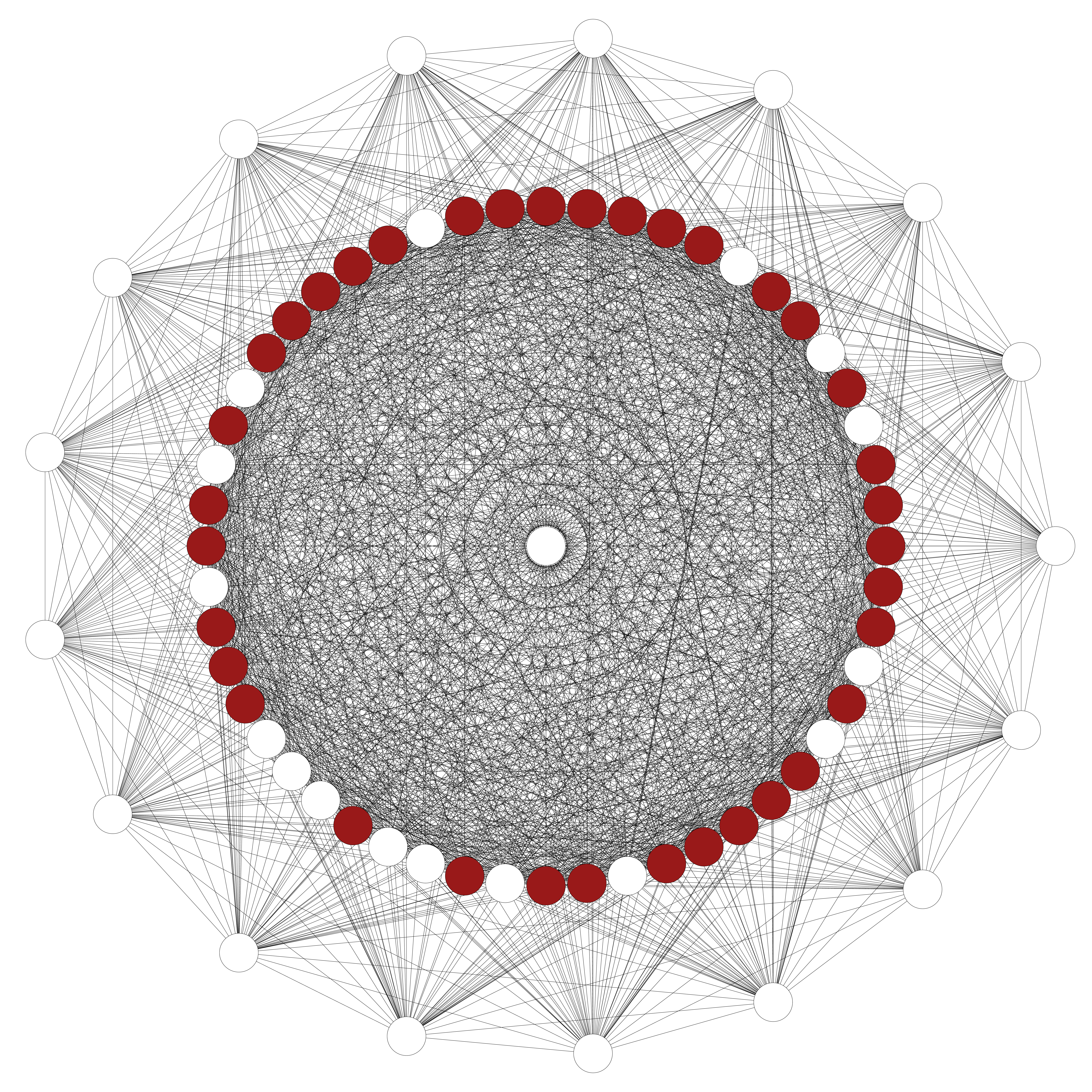}
\caption{Hilbert space graphs for $L = 8$ shown in the spin basis (left) and the Anderson basis (right). The central vertex represents a random initial condition in the middle of the energy spectrum, that is a basis state of the Hamiltonian. All vertices of the graph are connected with black edges denoting the distance in the Hilbert space graph, given by number of applications of the Hamiltonian. In the spin basis we use concentric blue-dashed circles to aid the identification of these equidistant vertices, as the structure in this case is more irregular. The red colored vertices correspond to basis states belonging to the equator set $\mathcal E$ in both bases. 
}
\label{fig:Graphs}
\end{figure*} 
Additionally, the spin basis systematically predicts finite-size transitions at stronger disorder values compared to the Anderson basis. This leads to a substantial quantitative difference in the width of the regime where delocalization is driven by rare events. The origin of this discrepancy appears to be twofold.

First, as discussed above, the nature of the initial state plays a crucial role and depends strongly on the basis in which it is defined. In the Anderson basis, each basis state corresponds to a specific configuration of localized single-particle orbitals and is therefore a local modification of an eigenstate of the local integrals of motion (LIOMs). As such, an initial condition prepared in an Anderson basis state is expected to exhibit a small spreading in the MBL phase. 
In contrast, an initial state defined in the spin basis is expected to spread over many other vertices in Hilbert space---specifically, over all those LIOM eigenstates on which the initial state has significant projection---even in the MBL phase. This initial spreading leads to pronounced finite-size effects whenever $L \sim \xi_{\rm loc}$. The distinct spreading observed between different bases can also be understood from the multifractal properties of the eigenstates: in the MBL regime, Ref.~\cite{mace_multifractal_2019} reports a significantly smaller fractal dimension in the Anderson basis as compared to the spin basis.
Secondly, there is a mismatch between the equator sets $\mathcal{A}_0(q = 0)$ and $\mathcal{S}_0(q = 0)$, used to identify target basis states at large distance from the initial state $\ket{0}$. In the Anderson basis, all target states $\ket{f} \in \mathcal{E} = \mathcal{A}_0(q = 0)$ lie at the same graph distance  from $\ket{0}$ on the Hilbert space graph, (defined as the length of the shortest path between $\ket{0}$ and $\ket{f}$, i.e., the minimum number of applications of the off-diagonal part of the Hamiltonian needed to connect them). In contrast, the target states in the spin basis, $\ket{f} \in \mathcal{E} = \mathcal{S}_0(q = 0)$, have varying distances from $\ket{0}$, as different vertices have different degrees. The degree depends on the number of domain walls present in each basis state, which can fluctuate from $2$ to $L$, as explained in~\autoref{sec:spinbasis}. 

This difference between the two basis is illustrated in \autoref{fig:Graphs}, where we present the complete Hilbert space graph for the case $L=8$ in both bases. The central vertex represents a randomly chosen initial condition with expectation value of energy near the middle of the many-body spectrum. All edges connecting the Hilbert states through applications of the off-diagonal part of the Hamiltonian are shown. Vertices are arranged radially outward according to their distance on the graph from the central vertex $\ket{0}$. In the spin basis, we overlay concentric circles to help identify the vertices that are equidistant from $\ket{0}$, as the structure in this case is more irregular due to the fluctuating connectivity of the nodes. While in the Anderson basis all target states lie at the same graph distance from the initial condition, the targets in the spin basis are more dispersed: a large fraction of them are at a distance $L/4$ on the graph, but others are found at larger distances. Some of these target states are even located at the maximal possible distance from $\ket{0}$, reflecting the broader distribution of distances between spin configurations at zero overlap in the spin basis. 

However, as the system size increases, the differences between the two computational bases are expected to decrease. On the one hand, due to the local nature of the LIOMs, when $L \gg \xi_{\rm loc}$, the ‘blobs’ representing LIOM eigenstates in the spin basis become effectively smaller than the total Hilbert space volume.
On the other hand, as $L$ increases, most target states tend to concentrate uniformly around a distance $L/4$ from the initial state, since the majority of nodes on the Hilbert space graph contain approximately $L/2$ domain walls. As a result, we expect the transition lines to the MBL regime to eventually converge to a common value, independent of the choice of computational basis.

Remarkably, even very weak interactions (e.g., $\Delta = 0.05$ in the Anderson basis, see \autoref{fig:phase_diagram}) yield a finite typical value of $\mathcal{T}_0$, signaling delocalization for the finite system sizes considered. In other words, both transition lines, $W_{\rm MBL}$ and $W_{\rm MBL}^{\rm typ}$, remain finite at small $\Delta$, indicating a discontinuous departure from the Anderson insulator at $\Delta = 0$, where localization persists at arbitrarily small disorder $W$. This provides further numerical evidence of the non-perturbative effect of interactions, consistent with the spin–spin correlation analysis and the updated XXZ phase diagram reported in Refs.~\cite{colbois_interaction-driven_2024, colbois_statistics_2024}. The interaction-driven delocalization instability is a subtle effect that has recently been confirmed to be non-perturbative in Ref.~\cite{jiang2025quasiconservationlawssuppressedtransport}, where first-order corrections to the noninteracting local integrals of motion were computed, and no indication of instability was found.

\subsection{The distributions of the Hilbert space Landauer transmissions}
\label{sec:DistributionOfLandauer}

\begin{figure*}[!ht]
    \centering
    \includegraphics[width=\linewidth]{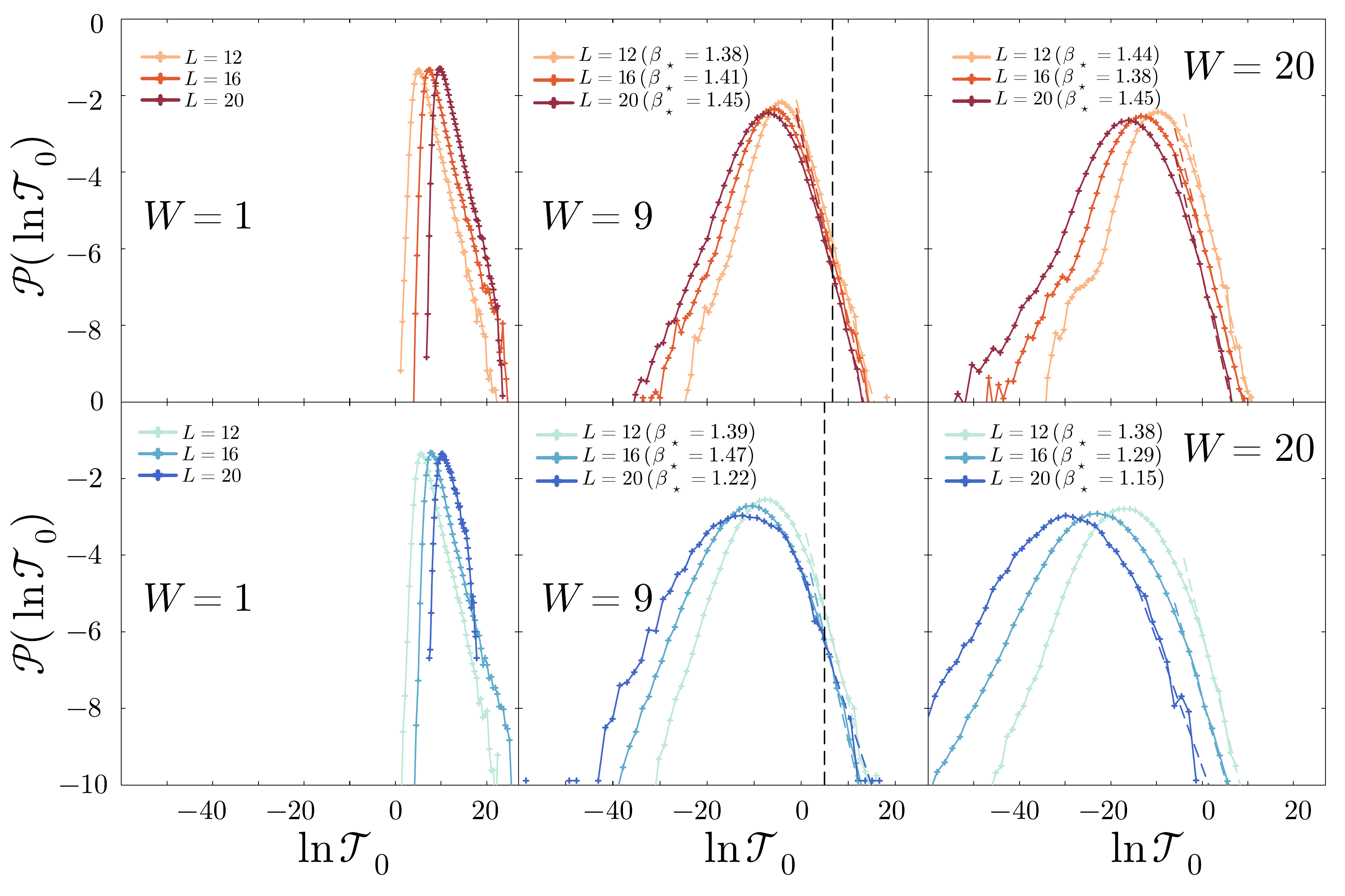}
    \caption{
       Probability distributions for the Hilbert space Landauer transmission $\mathcal{T}_0$, for the different sizes displayed in the legend. For small (left), mid (center), and strong (right) disorder strengths, in both spin (top panels) and Anderson (bottom panels) bases. These plots are the counterpart of the probability distributions of Fig.~\ref{fig:violins}, where the probability to delocalize to the states at zero overlap from the random initial state is replaced by our proxy $\mathcal{T}_0$. 
       The vertical black-dashed line in the middle panel corresponds to the Hilbert space Landauer transmission,  from a disorder realization that give rise to cat states with system-wide resonances, for $L = 20$ (see the main text for a detailed explanation).}
    \label{fig:pdf_conductivities}
\end{figure*}

In this section, we will explicitly examine the probability distributions of the Hilbert space Landauer transmissions $\mathcal{T}_0$, for several disorder strengths $W$ and system sizes $L$.  In \autoref{fig:pdf_conductivities}, we show the results for three different system sizes ($L = 12,\, 16,\, 20$) and three disorder strengths ($W = 1,\, 9,\, 20$). 

At weak disorder, $W = 1$, the peak of the distribution shifts rightward with increasing $L$, and the distributions develop a sharper cutoff. This behavior indicates the absence of rare outliers and suggests that the typical and average values of $\mathcal{T}_0$ are proportional to each other, growing as a power of $L$ due to the increasing number of outgoing channels $\mathcal{N}_\mathcal{E}$. This reflects the fact that $\phi_{q,a}(\beta = 2) > 0$, signaling an ergodic regime in which the typical value of $\mathcal{T}_0$ receives contributions from an exponential number of target states $\ket{f}$.

In contrast, for stronger disorder values, $W = 9$ and $W = 20$, the peak of the distribution shifts leftward, while the tails remain broad. This indicates the presence of significant fluctuations, causing the average and typical values of $\mathcal{T}_0$ to differ substantially. According to Derrida's theory of the freezing transition of directed polymers and its generalizations~\cite{derrida_polymers_1988, Gardner1989}, the exponent governing the tail of the probability distribution of $\mathcal{T}_0(\beta)$ is related to the freezing inverse temperature $\beta_\star$ via:
\begin{equation}
 \mathcal P (\mathcal T_0(\beta) ) \simeq \frac{e^{-\beta_\star^2 \ln \mathcal{N}_\mathcal{E} \phi_q / \beta}}{\mathcal{T}_0^{1+\beta_\star/\beta}} \;, 
\end{equation}
This behavior has a clear intuitive origin: for $\beta < \beta_\star$, the typical and average values of $\mathcal{T}_0$ remain proportional, since the average is dominated by the bulk of the distribution rather than its tail. In contrast, for $\beta > \beta_\star$, the average becomes dominated by rare, large fluctuations, as the tail of the distribution decays with an exponent smaller than 2. We test this prediction by extracting the value of $\beta_\star$ from the power-law fit of the tails of the distribution $\mathcal{P}(\mathcal{T}_0(\beta = 2) )$. The values of $\beta_\star$ found from the fits of the power-law tails of the distributions are displayed in the key of \autoref{fig:pdf_conductivities} and shown in \autoref{fig:minima_Delta1} with a star symbol for the largest system size $L = 20$ and disorder strength $W = 9$, showing consistency with the value of $\beta_\star$ found from the position of the minimum of the annealed free-energy.

Furthermore, in a recent study~\cite{Laflorencie2025-mv}, the existence of an intermediate, non-ergodic phase in the disordered Heisenberg chain was linked to the emergence of unusual high-energy eigenstates exhibiting anomalously strong longitudinal spin–spin correlations~\cite{colbois_statistics_2024, colbois_interaction-driven_2024}. These eigenstates appear in nearly degenerate pairs, sparsely distributed across the exponentially large many-body spectrum. Remarkably, their properties are accurately captured by a simple toy model of \emph{cat states}. These cat states take approximatively the form $\ket{\psi}_\pm \sim \ket{I_1} \pm \ket{I_2}$, where $\ket{I_1}$ and $\ket{I_2}$ are spin-basis states.

The occurrence of such cat eigenstates is frequent in the intermediate disorder regime ($W \sim 10$), but they become increasingly rare at stronger disorder ($W \gtrsim 20$). Intuitively, the presence of a resonant cat state in a localized system can enhance the probability of delocalization: if a random initial spin basis state has a significant projection with one of these cat states, a resonance between $\ket{I_1}$ and $\ket{I_2}$ enables the system to explore both configurations, thus promoting delocalization. However, such states are rare in the spectrum, and their impact on the decorrelation from a typical random initial state—and their direct connection to delocalization via rare resonances—remains unestablished.

To fill this gap, we test this picture with our observables defined in Hilbert space, and we compute the Hilbert space Landauer transmission $\mathcal{T}_0$---averaged over several initial conditions---for a specific disorder realization at system size $L = 20$ and disorder strength $W = 9$. This particular realization hosts pairs of nearly degenerate eigenstates exhibiting strong spin–spin correlations, consistent with the cat-state scenario~\cite{Laflorencie2025-mv}. The corresponding value of $\mathcal{T}_0$ is indicated by the black dashed line in the middle panel of \autoref{fig:pdf_conductivities} for $W = 9$. Interestingly, this sample is also classified as a rare event in terms of the Hilbert space Landauer transmission, as its corresponding $\mathcal{T}_0$ lies within the tail of the distribution, where $P(\mathcal{T}_0(\beta=2)) \approx 10^{-6}$. This observation suggests a potential link between atypical values of Hilbert space observables---such as $\mathcal{T}_0$---and real space features like longitudinal spin–spin correlations. 

\subsection{Dependence on the target states distance}
\label{sec:new_targets}
In this section, we modify the selection of target states, which so far has been restricted to the equator set $\mathcal{E}$. Specifically, to investigate the progressive spreading of the wave packet onto configurations at increasing distances, we vary the parameter $q$ in the sets $\mathcal{S}_0(q)$ and $\mathcal{A}_0(q)$, defined in~\autoeqref{eq:set_targets}, for the spin and Anderson bases, respectively. We reintroduce the superscript on the overlap value, $q \to q^{S, A}$, to explicitly indicate the basis—spin (S) or Anderson (A)—in which the overlap is being measured. We reparametrize these overlaps to define a \emph{correlation distance} $\zeta^{S,A}$, which serves as an ultrametric distance in Hilbert space, and is defined by
\begin{equation}
    \zeta^{S,A} = 1 - q^{S,A} \;.
    \label{eq:re-scaled-overlaps}
\end{equation} where $\zeta^{S,A} = 0$ trivially corresponds to the initial random basis state itself, while $\zeta^{S,A} = 1$ represents completely uncorrelated states, where half of the spins have been flipped in terms of the spin basis or, equivalently, where half of the spinless-fermions have hopped to different orbital occupations, in the Anderson basis. We will omit the cases $1 < \zeta^{S,A} < 2$, that correspond to anti-correlated basis states with respect to the initial condition $\ket{0}$. 
In what follows we restrict the analysis to $\Delta = 1$.

\begin{figure}[!ht]
\centering
  \centering
  \includegraphics[width=\linewidth]{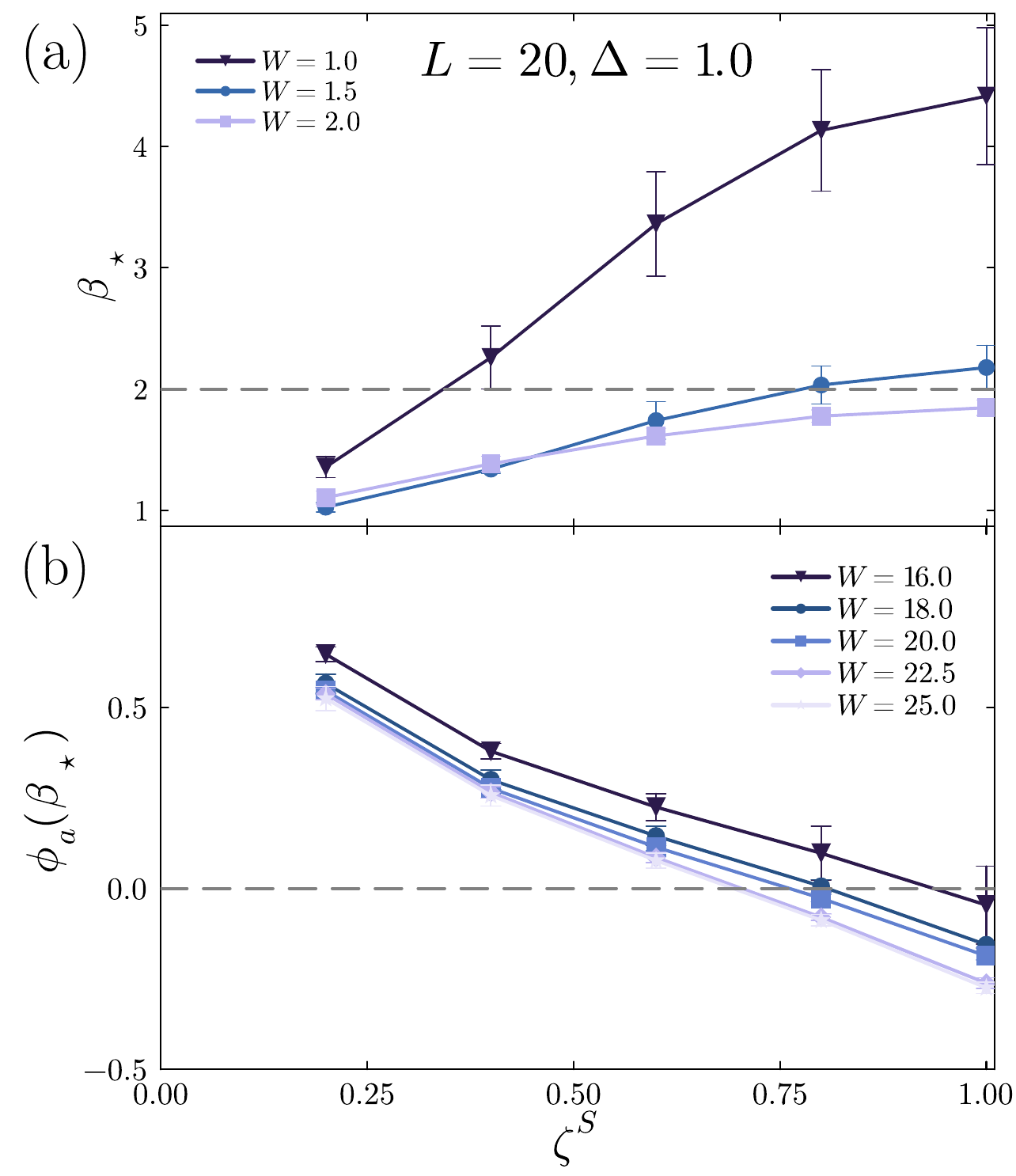}
\caption{Calculation of (a) $\beta_\star$ and (b) $\phi_a(\beta_\star)$ as functions of the correlation distance $\zeta^S$, for $\Delta = 1$ and $L = 20$. The values of the disorder strengths considered are shown in the legend. Horizontal gray dashed lines indicate the reference values $\beta_\star = 2$ and $\phi_a(\beta_\star) = 0$.}
\label{fig:MinimaNewTargets}
\end{figure}

We have computed $\mathcal{T}_0(\beta)$ while varying the correlation distance $\zeta^{S,A}$ between the target nodes and the initial condition, for several values of the disorder strength $W$ and system sizes $L$, averaging over many disorder realizations and many initial states $\ket{0}$ randomly chosen close to the middle of the many-body spectrum. The procedure follows the same steps as in previous sections, with the only difference being the new dependence of the target states on $\zeta^{S,A}$. We evaluate the annealed free-energy $\phi_a(\beta, L, \zeta^{S,A})$ and extract the corresponding values of $\beta_\star$ and $\phi_a(\beta_\star)$ for different values of the disorder close to the ergodic and MBL transitions, respectively. From these, we determine the characteristic distances that fulfill $\beta_\star(\zeta^{S,A}) = 2$ and $\phi_a(\beta_\star, \zeta^{S,A}) = 0$ by performing linear interpolations. An example of this procedure is shown in~\autoref{fig:MinimaNewTargets} for the spin basis, with $L = 20$ and $\Delta = 1$, an equivalent example for the Anderson basis is shown in~\appref{sec:AndersonResults}.

\begin{figure*}[!ht]
\centering
  \centering
  \includegraphics[width=\linewidth]{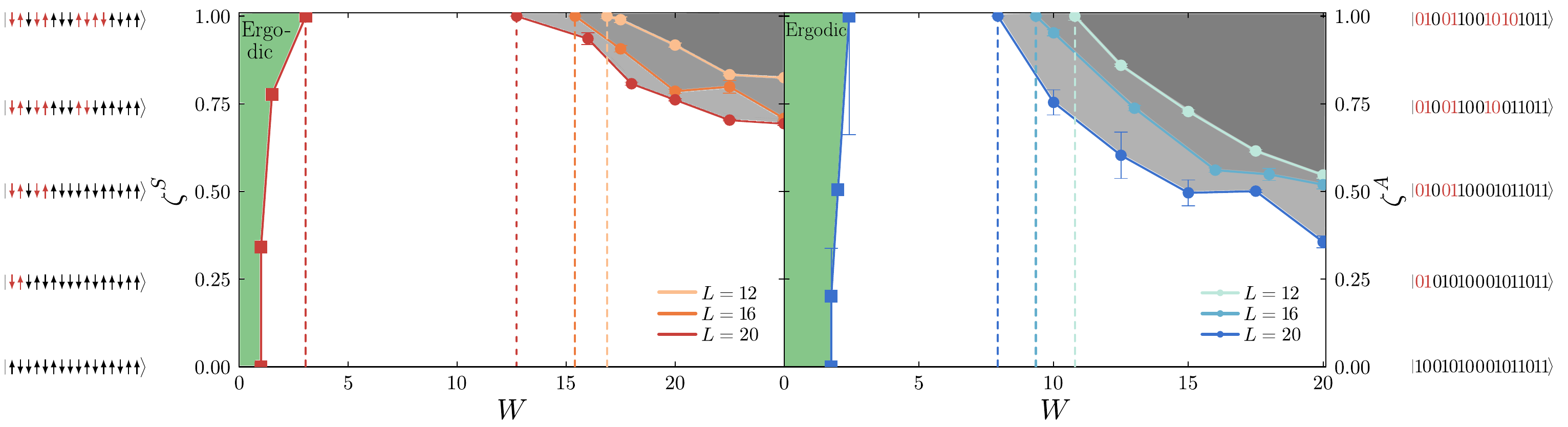}
\caption{Relevant regions in the $W$-$\zeta^{S,A}$ plane for spin (left) and Anderson (right) bases, with $\Delta = 1$. The transition lines to the ergodic region (green) is determined by the condition $\beta_\star(\zeta^{S,A}) = 2$, while the inaccessible regions (shades of gray) are determined by $\phi_a(\beta_\star, \zeta^{S,A}) = 0$, for each system size used. The critical disorder strengths $W_{\rm ergo}$ and $W_{\rm MBL}$ are identified with dashed lines, colored according to their respective system size. On the sides of each vertical axis, a random initial basis state is shown (for $\zeta^{S,A} = 0$ and $L=16$), followed by one of the possible basis states at increasing distances $\zeta^{S,A}$. The equator states, corresponding to $\zeta^S = 1$, represent a completely uncorrelated state with respect to the initial one. 
}
\label{fig:PD-distance}
\end{figure*} 

These characteristic distances define crossover lines that separate different regimes in the $W$-$\zeta$ plane, as shown in~\autoref{fig:PD-distance} for both the spin (left) and Anderson (right) bases. For the condition $\beta_\star(\zeta^{S,A}) = 2$, we used the largest system size, $L = 20$, where such a value of $\zeta^{S,A}$ is obtained. In contrast, for the condition $\phi_a(\beta_\star, \zeta^{S,A}) = 0$, we display the results for three different system sizes: $L = 12$, $16$, and $20$.  

In the first region (shaded green), the system remains ergodic within the distance defined by $\beta_\star(\zeta^{S,A}) = 2$: A randomly chosen initial basis state has a high probability of spreading uniformly to any other state in Hilbert space within the corresponding correlation distance. 
Interestingly, the curve $\beta_\star(\zeta^{S,A}) = 2$ bends to the right (i.e., toward larger disorder values) as the distance increases. This indicates the existence of a disorder window ($1.5 \lesssim W \lesssim 3$ approximately) where the spreading of the wave packet is inhomogeneous and driven by rare resonances at short distances, but recovers a uniform, ergodic-like spreading over an exponential number of configurations at larger distances. A similar behavior is observed on the metallic side of the Anderson model on the Bethe lattice~\cite{Biroli2017,biroli2020anomalous}.

In the white region, the spreading of the wave packet from the initial state $\ket{0}$ is highly inhomogeneous and dominated by a few rare resonances in the broad tails of the propagator distribution at the corresponding distances. As explained above, at weak disorder, for $W < W_{\rm ergo}$, the system eventually recover an ergodic behavior at large distances, after crossing the crossover line separating the white from the green region.
In the intermediate regime, $W_{\rm ergo} < W < W_{\rm MBL}$, the dynamics is still driven by rare resonances up to the distances corresponding to states with zero overlap with $\ket{0}$. As a result, the spreading remains inhomogeneous across the entire Hilbert space. For $W > W_{\rm MBL}$, the transport is dominated by a few $O(1)$ resonances at short distances; beyond that, i.e., for distances within the gray regions, they become so rare that the probability for a random initial state to reach such distant configurations decays exponentially with system size---even when accounting for the statistical weight of rare events. The crossover distance, determined by the condition $\phi(\beta_\star, \zeta^{S,A}) = 0$ and marking the boundary of the gray regions, represents the maximal Hilbert space correlation length accessible under unitary dynamics from a typical initial state. Consequently, in the large-$L$ limit, the gray regions become asymptotically inaccessible for typical initial conditions and disorder realizations. 

As the disorder strength grows, the system becomes progressively confined to a smaller portion of Hilbert space near the initially prepared configuration. In other words, resonant transmissions become increasingly short-ranged. This behavior becomes more pronounced with larger system sizes, as the inaccessible region of Hilbert space expands, restricting more and more the set of basis states available for delocalization. This behavior reflects the persistent memory of the initial condition at strong disorder, where the values along the transition line $\phi_a(\beta_\star) = 0$ define the maximum correlation distance (i.e., the minimal overlap) the system can explore.

Note that even deep within the MBL phase (e.g. $W=20$) there still exist rare resonances in the Hilbert space that allows the system to flip a finite fraction of the spins and partially decorrelate from the initial condition. This is consistent with the picture of resonant cat states~\cite{Laflorencie2025-mv}, formed by two nearly degenerate spin basis states that differ by a fraction of spin flips. These rare events can drive partial delocalization even in strongly disordered regimes, within the system sizes considered.

\subsection{Rare {\it vs} typical samples}
\label{sec:decays}
 \begin{figure*}[!ht]
 \vspace{0.5cm}
    \centering
    \includegraphics[width=\linewidth]{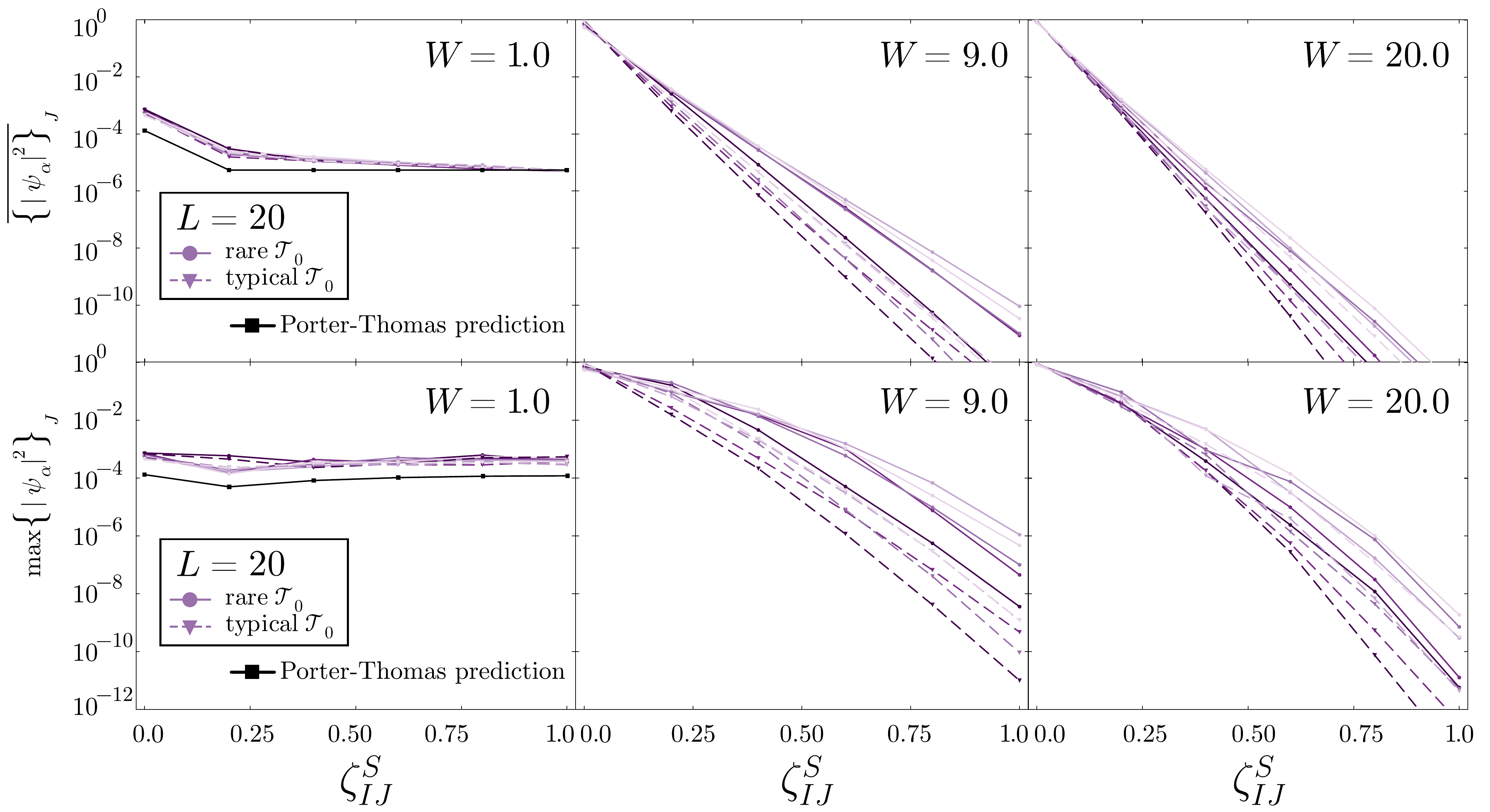}
     \caption{Basis state amplitudes, within a given eigenstate, as a function of the correlation distance $\zeta_{IJ}^{S}$ from the basis state maximal amplitude. The top row of panels shows the average amplitudes over all spin states that share the same distance  $\zeta^S$, while the bottom panel displays the maximum amplitudes among those same basis states.  Different color tones correspond to five distinct eigenstates. Dashed lines with triangular markers represent typical disorder realizations, whereas solid lines with circular markers denote rare realizations.} 
    \label{fig:EVdecays_bosons}
\end{figure*}

One of the key features of our proxy observable for the probability to decorrelate from a random initial state, $\mathcal{T}_0$, is that it naturally allows us to distinguish between typical and rare samples. In this section, we take advantage of this property to explore how the spectral and transport features differ when the quenched disorder corresponds to rare versus typical realizations. Specifically, rare samples are defined as disorder realizations of the random fields $h_i$ for which the corresponding value $\mathcal{T}_0$, averaged over multiple initial conditions, lies in the tails of the distributions shown in \autoref{fig:pdf_conductivities}. In contrast, typical samples are selected from the vicinity of the peak of the $\mathcal{T}_0$ distribution, representing the most probable values. We denote these rare and typical realizations of random fields as $\{ h_i\}_{\rm rare}$ and $\{ h_i\}_{\rm typ}$, respectively.
  
\subsubsection{The structure of the eigenstates' amplitudes}
We first probe these typical and rare samples by diagonalizing their associated Hamiltonian, through an implementation of Chebyshev filter diagonalization~\cite{pieper_high-performance_2016} that finds eigenpairs within the center of the many-body spectrum~\cite{Andreanov2025}. We perform this procedure for small ($W = 1$), intermediate ($W = 9$) and strong ($W = 20$) disorders, for a fixed system size $L = 20$. 

For each of the several eigenstates found---here numbered with the sub-index $\alpha$ ---we extract the basis state with largest wavefunction amplitude, i.e. $I \equiv \arg \max_{I'} |\psi_\alpha(I')|^2$.  For a given eigenstate $\alpha$, we order the amplitudes of all other basis states $|\psi_\alpha(J)|^2_{J \neq I}$, in terms of the correlation distance to the most probable state $I$. Similar to Eq.~\eqref{eq:re-scaled-overlaps}, this correlation distance is given by
\begin{equation}
    \zeta^{S,A}_{IJ} = 1 - q_{IJ}^{S,A} \;,\label{eq:correlation_distance_decays}
\end{equation} for the spin and Anderson bases, respectively. The basic idea is to study the decay of many-body eigenstates from their main peak, by averaging the wavefunction amplitudes of all states $\{ \ket J \}$ at equal correlation distances $\zeta^S_{IJ}$. The results of this procedure are shown in the top panel of \autoref{fig:EVdecays_bosons}, where we have calculated the corresponding eigenstates for both rare, $\{h_i\}_{\rm rare}$, and typical, $\{h_i\}_{\rm typ}$, realizations of the random fields. We present the results associated with the Anderson basis in~\appref{sec:AndersonResults}. 

At weak disorder ($W = 1$), the amplitudes of the eigenstates remain nearly constant across the Hilbert space graph, showing no decay with distance. Fluctuations between eigenstates are minimal, even across different disorder realizations. In contrast, at stronger disorder ($W = 9$ and $W = 20$), the eigenstates exhibit exponentially fast decay with the correlation distance $\zeta^{(S)}$, becoming strongly peaked around the reference basis state $I$ with amplitudes of order 1. In this regime, differences between disorder realizations become apparent: the fluctuations in eigenstate amplitudes are meaningful---specially for the the intermediate disorder strength of $W = 9$---and eigenstates from rare disorder realizations tend to decay more slowly, by $2$ orders of magnitudes close to $\zeta^{(S)}=1$, compared to those from typical ones.

To further explore this behavior, we modify the previous analysis by taking the maximum amplitude---rather than the average---among all basis states at a given correlation distance $\zeta^S$. This modified approach is illustrated in the bottom panel of \autoref{fig:EVdecays_bosons}. The basic idea is to investigate whether rare, disorder-dependent resonances occurring at specific points in the Hilbert space graph lead to an inhomogeneous decay of the wave functions along different paths, resulting in a strong disparity between the average decay and the decay along the path corresponding to the maximal amplitude.

For weak disorder ($W = 1$), the change is minimal: the decay remains largely unaffected, as the eigenstates display ergodic behavior. In this regime, amplitude fluctuations are small, and the wavefunction remains nearly uniform across the entire Hilbert space graph, equivalent to its averaged counterpart. In contrast, at strong disorder ($W = 9$ and $W = 20$), the use of the maximum amplitude reveals significant fluctuations between eigenstates, particularly between typical and rare disorder realizations. In these cases, eigenstates from typical realizations decay faster than those from rare ones. Moreover, examining individual eigenstates shows that the decay away from the maximum amplitude is highly anisotropic across the Hilbert space graph: Specific directions,  aligned with the largest amplitudes exhibit a much slower decay (by 3 or 4 ourders of magnitudes close to $\zeta^S=1$, than the average one. In other words, there exist spin basis states at large Hilbert space distances whose amplitudes are anomalously large compared to the typical amplitudes at the same distance. 

This effect is especially pronounced at intermediate disorder. It suggests that, in this regime, delocalization proceeds in a highly heterogeneous manner, along rare, disorder-dependent paths through the Hilbert space graph. These paths are determined by the presence of long-range resonances that connect distant basis states and dominate the eigenstate structure in certain realizations. We will further investigate this structure by explicitly examining the presence of resonant paths in the Hilbert space graph, as generated by both typical and rare realizations of the random disorder fields.

\subsubsection{Rarefaction of paths on the Hilbert space graph}
\label{sec:paths}

Our results so far suggest that there is a broad intermediate regime in which delocalization occurs via a small number of rare long-range resonances on the Hilbert space graph. This interpretation is also supported by the analysis of the eigenstate structure presented above. In this section, we further explore this scenario by directly investigating the paths on the Hilbert space graph that contribute most significantly to the delocalization probability from a random initial configuration. We adopt an approach that has previously been used by Lemari\'e in Ref.~\cite{lemarie_glassy_2019} to study the zero temperature properties of single-particle Anderson localization in a two dimensional geometry. 

In the standard setup of quantum transport involving a scattering geometry, as illustrated schematically in~\autoref{fig:schematic_proxy}, electrons are injected from the leads on the left and extracted through those on the right. In the presence of strong disorder within the scattering region, electron transport becomes highly inhomogeneous. Rather than spreading uniformly, an electron at zero temperature follows a narrow, meandering path through the disordered potential landscape—effectively forming a 'trajectory' or conducting channel connecting the leads~\cite{Datta2013-vq}. This behavior contrasts with the weak-disorder, diffusive regime, in which the electron’s probability distribution is approximately uniform across the sample. A central challenge, therefore, is how to visualize or reconstruct these hidden transmission paths in the localized regime.

As introduced in Refs.~\cite{Pichard1991, Marko2010}, one can devise a clever numerical perturbation technique, inspired by experimental scanning gate microscopy methods~\cite{Abbout2011, Gorini2013} to visualize these dominant paths. In such experiments, a movable tip locally modifies the potential landscape of a nanoscale conductor, and the resulting changes in conductance reveal the regions through which current flows. The numerical analog operates as follows: for a given disordered sample, one slightly perturbs the on-site disorder at a specific location and measures how much the conductance $g_0$ between the incoming and outgoing leads changes. If the perturbed site lies along a main transmission path, even a local modification will significantly affect the coherent transport, resulting in a noticeable change in $g_0$. Conversely, if the site is far from the dominant path (i.e., weakly visited by the electron's wavefunction), the conductance remains essentially unchanged. By systematically applying this 'poke test' across all sites, one obtains a spatial map of conductance sensitivity: regions where $g_0$ is highly responsive to local perturbations directly identify the dominant current-carrying pathways.

Ref.~\cite{lemarie_glassy_2019} builds on the same principle, applied to quantum transport of non-interacting electrons at zero temperature in $2d$. In this approach, the on-site disorder potential $\varepsilon_i$ of the single-particle Anderson model is locally perturbed according to $\varepsilon_i \to -\varepsilon_i$.  
The resulting conductance $g_i$ is then computed for each perturbed site. To quantify the impact of the local perturbation, the relative conductance response at site $i$ is defined as
\begin{equation}
    \delta g(i) \equiv \frac{|g_i - g_0|}{g_0}
    \label{eq:lemariedeltag}
\end{equation}
where $g_0$ is the original conductance of the sample, and $g_i$ is the conductance after flipping the disorder at site $i$. 

As in Anderson localization, we argue that in the strong disorder regime, the propagation of a fictitious particle in Hilbert space becomes highly inhomogeneous, dominated by an $O(1)$ number of transmission paths on the Hilbert space graph. To probe this structure, we adopt an analogous approach to the one outlined above.  
To this aim, we formally reintroduce the semi-infinite leads: an incoming lead is connected to the initial basis state $\ket{0}$---with energy in the middle of the many-body spectrum---and several outgoing leads connected to each of the target states $\ket{f} \in \mathcal{E}$. The fictitious particle is then injected at $\ket{0}$ and may propagate through the network of allowed transitions in Hilbert space before being absorbed at one of the target states. This defines an effective transport setup in Hilbert space, where the network of paths connecting $\ket{0}$ to the various $\ket{f} \in \mathcal{E}$ plays the role of the scattering geometry. For simplicity we only perform this analysis in the spin basis. 

\begin{figure}[h]
    \centering
    \includegraphics[width = \linewidth]{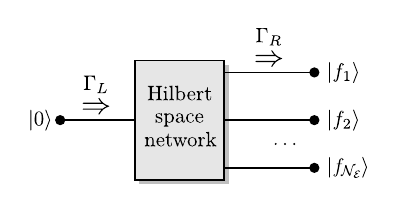}
    \caption{Scattering geometry to  measure the reaction of the Hilbert space conductance and responses to small perturbations. The initial basis state at the center of the many-body spectrum is  connected to  a semi-infinite lead through which fictitious particles are injected. Similarly, semi-infinite right leads are connected to the target vertices $\ket{f}$, that belong to the equator $\mathcal{E}$.}
    \label{fig:FockSpaceLeads}
\end{figure}

The effect of the leads is incorporated by adding a self-energy term, $-i\Sigma_{I}$, to the diagonal element of the Hamiltonian in Eq.~(\ref{eq:TBH}) for the respective basis state at which the lead is connected. This is implemented via the replacement
\begin{equation}
 \hat{\mathcal H} \to \hat{\mathcal H} - i \Sigma_I \ket I \bra I   \;,
\end{equation} with $I$ corresponding to both the inital basis state $\ket 0$, and its respective targets $\ket f \in \mathcal{E}$. The addition of an imaginary part of the self-energies of the leads is equivalent to inserting dissipation to the Hilbert space-graph at the vertices of interest. A pictorial representation of this construction can be seen in \autoref{fig:FockSpaceLeads}.

In our case the local perturbations to the disordered potential are introduced indirectly, by modifying the diagonal elements of the sparse Hamiltonian, which encodes the quenched disorder from a specific realization of the random fields. Specifically, we alter the diagonal element associated with the basis state $\ket{I}$---a vertex in the Hilbert space graph---through:
\begin{equation}
 {\mathcal{H}}'_{II} = {\mathcal{H}_{II}} + 2(\overline{E}-\mathcal{H}_{II}) \ket{I} \bra{I}\;,   
\label{eq:perturbation_hamiltonian}
\end{equation} 
where $\overline{E}$ is the average energy over all basis states---or equivalently, the average of the diagonal elements of the Hamiltonian matrix $\mathcal H$. In other words, this perturbation flips the value of the energy associated to the state $\ket{I}$ with respect to the mean $\overline{E}$. We repeat this procedure for each of the $\mathcal{N}$ diagonal elements, and calculate the associated resolvent matrix each time. Instead of doing this exhaustively, we calculate it using perturbation theory, which becomes exact in this case and reduces to the Sherman-Morrison formula for inverting matrices under $1$-rank perturbations. Using this formula 
\begin{equation}
    \mathcal G' = \mathcal G - \frac{2 (\mathcal{H}_{II}-\overline{E})\, \mathcal{G} \ket{I} \bra{I} \mathcal{G}}{1+2 (\mathcal{H}_{II}-\overline{E}) \, \mathcal G_{II}} \;,
\end{equation} 
we can recompute the Hilbert space Landauer transmissions perturbed at each vertex $I$,
which we call $\mathcal{T}^{(p)}_0(I)$, and we measure the (normalized) response defined as
\begin{equation} \label{eq:normalised_response}
    \delta g(I) \equiv \frac{|\mathcal{T}^{(p)}_0(I)-\mathcal{T}_{0}|}{\sum\limits_I|\mathcal{T}^{(p)}_0(I)-\mathcal{T}_{0}|} \;.
\end{equation} 

We have calculated the response $\delta g$ at low, intermediate, and large disorder strengths. We have repeated the calculations for typical samples (for which $\mathcal{T}_0$ is in the bulk of the probability distribution) and for rare samples (corresponding to disorder realizations that produce values of $\mathcal{T}_0$ in the tails of the distribution). In order to visualize the effect of the perturbation---that can be interpreted as the conductance generated by the incoming fictitious particles in the Hilbert space network---we normalize $\delta g$ in the following way:  
\begin{equation}
\delta g'(I) =
\begin{cases}
\displaystyle 
  0.5\,\frac{\delta g(I) - \delta g_{\min}}{\delta g(0) - \delta g_{\min}},
  & \text{if } \delta g(I) < \delta g(0) \\ 
\displaystyle 
  0.5 + 0.5\,\frac{\delta g(I) - \delta g(0)}{\delta g_{\max} - \delta g(0)},
  & \text{if } \delta g(I) > \delta g(0) \\ 
  0.5, & \text{otherwise.}
\end{cases}
\label{eq:normalization_deltag}
\end{equation} 
In words, we normalize the response $\delta g(I)$ onto a scale from $0$ to $1$. The initial basis state $| 0\rangle$ (to which the incoming semi-infinite lead is connected) defines the reference of this scale, because we are interested in probing how strong or weak each other vertex’s response is relative to the initial signal at $\ket 0$ here denoted as $\delta g (0)$.

\begin{figure*}[!ht]
\vspace{0.5cm}
    \centering
    \includegraphics[width=\linewidth]{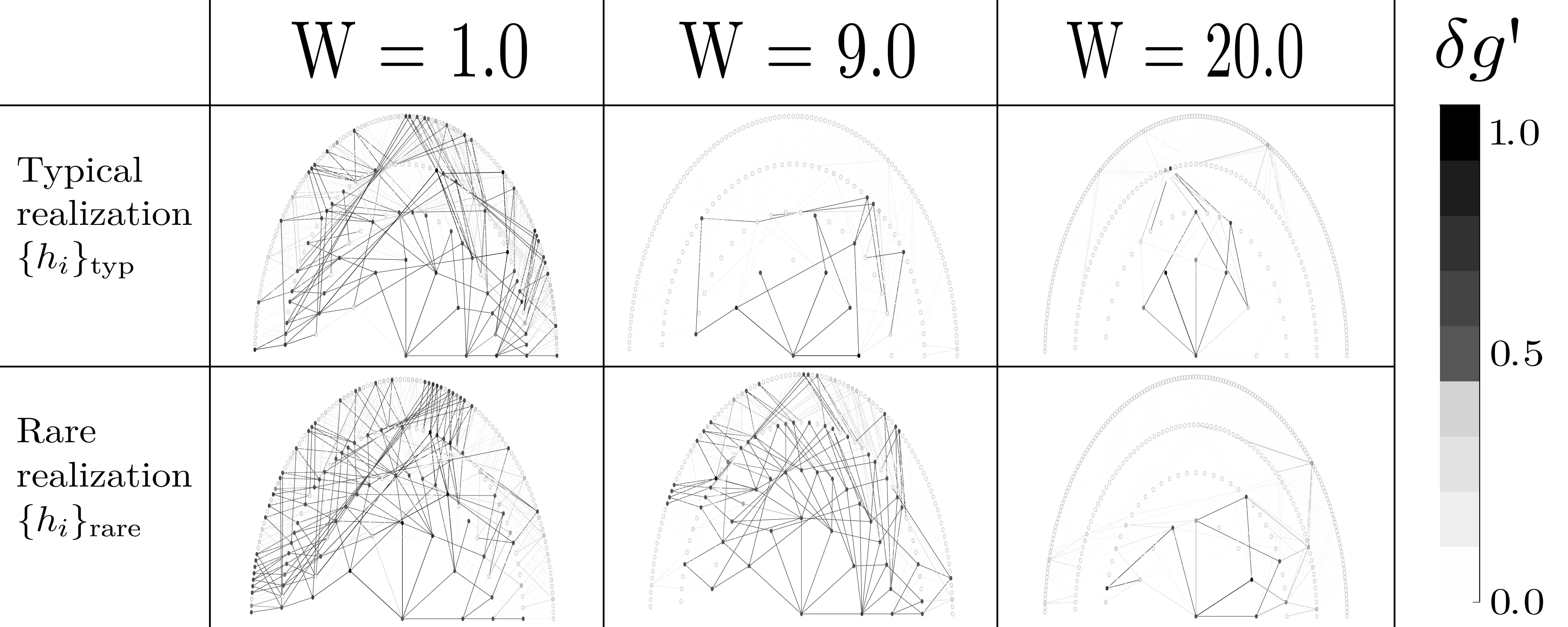}
     \caption{Rarefaction of paths in the Hilbert space graph (for the spin basis) for $L = 16$, for three disorder strengths. The color scale is set according to the central vertex $\delta g' = 0.5$. }
    \label{fig:polymers}
\end{figure*}

We visualize the construction by focusing on the induced subgraph of the Hilbert-space graph that contains the initial state $\ket{0}$ (where the incoming semi-infinite lead is attached) and all basis states connected to it, extending out to the `far' states $\ket{f} \in \cal E$ (where the other semi-infinite leads are attached). We show this visualization in~\autoref{fig:polymers} for $L=16$. The initial state $\ket 0$ is placed at the center in each of the six diagrams, and all other vertices connected to it are also shown. These vertices are arranged radially outward, ordered according to their Hilbert space distance from $\ket{0}$: points on the same semi-circle have the same Hilbert-space distance.

The central vertex carries the incoming signal $\delta g(0)$. Under the normalization of Eq.~\eqref{eq:normalization_deltag} $\delta g (0)$ is mapped to $\delta g' = 0.5$---within each sub-network---and appears colored gray (see the center of the color bar in~\autoref{fig:polymers}). Responses above the reference value, $\delta g(I) > \delta g(0)$, are mapped linearly to $(0.5,1.0]$, with $\delta g_{\rm max} \to 1$ (black), while responses below the initial signal, $\delta g(I) < \delta g(0)$, are mapped to $[0, 0.5)$, with $\delta g_{\rm min} \to 0$ (white). Consequently, vertices are colored according to $\delta g'$: darker colors ($\delta g(I) \geq 0.5$, up to black at $\delta g' = 1$) denote stronger or equal responses to the signal $\delta g (0)$, indicating basis states that lie along dominant transmission paths from $\ket{0}$. Lighter shades ($\delta g' < 0.5$) indicate weaker responses, corresponding to basis states that exhibit a low probability for the incoming fictitious particle to reach or delocalize onto starting from the initial configuration. Although the reaction $\delta g'$ is a variable assigned to the vertices of the graph, we have colored the edges to aide visualization. Hence, the edges are colored according to the value of the preceding vertex connected to them.

For weak disorder strength ($W = 1$), both rare and typical disorder realizations exhibit a proliferation of transmission paths in Hilbert space, enabling the transport of the injected fictitious particles to distant vertices in a mostly uniform way, with most of the target states at large distance from $\ket{0}$ reached by dark paths. When disorder is increased ($W = 9$), these reactions become smaller and dark conducting paths are strongly rarefied. For typical samples these rarefied conducting paths do not reach far away spin configurations, corresponding to the suppression of wave-packet spreading and localization. Yet, rare disorder realizations from the tails of the distribution of $\mathcal{T}_0$ exhibit much more reactive paths than typical samples, some of them extending to several basis states at large distances. This corresponds to a strongly inhomogeneous spreading of the wave-packet starting at $\ket{0}$, and delocalization along specific disorder-dependent paths, occurring only for rare disorder realizations.
At very strong disorder $W = 20$, finding subgraphs with highly-reactive paths is even more rare---although possible---but even these rare paths are unable to reach distant spin configurations, even for rare disorder realizations, corresponding to complete suppression of long-distance spreading of the wave-packet and a hallmark of genuine MBL. 

\begin{figure}[!ht]
\vspace{0.5cm}
\centering
 \includegraphics[width=\linewidth]{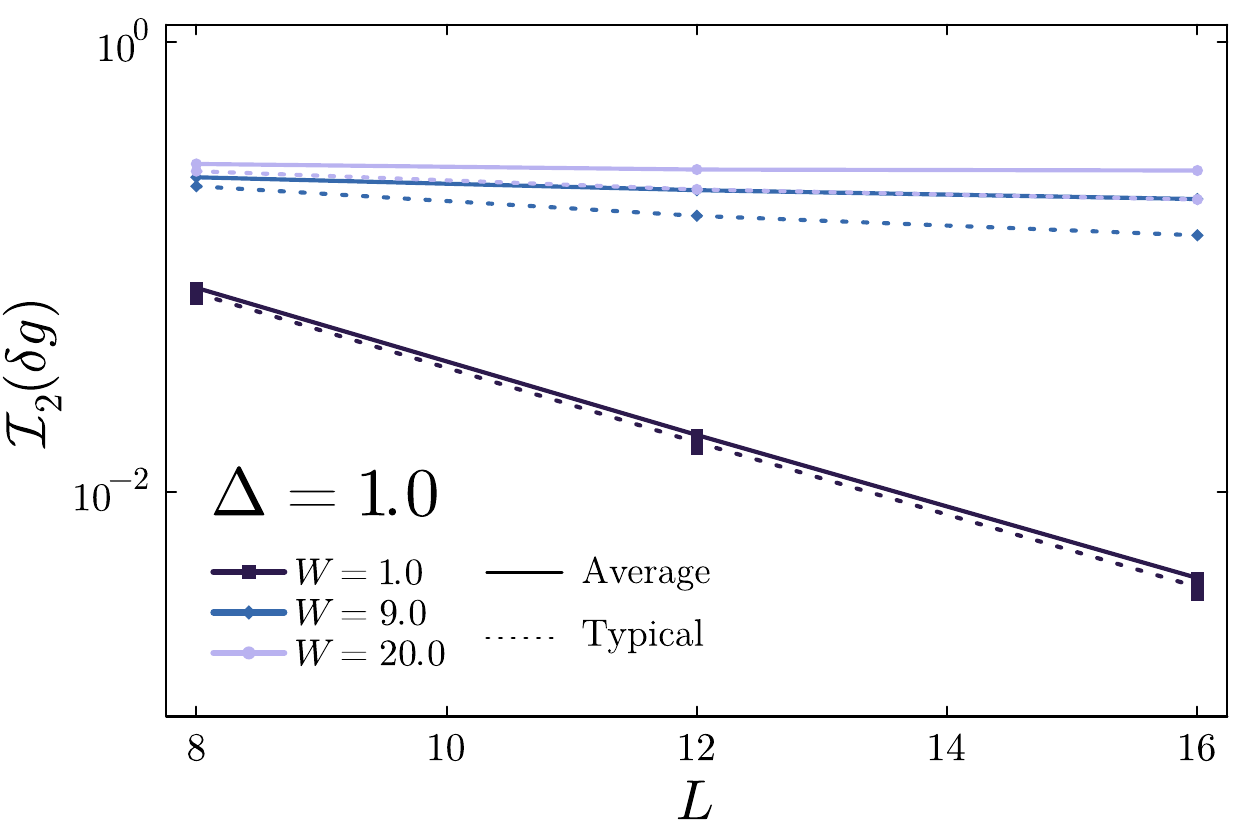}
 \caption{Inverse participation ratios ($\mathcal{I}_2$) associated to the reaction term $\delta g$ as a function of the system size $L$. The different curves correspond to average and typical IPR as in Eq.~\eqref{eq:IPR}.
 }
\label{fig:IPRS}
\end{figure}

More quantitatively, the contribution of these strongly reactive paths can be characterized using the (average and typical) inverse participation ratio (IPR) associated with the reaction amplitudes, they are defined as 
(recall that according to our definition~\eqref{eq:normalised_response} the response is normalized to one, i.e. $\sum_I \delta g (I) = 1$): 
\begin{equation}
\begin{aligned}
    \mathcal{I}_2(\delta g) & = \mathbb{E}\left[\sum_I \delta g (I)^2\right] \;, \\
    \mathcal{I}_2^{\rm typ} (\delta g) & = \exp \left( \mathbb{E}  \left [ \log \sum_I \delta g (I)^2  \right ] \right) \;,
    \end{aligned}
    \label{eq:IPR}
\end{equation} where $\mathbb{E}[\cdots]$ is taken for several initial conditions and disorder realizations. In \autoref{fig:IPRS}, we show the behaviour of the IPR as a function of the system size $L$. The definition of the typical IPR suppresses the influence of rare events, thereby capturing the behavior of typical disorder realizations.

For weak disorder ($W = 1$), the response values are approximately uniform across all $\mathcal{N}$ basis states. In this regime, the inverse participation ratio (IPR) scales as $1/\mathcal{N}$, decreasing exponentially with system size $L$ and reflecting the contribution of an extensive number of transmission paths. In contrast, for stronger disorder ($W = 9$, $W = 20$), only a small $O(1)$ number of sites---not growing with the size of the Hilbert space---exhibit significant responses, leading to an IPR that saturates to a finite plateau over the system sizes studied.

At small disorder ($W = 1$), the typical and average IPR are essentially equivalent. At large disorder ($W = 9$, $W = 20$), the typical IPR is consistently slightly smaller than the average one. This small difference highlights the presence of rare disorder realizations with exceptionally reactive dominant paths, which significantly affect the arithmetic average.

These results are consistent with the picture developed in~\autoref{sec:decays}, which highlights the strong fluctuations in the structure of eigenstates near the middle of the spectrum. In the present context, this heterogeneity manifests as a broad distribution of dominant transmission paths in the Hilbert space graph, particularly pronounced in the intermediate disorder regime ($W = 9$).

\section{Conclusions and outlook}

In this work we have developed an approach based on an analogy with a class of mean-field disordered glassy systems that allows one to take into account the statistical weights of rare events for the MBL transition. 
We have extended the analysis of Ref.~\cite{biroli_large-deviation_2024} by applying this method to the XXZ model varying the strength of the interaction $\Delta$, considering both the spin and Anderson bases. Our study emphasizes the role of rare long-range resonances---arising in rare disorder realizations---in destabilizing the MBL transition at finite sizes in a broad intermediate disorder range. These rare system-wide resonances are identified as the outliers in the probability distribution of transition amplitudes between distant configurations of the system in Hilbert space, expressed in the chosen computational basis. Concretely, we use the propagators $|\mathcal{G}_{0f}|^2$ as a proxy for the probability that a system initialized in a random configuration $\ket{0}$ at time $t = 0$ is found in the configuration $\ket{f}$---located at large distance from $\ket{0}$---at infinite time, $\sum_n |\langle f | n \rangle  \langle n | 0\rangle|^2$. Our study complements previous studies focused on real space observables and spectral signatures of rare events~\cite{Khemani2017, Crowley2020, Garratt2021,  morningstar_avalanches_2022, long_phenomenology_2023, Ha2023, colbois_statistics_2024, colbois_interaction-driven_2024},  by providing a Hilbert space-based perspective.

To correctly evaluate the statistical weight of rare resonances in the asymptotic limit of large system size $L$, we exploited an analogy with classical disordered systems and introduced an auxiliary parameter $\beta$ that plays the role of an effective temperature. This extension of the parameter space allows us to tune the influence of extreme outliers in the heavy-tailed distribution of propagators, and to identify, for each given disorder strength $W$, the value of $\beta$ at which rare events begin to dominate the statistical measure.

This method reveals the existence of three distinct regimes: (i) an ergodic phase, (ii) an intermediate regime in which delocalization is driven by rare long-range resonances in an otherwise localized background, and (iii) a genuinely many-body localized phase, which remains stable even in the presence of anomalously large outliers emerging from the tails of the distributions of the propagators. We show that typical samples that we can probe numerically lack the system-wide resonances that ultimately lead to delocalization in the asymptotic limit in the intermediate regime. Yet, our approach inspired by the analogy with mean-field glassy systems captures their asymptotic contribution.

It is important to acknowledge the potential limitations and drawbacks of our approach. As explained in Sec.~\ref{sec:proxy}, for numerical convenience we do not perform a systematic study of the statistics of the `true' order parameter for delocalization---namely, the typical value of the probability to delocalize from a random initial state after infinite time, $\mathbb{P}_\mathcal{E}$---but instead study the typical value of the unregularized Hilbert space transmission $\mathcal{T}_0$. This approximation is partially justified by the fact that the typical values of $\mathcal{T}_0$ and $\mathbb{P}_\mathcal{E}$ exhibit the same scaling with $L$ when both decay exponentially with $L$ (see Sec.~\ref{sec:proxy} and Fig.~\ref{fig:sketch}), since in this regime the regularization of poles becomes asymptotically unnecessary. However, at finite $L$, the typical values of $\mathbb{P}_\mathcal{E}$ and $\mathcal{T}_0$---obtained by computing $\mathbb{E}[ \ln \mathbb{P}_\mathcal{E} ]$ and $\mathbb{E}[ \ln \mathcal{T}_0 ]$---begin to decrease exponentially with $L$ well before the MBL transition. In fact, throughout almost the entire intermediate region (ii), which is dominated by rare resonances, both $e^{\mathbb{E}[ \ln \mathbb{P}_\mathcal{E} ]}$ and $e^{\mathbb{E}[ \ln \mathcal{T}_0 ]}$ decay exponentially with $L$. We then employ a large-deviation approach to determine whether rare events in the tail of the probability distribution might alter this scaling when properly accounted for at large $L$. We find that in this intermediate regime, the system is indeed delocalized despite the exponential decay of $e^{\mathbb{E}[ \ln \mathbb{P}_\mathcal{E} ]}$ and $e^{\mathbb{E}[ \ln \mathcal{T}_0 ]}$ at accessible system sizes.

A legitimate question then arises: could the large outliers that we suggest destabilize localization be overestimated due to the lack of regularization in $\mathcal{T}_0$? In this sense, our MBL threshold should be viewed as an upper bound---some of the rare events we consider crucial for MBL destabilization might be spurious artifacts that would disappear under proper regularization.
To address this concern, we note that when applied to benchmark cases, our method performs well in predicting phase boundaries. Specifically, we have tested it on the Anderson model on the RRG~\cite{Tarzia2026} and on the Rosenzweig-Porter model~\cite{biroli_large-deviation_2024}. For the Anderson model on the RRG, the method locates the transition at $W_c \approx 18$, very close to the exact value. We also identify an intermediate region at accessible system sizes where $e^{\mathbb{E}[ \ln \mathcal{T}_0 ]}$ decreases exponentially but the system will eventually delocalize due to rare outliers of $\mathcal{T}_0$. Importantly, we observe a drift of this intermediate region toward larger disorder values as $L$ increases, consistent with the drift observed in numerical exact diagonalizations. For the Rosenzweig-Porter model, the method accurately identifies all three phases---localized, fractal, and delocalized---even at relatively small system sizes.
In summary, while our method might in principle overestimate the effect of large resonances in the intermediate phase, this issue does not appear in the two benchmark cases we have studied.

By studying the model in two different computational bases provides another way to probe the limitations of our method. In particular, we examined how the structure of random initial conditions differs in the spin and Anderson bases. This difference gives rise to significant finite-size effects at intermediate disorder. Nevertheless, the overall physical picture remains robust and consistent with recent findings. Notably, the finite-size phase diagram obtained in the Anderson basis closely matches the results of Refs.~\cite{colbois_interaction-driven_2024, colbois_statistics_2024}, where alternative approaches based on real space observables—such as longitudinal spin-spin correlation functions—were employed.

Surprisingly, in our case, finite-size effects manifest differently than in the random-field Ising model in a transverse field previously studied in Ref.~\cite{biroli_large-deviation_2024}. In that model, the method yields a critical disorder strength for the MBL transition that remains approximately stable with increasing system size. In contrast, in the present study, we observe a systematic drift of the critical disorder strength for the MBL transition towards lower values as the system size $L$ increases. Simultaneously, the apparent boundary of the ergodic phase shifts towards higher disorder strengths with growing $L$. This concomitant trend suggests that the two crossover lines may ultimately merge into a single critical line in the thermodynamic limit, signaling a direct transition from the ergodic phase to the MBL phase. In this scenario, the broad intermediate regime---where delocalization is mediated by rare, system-spanning resonances---would then correspond to a finite-size prethermal crossover that disappears at large $L$, as proposed in Ref.~\cite{morningstar_avalanches_2022}. The origin of the discrepancy between the finite-size behavior observed here and in random-field quantum spin chain studied in Ref.~\cite{biroli_large-deviation_2024} is both intriguing and not yet understood. At the level of microscopic ingredients, the most prominent distinction between these systems is the presence of a conserved $U(1)$ symmetry in the XXZ chain, which suggests that symmetry may play a role in shaping the structure of resonant processes and, consequently, the finite-size scaling behavior. Further investigation is necessary to clarify the mechanisms behind these differing trends. In particular, it would be highly interesting to investigate whether the finite-size effects reported here can be interpreted within the framework of the renormalization-group flow recently proposed in Refs.~\cite{RG_AM, RG_XXZ}.

To resolve the spatial structure and statistics of rare resonances, we investigate the progressive delocalization from a random initial state as a function of the distance between the initial and target configurations in Hilbert space. In the strong disorder regime, deep in the MBL phase, we find that the maximum reachable distance decreases with increasing disorder. As a consequence, at strong disorder, resonant transmission processes remain confined to a small region of Hilbert space around the initial state, allowing only for partial delocalization involving a limited fraction of the degrees of freedom. The extent of this region decreases with increasing system size, reflecting the progressive suppression of rare long-range resonances within the MBL phase.
At weak disorder, in the ergodic regime, we observe that uniform spreading of the wavepacket is recovered only at large distances, while at shorter distances, strong resonances responsible for delocalization do not uniformly cover Hilbert space at that scale.

At large disorder, the spatial structure of high-energy eigenstates in Hilbert space exhibits pronounced fluctuations across different disorder realizations, reflecting the difference between typical and rare configurations of the disorder. To probe this heterogeneity, we identified and visualized the dominant resonant paths that form on the Hilbert space graph. This analysis, inspired by approaches to inhomogeneous quantum transport in real space networks~\cite{Pichard1991, Marko2010, lemarie_glassy_2019}, offers a novel perspective on the MBL instability: it stems from the inclusion of rare resonant paths, which, however, become progressively shorter and increasingly scarce deep within the MBL phase.

This picture of rarefied transmission paths in Hilbert space presents an intriguing direction for further exploration. The original analogy with Anderson localization in two dimensions~\cite{lemarie_glassy_2019} expands into how these dominant paths pinned by disorder can change suddenly and abruptly producing avalanches---as conceived in the classical setting of directed polymers---when the energy is varied. The depinning transition of the polymers through avalanches can be directly related to the singular behavior of the overlap correlation function between eigenstates at different energies which, in our case, corresponds to the correlation between the Hilbert space Landauer transmissions at different values of the energy---for a given disorder realization. It would be therefore interesting to investigate whether some signatures of these avalanches and shocks are present also in the quantum many-body problem.  

Although the importance of system-wide resonances in determining the properties of the MBL transition and the stability of the localized phase has been highlighted in this and several related works~\cite{Khemani2017, Crowley2020, VillalongaArxiv, Garratt2021, morningstar_avalanches_2022, Garratt2022, Ha2023, long_phenomenology_2023, biroli_large-deviation_2024, colbois_interaction-driven_2024, colbois_statistics_2024}, a proper characterization of the disorder realizations that lead to the formation of these resonances remains an open question. A first step in this direction was the proposal that such many-body resonances manifest as nearly degenerate \emph{cat states}~\cite{VillalongaArxiv, villalonga_characterizing_2020, crowley_constructive_2022}, a hypothesis recently tested in Ref.~\cite{Laflorencie2025-mv}. The anatomy of these nearly degenerate eigenstates reveals resonant events whose probability decreases with increasing disorder strength, consistent with the findings of the present work. Furthermore, rare disorder realizations at intermediate disorder, whose spectra contain such sparse, nearly degenerate eigenstate pairs, also exhibit probabilities to delocalize from a random initial state  that are classified as 'rare events' under our metric, appearing in the tails of the distribution of $\mathcal{T}_0$. 

However, a proper characterization of the structure of rare disorder realizations that give rise to anomalously large delocalization probabilities remains to be performed. This task is highly computationally demanding when using standard sampling techniques. A promising direction would be to employ importance sampling strategies~\cite{Korner2006-lv}: by biasing the sampling towards disorder configurations that enhance the likelihood of rare resonances, one could develop a genuine large-deviation framework and obtain a more accurate statistical characterization of the spatial structure of these rare events.

Similarly, understanding whether rare delocalization events are favored by the presence of extended regions with anomalously weak disorder---as suggested by the avalanche scenario~\cite{sels_bath-induced_2022, de_roeck_stability_2017, thiery_many-body_2018, luitz_how_2017, goihl_exploration_2019, Crowley2020, leonard_probing_2023, peacock_many-body_2023, szoldra_catching_2024}---remains an open problem. To make progress in this direction, it would be valuable to apply our method to study the system’s response when coupled to a thermal bath~\cite{Crowley2020, sels_bath-induced_2022, morningstar_avalanches_2022, Sierant2023-ba, Ha2023, szoldra_catching_2024, Colmenarez2024}. Such an analysis could help reveal the signatures in the Hilbert space propagators of rare ergodic bubbles in real space.

The methods proposed here can also be applied to other systems, for example, models of interacting fermions in a quasi-periodic potential, similar to the one realized in cold-atom experiments~\cite{Schreiber2015-qi, Bordia2017-qv, Luschen2017-or}. In this case, the only source of randomness comes from the choice of the initial state. It would be useful to compare the statistics of rare resonances found in the quasi-periodic case with those of uncorrelated random fields. This could help discriminate between the effects of rare resonances created by large segments with anomalously small values of the disorder in real space, and those due to rare paths with anomalously strong transmission amplitudes in Hilbert space. 

The data that support the findings of this article are openly available in Zenodo~\cite{ZenodoData}.

\section{Acknowledgments}
We are grateful to A. Mirlin for insightful comments and suggestions that led to significant improvements in this work. G.A.M. warmly thanks B. Douçot for stimulating discussions. The simulations were performed on the SACADO MeSU platform at Sorbonne Universit\'e. We acknowledge financial support from the ANR research grant ManyBodyNet
ANR-24-CE30-5851. FA and NL also benefited from the support of the Fondation Simone et Cino Del Duca.

\appendix
\section*{Appendices}
\section{Details of the numerical simulations}
\label{sec:SimulationDetails}
The task at hand reduces to calculate the $0f$ 
elements  for the resolvent matrix
\begin{equation}
\label{eq:Resolvent}
    \mathcal{G}_{0f}(\overline{ E}) = (\overline{  E} \mathbb{I} - \mathcal H)^{-1}_{0f} \;,
\end{equation} for the energy $\overline E$ in the middle of the spectrum. We have approximated this average energy by $\overline{ E} = \frac{\rm Tr {\mathcal H}}{\mathcal N}$. 
The initial conditions denoted by '$0$' are selected by choosing basis states with energy close to $\overline{ E}$. Numerically, we do this by selecting the states $\ket 0$ following 
\begin{equation}
    \{ | 0 \rangle\} \equiv \left\{| 0\rangle ~:~ E_{| 0 \rangle} = \mathcal{H}_{00} \in \left[\overline{ E} - \frac{\overline{ E}}{\eta}, \overline{ E} + \frac{\overline{ E}}{\eta} \right] \right\} \;,
\end{equation} where $\eta$ has been chosen to be $\eta = 64$ for intermediate and large sizes $L \geq 14$, and $\eta = 32$ for the smallest ones, $L \leq 12$. As seen in Eq. (\ref{eq:Resolvent}), we could invert the whole matrix $\overline{ E} \mathbb{I} - \mathcal H$ and extract the entries of interest i.e. the portion of columns '$0$' associated to the chosen initial conditions. However, inverting the full matrix of size $\mathcal N \times \mathcal N$ is computationally expensive. Instead, we directly compute portion of interest by solving the linear system:
\begin{equation} \label{eq:linear_system}
    (\overline{  E} \mathbb{I} - \mathcal{H}) \mathcal Y_{| 0\rangle} = \delta_{| 0\rangle} \;,
\end{equation} where $\delta_{| 0 \rangle}$ is a vector of zeros except for the entry corresponding to the chosen initial condition $| 0 \rangle$, which is set to one. This linear system is solved using the MUMPS~\cite{Amestoy2001-db, Amestoy2006-ni} or Pardiso~\cite{Schenk2004-pt} libraries, in their Julia interfaces \texttt{MUMPS.jl} and \texttt{Pardiso.jl}, respectively.
We have performed these calculations in both spin and Anderson bases. The number of initial conditions, $N_0$, and the number of samples, $N_S$, of the disordered fields $h_i$ are given in Tables \ref{tab:spin} and \ref{tab:anderson} for the spin and Anderson bases, respectively. They are presented as a function of the system-size $L$. The total number of samples, $N_{\rm total} = N_S \times N_{0}$, over which the average $\mathbb{E}[\cdots]$ is computed, is also shown—approximately—in the last column.
In certain cases—particularly for $\Delta = 1$ in the spin basis—we increased the number of samples where it was deemed necessary. 

\begin{table}[h!]
    \centering
    \begin{tabular}{|c|c|c|c|}
    \hline
    $L$ & $N_0$ & $N_S$ & $N_{\rm total}$\\
    \hline
    12 & $2^{L/2-2}$ & $125952$ & $2 \times 10^6$\\
    14 & $2^{L/2-2}$ & $16384$ & $5.25 \times 10^5$ \\
    16 & $2^{L/2-2}$ & $5120$ & $3.30 \times 10^5$\\
    18 & $2^{L/2-2}$ & $1280$ & $1.60 \times 10^5$\\
    20 & $2^{L/2-2}$ & $512$ & $1.30 \times 10^5$\\
    22 & $2^{L/2-1}$ & $128$ & $1.30 \times 10^5$\\
    \hline
    \end{tabular}
    \caption{Simulation values in the spin basis}
    \label{tab:spin}
\end{table}

\vspace{0.5cm}

\begin{table}[h!]
    \centering
    \begin{tabular}{|c|c|c|c|}
    \hline
    $L$ & $N_0$ & $N_S$ & $N_{\rm total}$\\
    \hline
    8  & $2^{L/2-2}$ & $1638400$ & $6.50 \times 10^6$\\
    12 & $2^{L/2-1}$ & $65536$ & $2 \times 10^6$\\
    16 & $2^{L/2-1}$ & $2048$ & $2.60 \times 10^5$\\
    20 & $2^{L/2-1}$ & $64$ & $3.30 \times 10^4$\\
    \hline
    \end{tabular}
    \caption{Simulation values in the Anderson basis}
    \label{tab:anderson}
\end{table}

Moreover, for system sizes that are not divisible by four (e.g., $L = 14,\, 18,\, 22$), there are no basis states that exactly satisfy the equator conditions, i.e., $q^{S,A} = 0$. Consequently, we perform an interpolation by averaging $\mathcal{T}_0$—and its logarithm, $\ln \mathcal{T}_0$—between the values computed using the nearest states to the equator. These target states correspond to basis states with overlaps 
\begin{equation}
q_{0f}^{S,A} = \pm \frac{2}{L} \;,
\end{equation} we select these two sets of target states, and average the quantities of interest between them. 
This procedure yields a consistent estimate of $\mathcal{T}_0$ (and $\ln \mathcal{T}_0$) at the equator for system sizes where exact equatorial states are not available.

\section{The computation of the error bars}
\label{sec:ErrorBars}

The error bars shown in \autoref{fig:free_energies} are directly extracted from the variances $\mathbb{E}[\mathcal{T}_0^2]-\mathbb{E}[\mathcal{T}_0]^2$ and $\mathbb{E}[(\ln \mathcal{T}_0)^2]-\mathbb{E}[\ln \mathcal{T}_0]^2$, and propagated accordingly for the functional forms of $\phi_a$ and $\phi_q$. For the annealed free-energy, the variance beyond $\beta_\star$ is ill-defined, making the error bars near $\beta_\star$ to be numerically unreliable. As a result, the propagated errors associated with the interpolated values of $\beta_\star$ and $\phi_a(\beta_\star)$ are extremely large and physically meaningless. Therefore, we have employed an alternative approach to assess the reliability of the numerical results produced by our method.

\begin{figure*}[!ht]
    \centering
    \includegraphics[width=\linewidth]{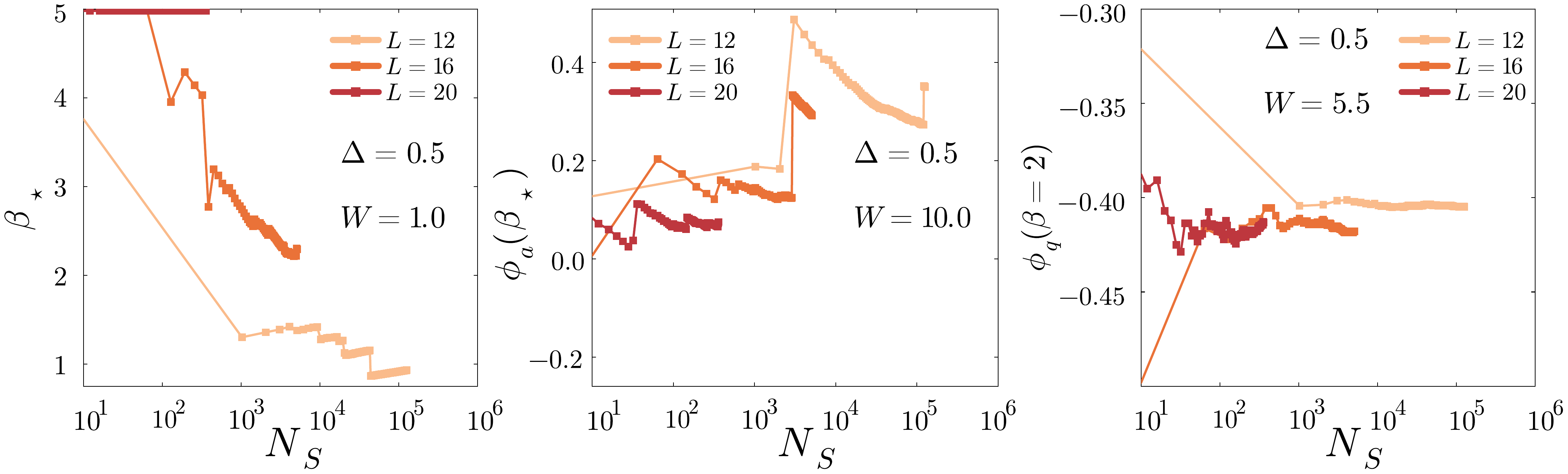}
    \includegraphics[width=\linewidth]{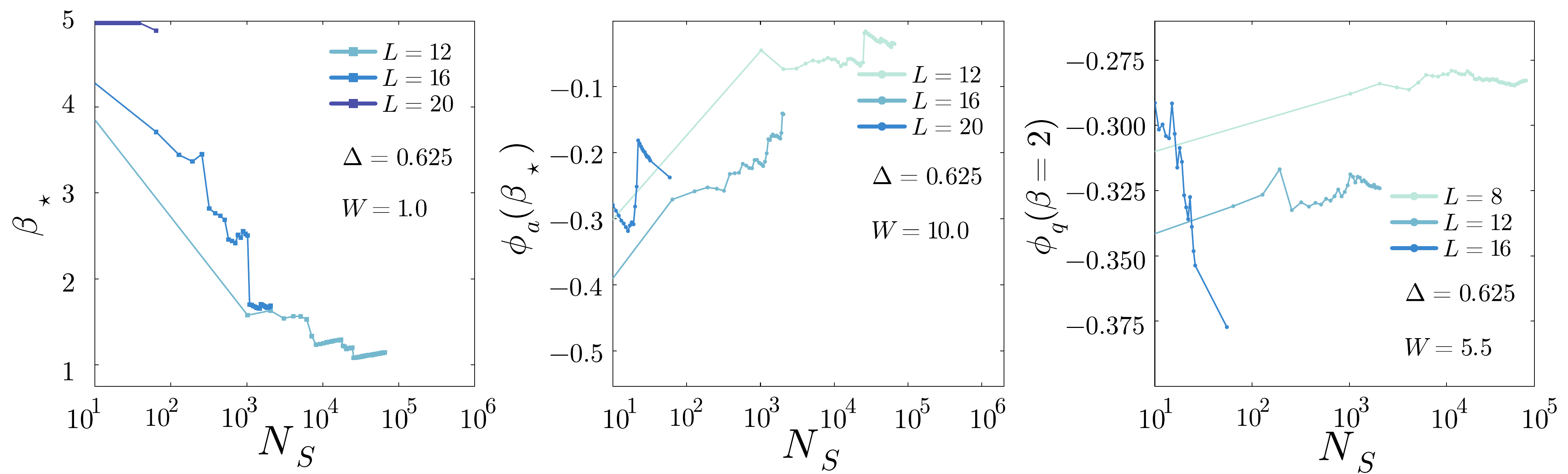}
    \caption{The values of $\beta_\star$, $\phi_a(\beta_\star)$, and $\phi_q(\beta = 2)$ as a function of the cumulative number of disorder realizations $N_S$ over which the average $\mathbb{E}[\cdots]$ is taken over. For both spin (top panels) and Anderson (bottom) bases.}
    \label{fig:CumulantErrors}
\end{figure*}

We assess the stability of the relevant quantities—$\beta_\star$, $\phi_a(\beta_\star)$, and $\phi_q(\beta = 2)$—under cumulative averaging. In other words, we calculate the relevant quantities with a cubic spline interpolation when averaged over $N_S$ disorder realizations---each of them with $N_0$ initial conditions---and we keep track of their behavior upon increasing $N_S$. This is shown in \autoref{fig:CumulantErrors} for both spin (top panels) and Anderson (bottom panels) bases, for the values of disorder strength, interaction parameter and sizes shown in the legend. 

We observe that the data exhibit jumps whose size and frequency decrease with increased sampling. The error bars are computed using the last range of values prior to the final average, which we have chosen to be the second half of the cumulant sample sequence. 

For example, in the case of $L = 20$ in the Anderson basis, where we use $N_S = 64$ disorder realizations, we store the values of $\beta_\star(N_S)$, $\phi_a(\beta_\star, N_S)$, and $\phi_q(\beta = 2, N_S)$ corresponding to the cumulative averages for $N_S = 1, 2, \dots, N_S = 64$. We then consider the second half of this sequence, i.e., from $N_S = 32$ to $N_S = 64$, and compute the error bars as the difference between the maximum and minimum values of $\beta_\star$, $\phi_a(\beta_\star)$, and $\phi_q(\beta = 2)$ within this range. Note that $\beta_\star$ and $\phi_a(\beta_\star)$ are obtained via cubic spline interpolation, for each average over the $N_S$ samples. This measure attempts to assess the stability of the values $\beta_\star$, $\phi_a(\beta_\star)$, and $\phi_q(\beta = 2)$ upon increasing the sampling of the averages $\mathbb{E}[\mathcal{T}_0(\beta)]$ and $\mathbb{E}[\ln \mathcal{T}_0(\beta)]$. 

After obtaining the associated error bars for the values $\beta_\star$, $\phi_a(\beta_\star)$, and $\phi_q(\beta = 2)$ in this way, we proceed to obtain their respective critical disorder strengths $W_{\rm ergo}$,  $W_{\rm MBL}$ and $W^{\rm typ}_{\rm MBL}$. The errors for $W_{\rm ergo}$ and  $W^{\rm typ}_{\rm MBL}$ are obtained from standard propagation of the errors. We perform a linear interpolation among the values closest to $\beta_\star = 2$ and $\phi_q(\beta = 2) = 0$, and propagate the errors accordingly. 

On the other hand, when determining $W_{\rm MBL}$, the variation of $\phi_a(\beta_\star, W)$ near the point where $\phi_a(\beta_\star, W) = 0$ is very small. This variation is negligible compared to the spacing along the $W$-axis, which is $\Delta W = 1.5$. As a result, the data do not effectively constrain the parameters. This issue is commonly referred to as a \emph{flat direction} in parameter space, or a degeneracy among parameters in non-linear statistical models.

Such degeneracies cause the covariance matrix derived from error propagation to have nearly zero eigenvalues, rendering it highly unstable and non-invertible. To address this, we instead estimate the uncertainty using a bootstrap Monte Carlo resampling approach. In this method, each data point of $\phi_a(\beta_\star, W)$ is randomly perturbed within its error bar, and for each perturbed dataset, the value of $W$ at which $\phi_a(\beta_\star, W) = 0$ is recalculated. The standard deviation of the new resulting data values for $W_{\rm MBL}(L)$ provides the error estimate. We use $10^4$ resampling iterations in this procedure.

\section{The DPRM correlations} \label{sec:ultrametric}
The mapping between the partition functions of directed polymers $\mathcal{Z}_N(\beta)$, defined in \autoeqref{eq:Zdprm}, and the biased Hilbert space Landauer transmissions $\mathcal{T}_0(\beta)$ is made explicit through the identification of
\begin{equation}
    E_\mathcal P = - \ln |\mathcal G_{0f} |\;,
    \label{eq:PolymerEnergy}
\end{equation} where $\mathcal P$ therefore corresponds to a 'path' defined between vertices $0$ and $f$ in the Hilbert space graph. As discussed in \autoref{sec:DPRM}, the presence of shared edges among different paths $\mathcal{P}$ and $\mathcal{P}'$ introduces correlations between their respective energies $E_\mathcal{P}$ and $E_{\mathcal{P}'}$. For the quantum many-body problem the connected correlation $\langle E_{\mathcal{P}} E_{\mathcal{P}'}\rangle_c$ is thus identified with $\langle \ln |\mathcal G_{0f}| \ln |\mathcal G_{0f'}|\rangle_c$, where the polymers $\mathcal{P}$ and $\mathcal{P}'$ are associated to the Hilbert space 'paths' between vertices $0$ to $f$, and $0$ to $f'$, respectively.
We measure this connected correlation as a function of the correlation distance---the rescaled overlaps---defined as  
\begin{equation}
    \zeta^{S,A}_{0f} = 1 - q_{0f}^{S,A} \;.
    \label{eq:re-scaled-overlaps-targets}
\end{equation} In this case, $\ket f$ will vary according to the correlation distance to the initial state $\ket 0$. The connected correlation is then computed as follows:
\begin{equation}
\begin{aligned}
     \langle \ln |\mathcal G_{0f}| & \ln |\mathcal G_{0f'}|\rangle_c \equiv \\ & \mathbb{E}[ \ln |\mathcal G_{0f}| \ln |\mathcal G_{0f'}|] - \mathbb{E}[ \ln |\mathcal G_{0f}| ]^2 \;,
\end{aligned}
\end{equation} where the average $\mathbb{E}[\cdots]$ is computed for several initial conditions, disorder realizations and among different nodes $\ket{f}$ within the same correlation distance from the chosen initial condition. 

\begin{figure}[!ht]
    \centering
    \includegraphics[width=\linewidth]{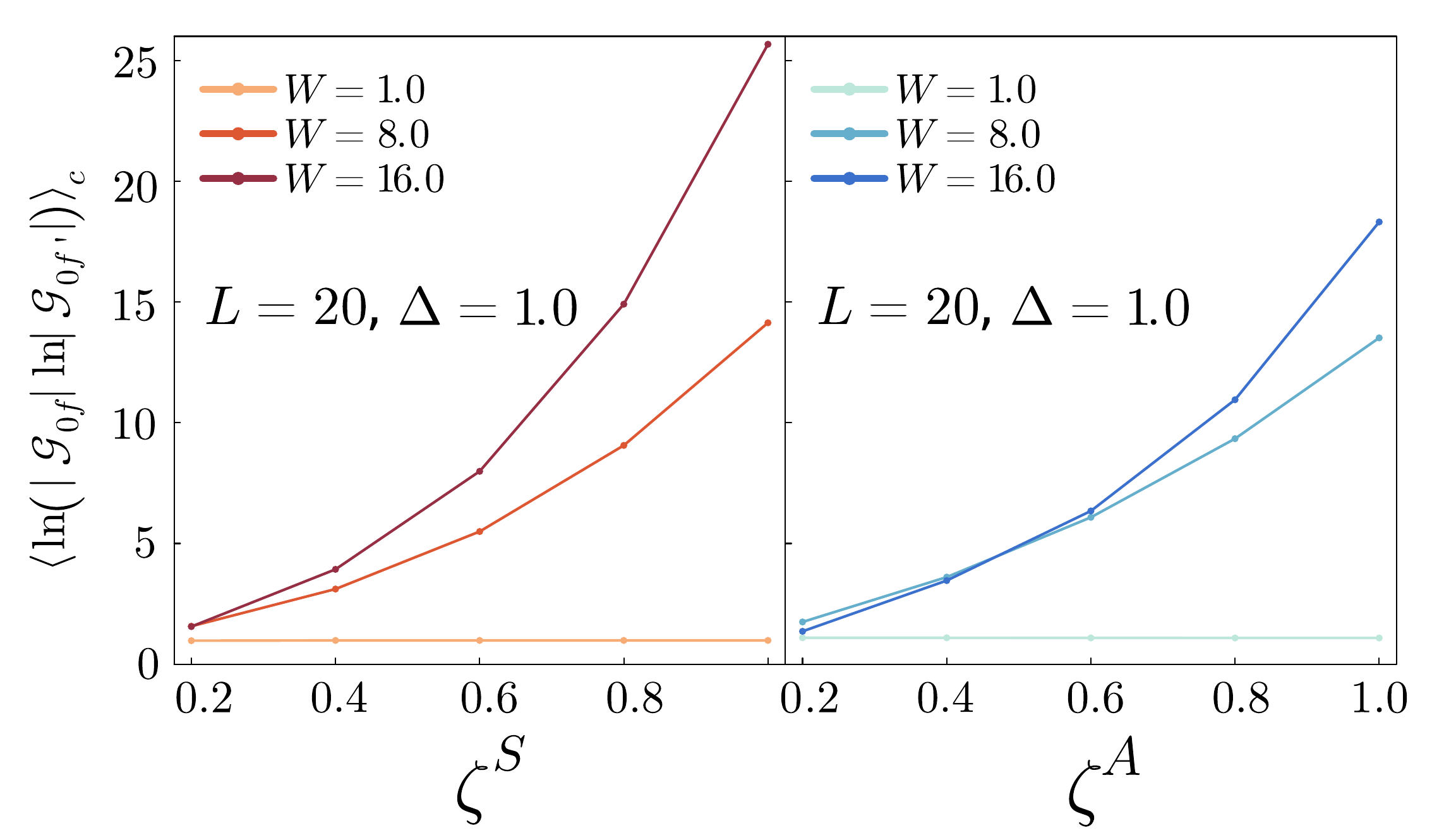} 
    \caption{Correlations of the equivalent polymer energy for the XXZ model. For both spin (left) and Anderson (right) bases, as a function of their respective correlation discante $\zeta^{S, A}$.
    }
    \label{fig:Correlations}
\end{figure}

In \autoref{fig:Correlations} we show the connected 'energy' correlations $\langle \ln |\mathcal G_{0f}| \ln |\mathcal G_{0f'}|\rangle_c$ as a function of the correlation distance $\eta^{S, A}_{0f}$. For small disorder ($W = 1$), the correlations remain uniform, appearing as a plateau in both bases. This behavior reflects the fact that, at low disorder, the system is ergodic and there is a proliferation of Hilbert space paths that enable delocalization. At stronger disorders ($W = 8$ and $W = 16$), the connected correlations between 'energies' increase significantly, reflecting the $O(1)$ preferred paths that extend far away in Hilbert space and allow for transmission events. 

In the original classical problem of directed polymers in infinite dimensional graphs, this correlation grows linearly with the real space distance of the extending polymer. The exact behaviour in our case is difficult to assess definitively, as the analogous 'polymer' in Hilbert space has a length $L/4$,  that for the largest system size with no interpolation ($L =20$) corresponds to five spin-flips events in the chain.

\section{A simple model for the IPR of the eigenstates of $L/2$ non-interacting spinless fermions} \label{app:IPR}

In this appendix we present a simple heuristic argument, used in Sec.~\ref{sec:phasediagrams}, to illustrate the different structures of the many-body eigenstates when expressed in the spin basis or in the Anderson basis.
We consider $L/2$ non-interacting spinless fermions on a one-dimensional chain described by the Hamiltonian~\eqref{eq:XXZ_FermionsHamiltonian} for $\Delta = 0$.
The many-body eigenstates of this system are Slater determinants constructed from the tensor products of single-particle localized orbitals.

For simplicity, we assume that the amplitudes of these single-particle orbitals decay exponentially over a characteristic localization length $\xi_{\rm loc}$ around a localization center $r_0$:
\begin{equation}
|\psi_{\alpha}(r)|^2 = C \, e^{-|r - r_0| / \xi_{\rm loc}} \, .
\end{equation}
The localization length $\xi_{\rm loc}$ depends on the disorder strength $W$ as $\xi_{\rm loc} \sim 1 / \ln W$ for sufficiently large $W$, and also varies with the energy $E_\alpha$ of the single-particle eigenstate (being smaller near the band edges of the one-dimensional tight-binding model and larger near the band center~\cite{Colbois2023}).
In the following, we neglect this energy dependence for simplicity.

Assuming periodic boundary conditions, the inverse participation ratio of a single-particle eigenstate is
\begin{equation}
I_2^{(1)} = \sum_i |\psi_{\alpha}(r)|^4 = 
\frac{1 + \sum_{i=1}^{L/2} e^{-2i / \xi_{\rm loc}}}{
\left( 1 + \sum_{i=1}^{L/2} e^{-i / \xi_{\rm loc}} \right)^2 } \, .
\end{equation}
For $L \gg \xi_{\rm loc}$ and in the large $W$ regime, this expression simplifies to
\begin{equation}
I_2^{(1)} \simeq
\frac{(1 + e^{2/\xi_{\rm loc}})(e^{2/\xi_{\rm loc}} - 1)}{(1 + e^{1/\xi_{\rm loc}})^3}
\simeq
\frac{(1 + W^2)(W - 1)}{(1 + W)^3} \, .
\end{equation}

The inverse participation ratio of a many-body eigenstate composed of $L/2$ localized orbitals, forming a Slater determinant of the corresponding single-particle states, is then given by
\begin{equation}
I_2^{(L/2)} \approx \left( I_2^{(1)} \right)^{L/2} \, .
\end{equation}
The total volume of the many-body Hilbert space is ${\cal N} = {L \choose L/2} \simeq \sqrt{\frac{2}{\pi L}} 2^L$.
The fractal dimension $D_2$ is defined by the scaling relation
\begin{equation}
I_2^{(L/2)} \propto {\cal N}^{-D_2} \, .
\end{equation}
From this definition, we obtain
\begin{equation} \label{eq:D2}
D_2 = - \frac{\ln I_2^{(1)}}{2 \ln 2}
\approx \frac{2}{W \ln 2} + o(W^{-1}) \, .
\end{equation}
Hence, the fractal dimension remains strictly positive for any finite disorder. (Of course, however, this expression is only valid in the strong-disorder regime where $D_2 < 1$.)

This simple argument thus shows that a many-body eigenstate constructed as a tensor product of single-particle localized orbitals necessarily occupies an exponentially large volume of the Hilbert space in the spin or particle basis. On the contrary, the same eigenstate is localized on a single node of the Hilbert-space graph when expressed in the Anderson basis by construction.

\section{Supplementary results for the Anderson basis}
\label{sec:AndersonResults}
In this Section, we present results for the Anderson basis that were omitted from the main text. Specifically, they include: (i) an example of the curves where the characteristic length is extracted for $\beta_\star(\zeta^A) = 2$ and $\phi_a(\beta_\star(\zeta^A)) = 0$, at the largest size $L = 20$. This is equivalent to \autoref{fig:MinimaNewTargets} for the Anderson basis. (ii) the eigenstate decays with respect to the most probable basis state, presented in \autoref{sec:decays}.

\subsection{Dependence on the target basis states}

Here, we present the dependence of $\beta_\star$ and $\phi_a(\beta_\star)$ on the correlation distance $\zeta^A$ for $L=20$. The conditions $\beta_\star(\zeta^A)=2$ and $\phi_a(\beta_\star,\zeta^A)=0$ define the characteristic correlation distances shown in the right panel of~\autoref{fig:PD-distance} for $L=20$. The same analysis was carried out for $L=12$ and $L=16$. While the corresponding results for the spin basis are presented in the main text (see~\autoref{fig:MinimaNewTargets}), Fig.~\autoref{fig:MinimaNewTargets2} below shows the analogous results for the Anderson basis.
\begin{figure}[h!]
\centering
  \centering
  \includegraphics[width=\linewidth]{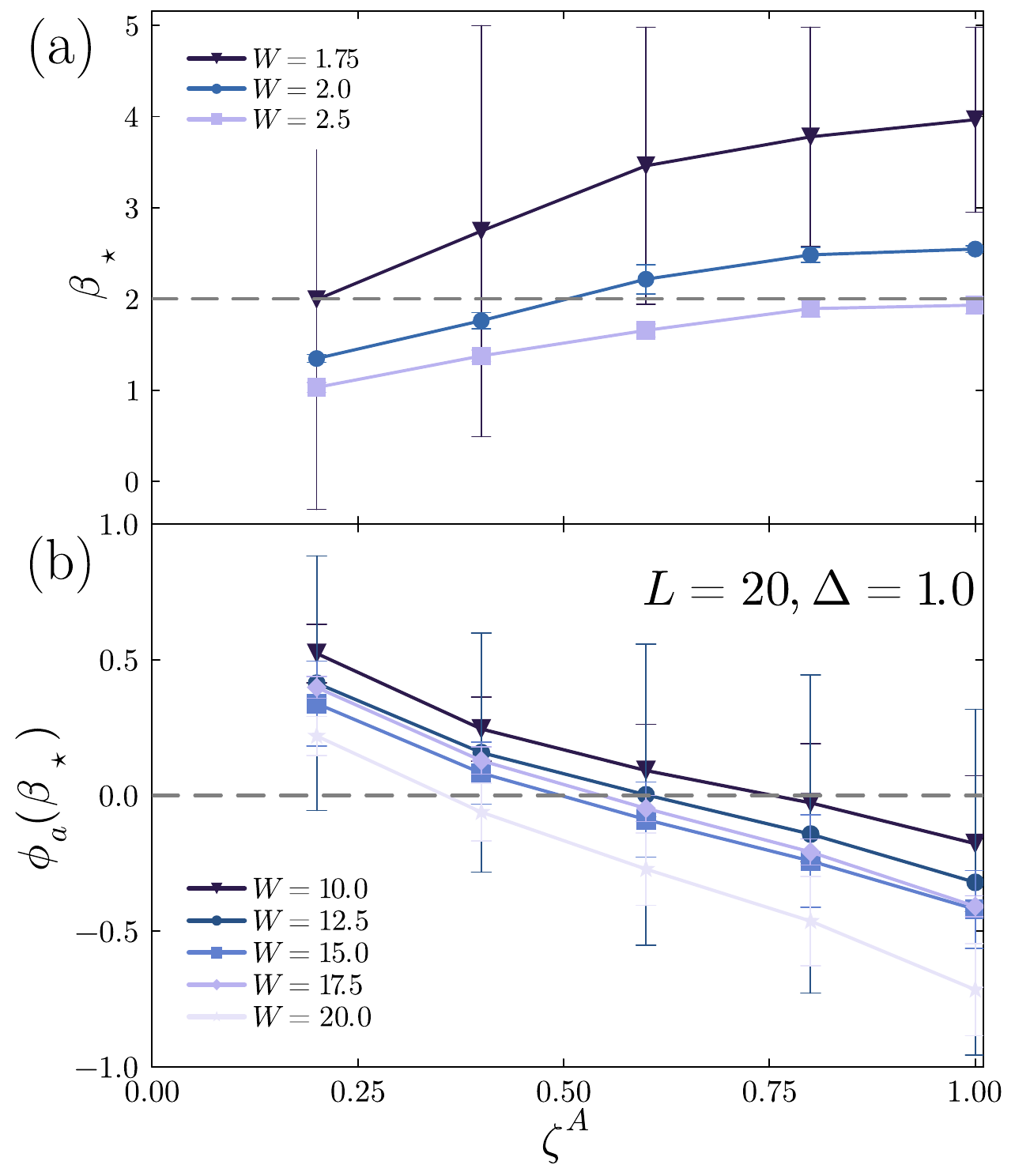}
\caption{Calculation of (a) $\beta_\star$ and (b) $\phi_a(\beta_\star)$ as functions of the correlation distance $\zeta^S$, for $\Delta = 1$ and $L = 20$. The values of the disorder strengths considered are shown in the legend. Horizontal gray dashed lines indicate the reference values $\beta_\star = 2$ and $\phi_a(\beta_\star) = 0$. These lines are used to extract the corresponding characteristic correlation distances.
}
\label{fig:MinimaNewTargets2}
\end{figure} 
\subsection{The eigenstate amplitude decays}
\begin{figure}[!ht]
    \centering
    \includegraphics[width=\linewidth]{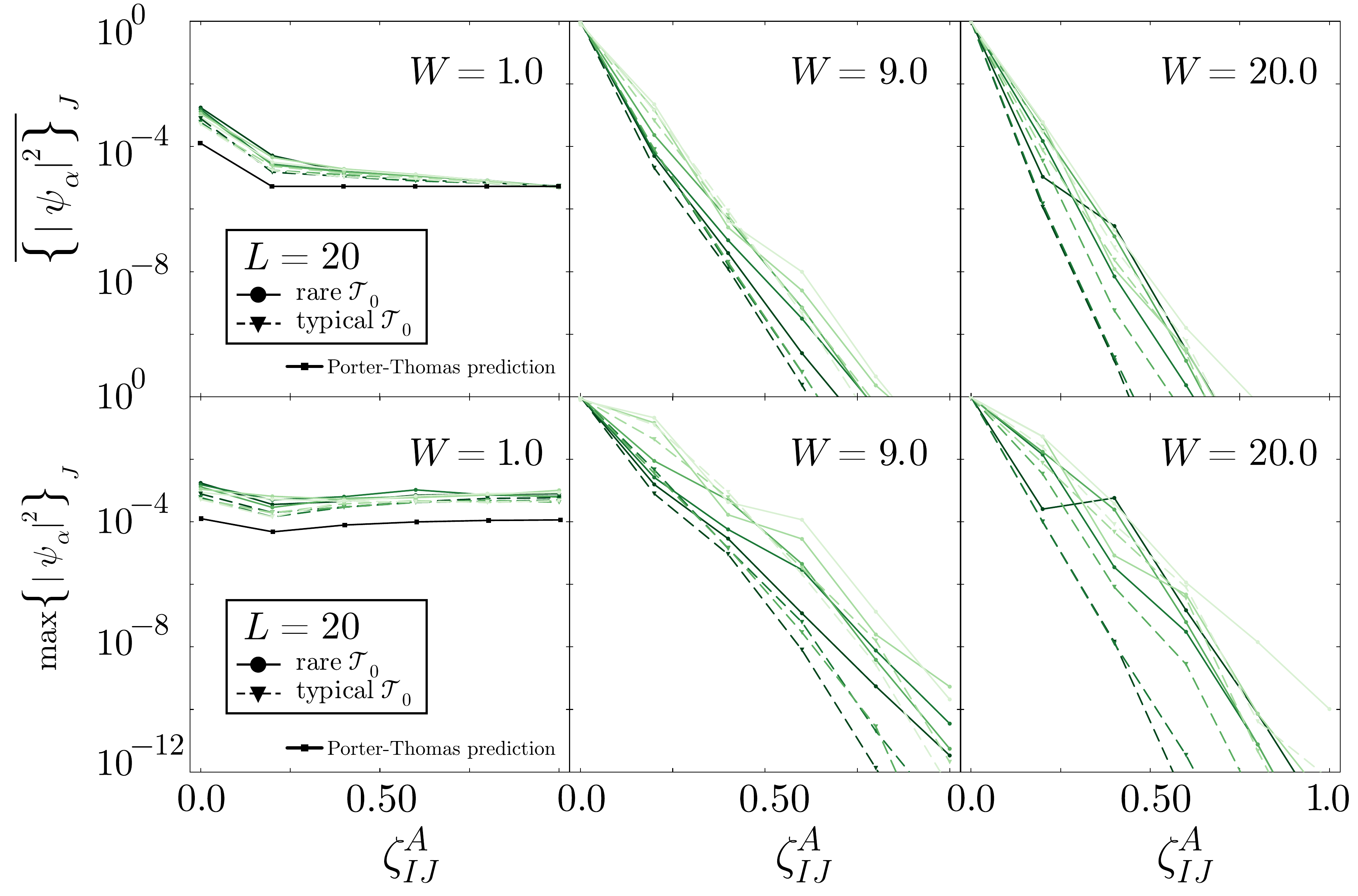}
     \caption{Decay of the basis state probability as a function of the correlation distance $\zeta_{IJ}^{A}$ between the most probable state $I$ and the other states $J$ that share the same correlation distance $\zeta^A$. Equivalent construction as in~\autoref{fig:EVdecays_bosons} but using the Anderson basis.
     }
    \label{fig:EVdecays_fermions}
\end{figure}

For this computation, we did not perform an exact diagonalization in the Anderson basis. Instead, we extracted the same eigenstates shown in~\autoref{fig:EVdecays_bosons} and rotated them using the transformation to the Anderson basis (see~\autoeqref{eq:AndersonNI-basis-states}). The eigenstate decay results for the Anderson basis are shown in~\autoref{fig:EVdecays_fermions}.

\newpage

\bibliography{references}

@article{anderson_absence_1958,
  title = {Absence of {Diffusion} in {Certain} {Random} {Lattices}},
  volume = {109},
  copyright = {http://link.aps.org/licenses/aps-default-license},
  issn = {0031-899X},
  url = {https://link.aps.org/doi/10.1103/PhysRev.109.1492},
  language = {en},
  number = {5},
  urldate = {2025-04-18},
  journal = {Phys. Rev.},
  author = {Anderson, P. W.},
  year = {1958},
  pages = {1492-1505},
  doi = {10.1103/PhysRev.109.1492}
}

@article{Gornyi2005,
  title = {Interacting Electrons in Disordered Wires: Anderson Localization and Low-$T$ Transport},
  author = {Gornyi, I. V. and Mirlin, A. D. and Polyakov, D. G.},
  journal = {Phys. Rev. Lett.},
  volume = {95},
  issue = {20},
  pages = {206603},
  numpages = {4},
  year = {2005},
  month = {Nov},
  publisher = {American Physical Society},
  doi = {10.1103/PhysRevLett.95.206603},
  url = {https://link.aps.org/doi/10.1103/PhysRevLett.95.206603}
}

@article{Basko2006,
  title = {Metal–insulator transition in a weakly interacting many-electron system with localized single-particle states},
  volume = {321},
  ISSN = {0003-4916},
  url = {http://dx.doi.org/10.1016/j.aop.2005.11.014},
  number = {5},
  journal = {Ann. Phys. (N. Y.)},
  publisher = {Elsevier BV},
  author = {Basko,  D.M. and Aleiner,  I.L. and Altshuler,  B.L.},
  year = {2006},
  month = may,
  pages = {1126–1205}
}

@article{Nandkishore2015,
  title = {Many-Body Localization and Thermalization in Quantum Statistical Mechanics},
  volume = {6},
  issn = {1947-5462},
  url = {http://dx.doi.org/10.1146/annurev-conmatphys-031214-014726},
  number = {1},
  journal = {Annu. Rev. Condens. Mat. Phys.},
  publisher = {Annual Reviews},
  author = {Nandkishore,  Rahul and Huse,  David A.},
  year = {2015},
  pages = {15–38}
}

@article{abanin_colloquium_2019,
  title = {\textit{{Colloquium}} : {Many}-body localization, thermalization, and entanglement},
  volume = {91},
  issn = {0034-6861, 1539-0756},
  shorttitle = {\textit{{Colloquium}}},
  url = {https://link.aps.org/doi/10.1103/RevModPhys.91.021001},
  language = {en},
  number = {2},
  urldate = {2022-05-19},
  journal = {Rev. Mod. Phys.},
  author = {Abanin, Dmitry A. and Altman, Ehud and Bloch, Immanuel and Serbyn, Maksym},
  year = {2019},
  pages = {021001}
}

@article{alet_many-body_2018,
  title = {Many-body localization: {An} introduction and selected topics},
  volume = {19},
  issn = {16310705},
  shorttitle = {Many-body localization},
  url = {https://linkinghub.elsevier.com/retrieve/pii/S163107051830032X},
  language = {en},
  number = {6},
  urldate = {2022-05-26},
  journal = {C. R. Phys.},
  author = {Alet, Fabien and Laflorencie, Nicolas},
  year = {2018},
  pages = {498-525}
}

@article{sierant_many-body_2025,
  title = {Many-body localization in the age of classical computing$^{\textrm{*}}$},
  volume = {88},
  issn = {0034-4885, 1361-6633},
  url = {https://iopscience.iop.org/article/10.1088/1361-6633/ad9756},
  number = {2},
  urldate = {2025-03-02},
  journal = {Rep. Prog. Phys.},
  author = {Sierant, Piotr and Lewenstein, Maciej and Scardicchio, Antonello and Vidmar, Lev and Zakrzewski, Jakub},
  year = {2025},
  pages = {026502}
}

@article{Serbyn2013,
  title = {Local Conservation Laws and the Structure of the Many-Body Localized States},
  volume = {111},
  issn = {1079-7114},
  url = {http://dx.doi.org/10.1103/PhysRevLett.111.127201},
  number = {12},
  journal = {Phys. Rev. Lett.},
  publisher = {American Physical Society (APS)},
  author = {Serbyn,  Maksym and Papić,  Z. and Abanin,  Dmitry A.},
  year = {2013},
  pages = {127201}
}

@article{Huse2014,
  title = {Phenomenology of fully many-body-localized systems},
  volume = {90},
  issn = {1550-235X},
  url = {http://dx.doi.org/10.1103/PhysRevB.90.174202},
  number = {17},
  journal = {Phys. Rev. B},
  publisher = {American Physical Society (APS)},
  author = {Huse,  David A. and Nandkishore,  Rahul and Oganesyan,  Vadim},
  year = {2014},
  pages = {174202}
}

@article{Ros2015,
  title = {Integrals of motion in the many-body localized phase},
  volume = {891},
  ISSN = {0550-3213},
  url = {http://dx.doi.org/10.1016/j.nuclphysb.2014.12.014},
  journal = {Nucl. Phys. B},
  publisher = {Elsevier BV},
  author = {Ros,  V. and M\"{u}ller,  M. and Scardicchio,  A.},
  year = {2015},
  month = feb,
  pages = {420--465}
}

@article{Imbrie2017,
  title = {Local integrals of motion in many‐body localized systems},
  volume = {529},
  ISSN = {1521-3889},
  url = {http://dx.doi.org/10.1002/andp.201600278},
  number = {7},
  journal = {Ann. Phys.},
  publisher = {Wiley},
  author = {Imbrie,  John Z. and Ros,  Valentina and Scardicchio,  Antonello},
  year = {2017},
  month = may ,
  pages = {1600278}
}

@article{de_roeck_stability_2017,
  title = {Stability and instability towards delocalization in many-body localization systems},
  volume = {95},
  copyright = {http://link.aps.org/licenses/aps-default-license},
  issn = {2469-9950, 2469-9969},
  url = {http://link.aps.org/doi/10.1103/PhysRevB.95.155129},
  language = {en},
  number = {15},
  urldate = {2025-03-03},
  journal = {Phys. Rev.  B},
  author = {De Roeck, Wojciech and Huveneers, François},
  year = {2017},
  pages = {155129},  
}

@article{suntajs_quantum_2020,
  title = {Quantum chaos challenges many-body localization},
  volume = {102},
  issn = {2470-0045, 2470-0053},
  url = {https://link.aps.org/doi/10.1103/PhysRevE.102.062144},
  language = {en},
  number = {6},
  urldate = {2025-03-03},
  journal = {Phys. Rev.  E},
  author = {Šuntajs, Jan and Bonča, Janez and Prosen, Tomaž and Vidmar, Lev},
  year = {2020},
  pages = {062144}
}

@article{sierant_polynomially_2020,
  title = {Polynomially {Filtered} {Exact} {Diagonalization} {Approach} to {Many}-{Body} {Localization}},
  volume = {125},
  issn = {0031-9007, 1079-7114},
  url = {https://link.aps.org/doi/10.1103/PhysRevLett.125.156601},
  language = {en},
  number = {15},
  urldate = {2025-03-15},
  journal = {Phys. Rev. Lett.},
  author = {Sierant, Piotr and Lewenstein, Maciej and Zakrzewski, Jakub},
  year = {2020},
  pages = {156601}
}

@article{suntajs_ergodicity_2020,
  title = {Ergodicity breaking transition in finite disordered spin chains},
  volume = {102},
  issn = {2469-9950, 2469-9969},
  url = {https://link.aps.org/doi/10.1103/PhysRevB.102.064207},
  language = {en},
  number = {6},
  urldate = {2025-03-15},
  journal = {Phys. Rev.  B},
  author = {Šuntajs, Jan and Bonča, Janez and Prosen, Tomaž and Vidmar, Lev},
  year = {2020},
  pages = {064207},
}

@article{sierant_thouless_2020,
  title = {Thouless {Time} {Analysis} of {Anderson} and {Many}-{Body} {Localization} {Transitions}},
  volume = {124},
  issn = {0031-9007, 1079-7114},
  url = {https://link.aps.org/doi/10.1103/PhysRevLett.124.186601},
  language = {en},
  number = {18},
  urldate = {2025-03-03},
  journal = {Phys. Rev. Lett.},
  author = {Sierant, Piotr and Delande, Dominique and Zakrzewski, Jakub},
  year = {2020},
  pages = {186601}
}

@article{kiefer-emmanouilidis_evidence_2020,
  title = {Evidence for {Unbounded} {Growth} of the {Number} {Entropy} in {Many}-{Body} {Localized} {Phases}},
  volume = {124},
  issn = {0031-9007, 1079-7114},
  url = {https://link.aps.org/doi/10.1103/PhysRevLett.124.243601},
  language = {en},
  number = {24},
  urldate = {2025-03-03},
  journal = {Phys. Rev. Lett.},
  author = {Kiefer-Emmanouilidis, Maximilian and Unanyan, Razmik and Fleischhauer, Michael and Sirker, Jesko},
  year = {2020},
  pages = {243601}
}

@article{kiefer-emmanouilidis_slow_2021,
  title = {Slow delocalization of particles in many-body localized phases},
  volume = {103},
  issn = {2469-9950, 2469-9969},
  url = {https://link.aps.org/doi/10.1103/PhysRevB.103.024203},
  language = {en},
  number = {2},
  urldate = {2025-03-03},
  journal = {Phys. Rev.  B},
  author = {Kiefer-Emmanouilidis, Maximilian and Unanyan, Razmik and Fleischhauer, Michael and Sirker, Jesko},
  year = {2021},
  pages = {024203}
}

@article{sierant_challenges_2022,
  title = {Challenges to observation of many-body localization},
  volume = {105},
  issn = {2469-9950, 2469-9969},
  url = {https://link.aps.org/doi/10.1103/PhysRevB.105.224203},
  language = {en},
  number = {22},
  urldate = {2025-03-03},
  journal = {Phys. Rev.  B},
  author = {Sierant, Piotr and Zakrzewski, Jakub},
  year = {2022},
  pages = {224203}
}

@article{sels_bath-induced_2022,
  title = {Bath-induced delocalization in interacting disordered spin chains},
  volume = {106},
  issn = {2469-9950, 2469-9969},
  url = {https://link.aps.org/doi/10.1103/PhysRevB.106.L020202},
  language = {en},
  number = {2},
  urldate = {2025-03-03},
  journal = {Phys. Rev.  B},
  author = {Sels, Dries},
  year = {2022},
  pages = {L020202}
}

@article{sels_thermalization_2023,
  title = {Thermalization of {Dilute} {Impurities} in {One}-{Dimensional} {Spin} {Chains}},
  volume = {13},
  issn = {2160-3308},
  url = {https://link.aps.org/doi/10.1103/PhysRevX.13.011041},
  language = {en},
  number = {1},
  urldate = {2025-03-03},
  journal = {Phys. Rev.  X},
  author = {Sels, Dries and Polkovnikov, Anatoli},
  year = {2023},
  pages = {011041}
}

@article{luitz_many-body_2015,
  title = {Many-body localization edge in the random-field {Heisenberg} chain},
  volume = {91},
  copyright = {http://link.aps.org/licenses/aps-default-license},
  issn = {1098-0121, 1550-235X},
  url = {https://link.aps.org/doi/10.1103/PhysRevB.91.081103},
  language = {en},
  number = {8},
  urldate = {2025-03-03},
  journal = {Phys. Rev.  B},
  author = {Luitz, David J. and Laflorencie, Nicolas and Alet, Fabien},
  year = {2015},
  pages = {081103}
}

@article{mace_multifractal_2019,
  title = {Multifractal {Scalings} {Across} the {Many}-{Body} {Localization} {Transition}},
  volume = {123},
  issn = {0031-9007, 1079-7114},
  url = {https://link.aps.org/doi/10.1103/PhysRevLett.123.180601},
  language = {en},
  number = {18},
  urldate = {2025-03-11},
  journal = {Phys. Rev. Lett.},
  author = {Macé, Nicolas and Alet, Fabien and Laflorencie, Nicolas},
  year = {2019},
  pages = {180601}
}

@article{sels_dynamical_2021,
  title = {Dynamical obstruction to localization in a disordered spin chain},
  volume = {104},
  issn = {2470-0045, 2470-0053},
  url = {https://link.aps.org/doi/10.1103/PhysRevE.104.054105},
  language = {en},
  number = {5},
  urldate = {2025-03-03},
  journal = {Phys. Rev.  E},
  author = {Sels, Dries and Polkovnikov, Anatoli},
  year = {2021},
  pages = {054105}
}

@article{luitz_how_2017,
  title = {How a {Small} {Quantum} {Bath} {Can} {Thermalize} {Long} {Localized} {Chains}},
  volume = {119},
  copyright = {https://link.aps.org/licenses/aps-default-license},
  issn = {0031-9007, 1079-7114},
  url = {https://link.aps.org/doi/10.1103/PhysRevLett.119.150602},
  language = {en},
  number = {15},
  urldate = {2025-03-03},
  journal = {Phys. Rev. Lett.},
  author = {Luitz, David J. and Huveneers, François and De Roeck, Wojciech},
  year = {2017},
  pages = {150602}
}

@article{thiery_many-body_2018,
  title = {Many-{Body} {Delocalization} as a {Quantum} {Avalanche}},
  volume = {121},
  issn = {0031-9007, 1079-7114},
  url = {https://link.aps.org/doi/10.1103/PhysRevLett.121.140601},
  language = {en},
  number = {14},
  urldate = {2025-03-03},
  journal = {Phys. Rev. Lett.},
  author = {Thiery, Thimothée and Huveneers, François and Müller, Markus and De Roeck, Wojciech},
  year = {2018},
  pages = {140601}
}

@article{goihl_exploration_2019,
  title = {Exploration of the stability of many-body localized systems in the presence of a small bath},
  volume = {99},
  issn = {2469-9950, 2469-9969},
  url = {https://link.aps.org/doi/10.1103/PhysRevB.99.195145},
  language = {en},
  number = {19},
  urldate = {2025-03-03},
  journal = {Phys. Rev.  B},
  author = {Goihl, Marcel and Eisert, Jens and Krumnow, Christian},
  year = {2019},
  pages = {195145}
}

@article{Crowley2020,
  title = {Avalanche induced coexisting localized and thermal regions in disordered chains},
  volume = {2},
  ISSN = {2643-1564},
  url = {http://dx.doi.org/10.1103/PhysRevResearch.2.033262},
  number = {3},
  journal = {Phys. Rev. Res.},
  publisher = {American Physical Society (APS)},
  author = {Crowley,  P. J. D. and Chandran,  A.},
  year = {2020},
  month = aug,
  pages = {033262}
}

@article{leonard_probing_2023,
  title = {Probing the onset of quantum avalanches in a many-body localized system},
  volume = {19},
  issn = {1745-2473, 1745-2481},
  url = {https://www.nature.com/articles/s41567-022-01887-3},
  language = {en},
  number = {4},
  urldate = {2025-03-03},
  journal = {Nat. Phys.},
  author = {Léonard, Julian and Kim, Sooshin and Rispoli, Matthew and Lukin, Alexander and Schittko, Robert and Kwan, Joyce and Demler, Eugene and Sels, Dries and Greiner, Markus},
  year = {2023},
  pages = {481--485}
}

@article{peacock_many-body_2023,
  title = {Many-body delocalization from embedded thermal inclusion},
  volume = {108},
  issn = {2469-9950, 2469-9969},
  url = {https://link.aps.org/doi/10.1103/PhysRevB.108.L020201},
  language = {en},
  number = {2},
  urldate = {2025-03-03},
  journal = {Phys. Rev.  B},
  author = {Peacock, J. Clayton and Sels, Dries},
  year = {2023},
  pages = {L020201}
}

@article{szoldra_catching_2024,
  title = {Catching thermal avalanches in the disordered {XXZ} model},
  volume = {109},
  issn = {2469-9950, 2469-9969},
  url = {https://link.aps.org/doi/10.1103/PhysRevB.109.134202},
  language = {en},
  number = {13},
  urldate = {2025-03-03},
  journal = {Phys. Rev.  B},
  author = {Szołdra, Tomasz and Sierant, Piotr and Lewenstein, Maciej and Zakrzewski, Jakub},
  year = {2024},
  pages = {134202}
}

@article{Suntajs2022,
  title = {Ergodicity Breaking Transition in Zero Dimensions},
  volume = {129},
  issn = {1079-7114},
  url = {http://dx.doi.org/10.1103/PhysRevLett.129.060602},
  number = {6},
  journal = {Phys. Rev. Lett.},
  publisher = {American Physical Society (APS)},
  author = {Šuntajs,  Jan and Vidmar,  Lev},
  year = {2022},
  pages = {060602}
}

@article{morningstar_avalanches_2022,
  title = {Avalanches and many-body resonances in many-body localized systems},
  volume = {105},
  issn = {2469-9969},
  url = {http://dx.doi.org/10.1103/PhysRevB.105.174205},
  number = {17},
  journal = {Phys. Rev. B},
  publisher = {American Physical Society (APS)},
  author = {Morningstar,  Alan and Colmenarez,  Luis and Khemani,  Vedika and Luitz,  David J. and Huse,  David A.},
  year = {2022},
  pages = {174205}
}

@article{Ha2023,
  title = {Many-Body Resonances in the Avalanche Instability of Many-Body Localization},
  volume = {130},
  ISSN = {1079-7114},
  url = {http://dx.doi.org/10.1103/PhysRevLett.130.250405},
  number = {25},
  journal = {Phys. Rev. Lett.},
  publisher = {American Physical Society (APS)},
  author = {Ha,  Hyunsoo and Morningstar,  Alan and Huse,  David A.},
  year = {2023},
  month = jun,
  pages = {250405}
}

@article{Pawlik2024,
  title = {Many-body mobility edge in quantum sun models},
  volume = {109},
  issn = {2469-9969},
  url = {http://dx.doi.org/10.1103/PhysRevB.109.L180201},
  number = {18},
  journal = {Phys. Rev. B},
  publisher = {American Physical Society (APS)},
  author = {Pawlik,  Konrad and Sierant,  Piotr and Vidmar,  Lev and Zakrzewski,  Jakub},
  year = {2024},
  pages = {L180201}
}

@article{Colmenarez2024,
  title = {Ergodic inclusions in many-body localized systems},
  volume = {109},
  issn = {2469-9969},
  url = {http://dx.doi.org/10.1103/PhysRevB.109.L081117},
  number = {8},
  journal = {Phys. Rev. B},
  publisher = {American Physical Society (APS)},
  author = {Colmenarez,  Luis and Luitz,  David J. and De Roeck,  Wojciech},
  year = {2024},
  pages = {L081117}
}

@article{pawlik2025,
  title = {Unconventional thermalization of a localized chain interacting with an ergodic bath},
  author = {Pawlik, Konrad and Laflorencie, Nicolas and Zakrzewski, Jakub},
  journal = {Phys. Rev. B},
  volume = {113},
  issue = {18},
  pages = {L180203},
  numpages = {8},
  year = {2026},
  month = {May},
  publisher = {American Physical Society},
  doi = {10.1103/qggc-sp5t},
  url = {https://link.aps.org/doi/10.1103/qggc-sp5t}
}

@article{Pietracaprina2018,
  title = {Shift-invert diagonalization of large many-body localizing spin chains},
  volume = {5},
  url = {http://dx.doi.org/10.21468/SciPostPhys.5.5.045},
  number = {5},
  journal = {SciPost Phys.},
  author = {Pietracaprina,  Francesca and Macé,  Nicolas and Luitz,  David J. and Alet,  Fabien},
  year = {2018},
  pages = {045}
}

@article{Laflorencie2020,
  title = {Chain breaking and Kosterlitz-Thouless scaling at the many-body localization transition in the random-field Heisenberg spin chain},
  volume = {2},
  issn = {2643-1564},
  url = {http://dx.doi.org/10.1103/PhysRevResearch.2.042033},
  number = {4},
  journal = {Phys. Rev. Res.},
  publisher = {American Physical Society (APS)},
  author = {Laflorencie,  Nicolas and Lemarié,  Gabriel and Macé,  Nicolas},
  year = {2020},
  pages = {042033}
}

@article{imbrie_diagonalization_2016,
  title = {Diagonalization and {Many}-{Body} {Localization} for a {Disordered} {Quantum} {Spin} {Chain}},
  volume = {117},
  copyright = {http://link.aps.org/licenses/aps-default-license},
  issn = {0031-9007, 1079-7114},
  url = {https://link.aps.org/doi/10.1103/PhysRevLett.117.027201},
  language = {en},
  number = {2},
  urldate = {2025-03-07},
  journal = {Phys. Rev. Lett.},
  author = {Imbrie, John Z.},
  year = {2016},
  pages = {027201}
}

@misc{VillalongaArxiv,
      title={Eigenstates hybridize on all length scales at the many-body localization transition}, 
      author={Benjamin Villalonga and Bryan K. Clark},
      year={2020},
      eprint={2005.13558},
      archivePrefix={arXiv},
      primaryClass={cond-mat.dis-nn},
      url={https://arxiv.org/abs/2005.13558}, 
}

@article{Garratt2021,
  title = {Local resonances and parametric level dynamics in the many-body localized phase},
  volume = {104},
  ISSN = {2469-9969},
  url = {http://dx.doi.org/10.1103/PhysRevB.104.184203},
  number = {18},
  journal = {Phys. Rev. B},
  publisher = {American Physical Society (APS)},
  author = {Garratt,  S. J. and Roy,  Sthitadhi and Chalker,  J. T.},
  year = {2021},
  month = nov,
  pages = {184203}
}

@article{crowley_constructive_2022,
  title = {A constructive theory of the numerically accessible many-body localized to thermal crossover},
  volume = {12},
  issn = {2542-4653},
  url = {https://scipost.org/10.21468/SciPostPhys.12.6.201},
  number = {6},
  urldate = {2025-03-03},
  journal = {SciPost Phys.},
  author = {Crowley, Philip and Chandran, Anushya},
  year = {2022},
  pages = {201}
}

@article{Gopalakrishnan2015,
  title = {Low-frequency conductivity in many-body localized systems},
  volume = {92},
  issn = {1550-235X},
  url = {http://dx.doi.org/10.1103/PhysRevB.92.104202},
  number = {10},
  journal = {Phys. Rev. B},
  publisher = {American Physical Society (APS)},
  author = {Gopalakrishnan,  Sarang and M\"{u}ller,  Markus and Khemani,  Vedika and Knap,  Michael and Demler,  Eugene and Huse,  David A.},
  year = {2015},
  pages = {104202}
}

@article{Imbrie2016,
  title = {On Many-Body Localization for Quantum Spin Chains},
  volume = {163},
  issn = {1572-9613},
  url = {http://dx.doi.org/10.1007/s10955-016-1508-x},
  number = {5},
  journal = {J. Stat. Phys.},
  publisher = {Springer Science and Business Media LLC},
  author = {Imbrie,  John Z.},
  year = {2016},
  pages = {998–1048}
}

@misc{DeRoeck2024,
      title={Absence of Normal Heat Conduction in Strongly Disordered Interacting Quantum Chains}, 
      author={Wojciech De Roeck and Lydia Giacomin and Francois Huveneers and Oskar Prosniak},
      year={2025},
      eprint={2408.04338},
      archivePrefix={arXiv},
      url={https://arxiv.org/abs/2408.04338}
}

@Article{Lin2018,
	title={{Many-body localization of spinless fermions with attractive interactions in one dimension}},
	author={Sheng-Hsuan Lin and Björn Sbierski and Florian Dorfner and Christoph Karrasch and Fabian Heidrich-Meisner},
	journal={SciPost Phys.},
	volume={4},
	pages={002},
	year={2018},
	doi={10.21468/SciPostPhys.4.1.002},
	url={https://scipost.org/10.21468/SciPostPhys.4.1.002}
}

@article{Geraedts2016,
  title = {Many-body localization and thermalization: Insights from the entanglement spectrum},
  volume = {93},
  issn = {2469-9969},
  url = {http://dx.doi.org/10.1103/PhysRevB.93.174202},
  number = {17},
  journal = {Phys. Rev. B},
  publisher = {American Physical Society (APS)},
  author = {Geraedts,  Scott D. and Nandkishore,  Rahul and Regnault,  Nicolas},
  year = {2016},
  pages = {174202}
}

@article{Khemani2017,
  title = {Critical Properties of the Many-Body Localization Transition},
  volume = {7},
  ISSN = {2160-3308},
  url = {http://dx.doi.org/10.1103/PhysRevX.7.021013},
  number = {2},
  journal = {Phys. Rev. X},
  publisher = {American Physical Society (APS)},
  author = {Khemani,  Vedika and Lim,  S.P. and Sheng,  D.N. and Huse,  David A.},
  year = {2017},
  month = apr ,
  pages = {021013}
}

@article{Colmenarez2019,
  title = {Statistics of correlation functions in the random Heisenberg chain},
  volume = {7},
  issn = {2542-4653},
  url = {http://dx.doi.org/10.21468/SciPostPhys.7.5.064},
  number = {5},
  journal = {SciPost Phys.},
  publisher = {Stichting SciPost},
  author = {Colmenarez,  Luis A. and McClarty,  Paul A. and Haque,  Masud and Luitz,  David J.},
  year = {2019},
  pages = {064}
}

@misc{villalonga_characterizing_2020,
      title={Characterizing the many-body localization transition through correlations}, 
      author={Benjamin Villalonga and Bryan K. Clark},
      year={2020},
      eprint={2007.06586},
      archivePrefix={arXiv},
      primaryClass={cond-mat.dis-nn},
      url={https://arxiv.org/abs/2007.06586}, 
}

@article{Garratt2022,
  title = {Resonant energy scales and local observables in the many-body localized phase},
  volume = {106},
  issn = {2469-9969},
  url = {http://dx.doi.org/10.1103/PhysRevB.106.054309},
  number = {5},
  journal = {Phys. Rev. B},
  publisher = {American Physical Society (APS)},
  author = {Garratt,  Samuel J. and Roy,  Sthitadhi},
  year = {2022},
  pages = {054309}
}

@article{Kjll2018,
  title = {Many-body localization and level repulsion},
  volume = {97},
  issn = {2469-9969},
  url = {http://dx.doi.org/10.1103/PhysRevB.97.035163},
  number = {3},
  journal = {Phys. Rev. B},
  publisher = {American Physical Society (APS)},
  author = {Kj\"{a}ll,  Jonas A.},
  year = {2018},
  pages = {035163}
}

@article{De_Tomasi2021-ua,
  title = {Rare thermal bubbles at the many-body localization transition from the Fock space point of view},
  volume = {104},
  ISSN = {2469-9969},
  url = {http://dx.doi.org/10.1103/PhysRevB.104.024202},
  number = {2},
  journal = {Phys. Rev. B},
  publisher = {American Physical Society (APS)},
  author = {De Tomasi,  Giuseppe and Khaymovich,  Ivan M. and Pollmann,  Frank and Warzel,  Simone},
  year = {2021},
  month = jul,
  pages = {024202}
}

@article{long_phenomenology_2023,
  title = {Phenomenology of the Prethermal Many-Body Localized Regime},
  volume = {131},
  ISSN = {1079-7114},
  url = {http://dx.doi.org/10.1103/PhysRevLett.131.106301},
  number = {10},
  journal = {Phys. Rev. Lett.},
  publisher = {American Physical Society (APS)},
  author = {Long,  David M. and Crowley,  Philip J.D. and Khemani,  Vedika and Chandran,  Anushya},
  year = {2023},
  month = sep,
  pages = {106301}
}

@article{colbois_interaction-driven_2024,
  title = {Interaction-{Driven} {Instabilities} in the {Random}-{Field} {X} {X} {Z} {Chain}},
  volume = {133},
  issn = {0031-9007, 1079-7114},
  url = {https://link.aps.org/doi/10.1103/PhysRevLett.133.116502},
  language = {en},
  number = {11},
  urldate = {2025-03-03},
  journal = {Phys. Rev. Lett.},
  author = {Colbois, Jeanne and Alet, Fabien and Laflorencie, Nicolas},
  year = {2024},
  pages = {116502}
}

@article{colbois_statistics_2024,
  title = {Statistics of systemwide correlations in the random-field {XXZ} chain: {Importance} of rare events in the many-body localized phase},
  volume = {110},
  issn = {2469-9950, 2469-9969},
  shorttitle = {Statistics of systemwide correlations in the random-field {XXZ} chain},
  url = {https://link.aps.org/doi/10.1103/PhysRevB.110.214210},
  language = {en},
  number = {21},
  urldate = {2025-02-25},
  journal = {Phys. Rev.  B},
  author = {Colbois, Jeanne and Alet, Fabien and Laflorencie, Nicolas},
  year = {2024},
  pages = {214210}
}

@article{biroli_large-deviation_2024,
  title = {Large-deviation analysis of rare resonances for the many-body localization transition},
  volume = {110},
  issn = {2469-9950, 2469-9969},
  url = {https://link.aps.org/doi/10.1103/PhysRevB.110.014205},
  language = {en},
  number = {1},
  urldate = {2025-03-07},
  journal = {Phys. Rev.  B},
  author = {Biroli, Giulio and Hartmann, Alexander K. and Tarzia, Marco},
  year = {2024},
  pages = {014205}
}

@article{pal_many-body_2010,
  title = {Many-body localization phase transition},
  volume = {82},
  copyright = {http://link.aps.org/licenses/aps-default-license},
  issn = {1098-0121, 1550-235X},
  url = {https://link.aps.org/doi/10.1103/PhysRevB.82.174411},
  language = {en},
  number = {17},
  urldate = {2025-03-03},
  journal = {Phys. Rev.  B},
  author = {Pal, Arijeet and Huse, David A.},
  year = {2010},
  pages = {174411}
}

@article{DeLuca2013,
  title = {Ergodicity breaking in a model showing many-body localization},
  volume = {101},
  issn = {1286-4854},
  url = {http://dx.doi.org/10.1209/0295-5075/101/37003},
  number = {3},
  journal = {EPL},
  publisher = {IOP Publishing},
  author = {De Luca,  A. and Scardicchio,  A.},
  year = {2013},
  pages = {37003}
}

@article{Gray2018,
  title = {Many-body localization transition: Schmidt gap,  entanglement length,  and scaling},
  volume = {97},
  issn = {2469-9969},
  url = {http://dx.doi.org/10.1103/PhysRevB.97.201105},
  number = {20},
  journal = {Phys. Rev. B},
  publisher = {American Physical Society (APS)},
  author = {Gray,  Johnnie and Bose,  Sougato and Bayat,  Abolfazl},
  year = {2018},
  pages = {201105}
}

@article{RAbouChacra1973,
  title = {A selfconsistent theory of localization},
  volume = {6},
  ISSN = {0022-3719},
  url = {http://dx.doi.org/10.1088/0022-3719/6/10/009},
  DOI = {10.1088/0022-3719/6/10/009},
  number = {10},
  journal = {Journal of Physics C: Solid State Physics},
  publisher = {IOP Publishing},
  author = {R Abou-Chacra and D J Thouless and P W Anderson},
  year = {1973},
  month = may,
  pages = {1734–1752}
}

@unpublished{Tarzia2026,
  author = {{Tarzia}, Marco},
  note   = {in preparation}
}

@article{Oganesyan2007,
  title = {Localization of interacting fermions at high temperature},
  volume = {75},
  ISSN = {1550-235X},
  url = {http://dx.doi.org/10.1103/PhysRevB.75.155111},
  DOI = {10.1103/physrevb.75.155111},
  number = {15},
  journal = {Phys. Rev. B},
  publisher = {American Physical Society (APS)},
  author = {Oganesyan,  Vadim and Huse,  David A.},
  year = {2007},
  month = apr,
  pages = {155111}
}

@article{doggen_many-body_2018,
  title = {Many-body localization and delocalization in large quantum chains},
  volume = {98},
  issn = {2469-9950, 2469-9969},
  url = {https://link.aps.org/doi/10.1103/PhysRevB.98.174202},
  language = {en},
  number = {17},
  urldate = {2025-03-15},
  journal = {Phys. Rev.  B},
  author = {Doggen, Elmer V. H. and Schindler, Frank and Tikhonov, Konstantin S. and Mirlin, Alexander D. and Neupert, Titus and Polyakov, Dmitry G. and Gornyi, Igor V.},
  year = {2018},
  pages = {174202}
}

@article{abanin_distinguishing_2021,
  title = {Distinguishing localization from chaos: {Challenges} in finite-size systems},
  volume = {427},
  issn = {00034916},
  shorttitle = {Distinguishing localization from chaos},
  url = {https://linkinghub.elsevier.com/retrieve/pii/S000349162100021X},
  language = {en},
  urldate = {2025-03-03},
  journal = {Ann. Phys. (N. Y.)},
  author = {Abanin, D.A. and Bardarson, J.H. and De Tomasi, G. and Gopalakrishnan, S. and Khemani, V. and Parameswaran, S.A. and Pollmann, F. and Potter, A.C. and Serbyn, M. and Vasseur, R.},
  year = {2021},
  pages = {168415},
  doi = {10.1016/j.aop.2021.168415}
}

@book{Pichard1991,
  title = {Random Transfer Matrix Theory and Conductance Fluctuations},
  isbn = {9781489936981},
  issn = {0258-1221},
  url = {http://dx.doi.org/10.1007/978-1-4899-3698-1_24},
  booktitle = {Quantum Coherence in Mesoscopic Systems},
  publisher = {Springer US},
  author = {Pichard,  Jean-Louis},
  year = {1991}
}

@article{Marko2010,
   title={Electron transport in strongly disordered structures},
   volume={405},
   ISSN={0921-4526},
   url={http://dx.doi.org/10.1016/j.physb.2010.01.042},
   DOI={10.1016/j.physb.2010.01.042},
   number={14},
   journal={Physica B: Condensed Matter},
   publisher={Elsevier BV},
   author={Markoš, P.},
   year={2010},
   month=jul, pages={3029–3032} }

@article{lemarie_glassy_2019,
  title = {Glassy {Properties} of {Anderson} {Localization}: {Pinning}, {Avalanches}, and {Chaos}},
  volume = {122},
  issn = {0031-9007, 1079-7114},
  shorttitle = {Glassy {Properties} of {Anderson} {Localization}},
  url = {https://link.aps.org/doi/10.1103/PhysRevLett.122.030401},
  language = {en},
  number = {3},
  urldate = {2025-03-16},
  journal = {Phys. Rev. Lett.},
  author = {Lemarié, G.},
  year = {2019},
  pages = {030401}
}

@article{Altshuler1997,
  title = {Quasiparticle Lifetime in a Finite System: A Nonperturbative Approach},
  volume = {78},
  issn = {1079-7114},
  url = {http://dx.doi.org/10.1103/PhysRevLett.78.2803},
  number = {14},
  journal = {Phys. Rev. Lett.},
  publisher = {American Physical Society (APS)},
  author = {Altshuler,  Boris L. and Gefen,  Yuval and Kamenev,  Alex and Levitov,  Leonid S.},
  year = {1997},
  pages = {2803–2806}
}

@article{Biroli2017,
  title = {Delocalized glassy dynamics and many-body localization},
  volume = {96},
  issn = {2469-9969},
  url = {http://dx.doi.org/10.1103/PhysRevB.96.201114},
  number = {20},
  journal = {Phys. Rev. B},
  publisher = {American Physical Society (APS)},
  author = {Biroli,  G. and Tarzia,  M.},
  year = {2017},
  pages = {201114}
}

@article{Logan2019,
  title = {Many-body localization in Fock space: A local perspective},
  volume = {99},
  issn = {2469-9969},
  url = {http://dx.doi.org/10.1103/PhysRevB.99.045131},
  number = {4},
  journal = {Phys. Rev. B},
  publisher = {American Physical Society (APS)},
  author = {Logan,  David E. and Welsh,  Staszek},
  year = {2019},
  pages = {045131}
}

@article{Tikhonov2021,
  title = {From Anderson localization on random regular graphs to many-body localization},
  volume = {435},
  issn = {0003-4916},
  url = {http://dx.doi.org/10.1016/j.aop.2021.168525},
  journal = {Ann. Phys. (N. Y.)},
  publisher = {Elsevier BV},
  author = {Tikhonov,  K.S. and Mirlin,  A.D.},
  year = {2021},
  pages = {168525}
}

@article{roy_fock-space_2024,
doi = {10.1088/1361-648X/ad94c3},
url = {https://doi.org/10.1088/1361-648X/ad94c3},
year = {2024},
month = {dec},
publisher = {IOP Publishing},
volume = {37},
number = {7},
pages = {073003},
author = {Roy, Sthitadhi and Logan, David E},
title = {The Fock-space landscape of many-body localisation},
journal = {J. Phys. Condens. Matter},
}

@article{Tikhonov2016,
  title = {Anderson localization and ergodicity on random regular graphs},
  volume = {94},
  ISSN = {2469-9969},
  url = {http://dx.doi.org/10.1103/PhysRevB.94.220203},
  number = {22},
  journal = {Phys. Rev. B},
  publisher = {American Physical Society (APS)},
  author = {Tikhonov,  K. S. and Mirlin,  A. D. and Skvortsov,  M. A.},
  year = {2016},
  month = dec,
  pages = {220203}
}

@article{Roy2020,
  title = {Fock-space correlations and the origins of many-body localization},
  author = {Roy, Sthitadhi and Logan, David E.},
  journal = {Phys. Rev. B},
  volume = {101},
  issue = {13},
  pages = {134202},
  numpages = {23},
  year = {2020},
  month = {Apr},
  publisher = {American Physical Society},
  doi = {10.1103/PhysRevB.101.134202},
  url = {https://link.aps.org/doi/10.1103/PhysRevB.101.134202}
}

@article{Roy2021,
  title = {Fock-space anatomy of eigenstates across the many-body localization transition},
  author = {Roy, Sthitadhi and Logan, David E.},
  journal = {Phys. Rev. B},
  volume = {104},
  issue = {17},
  pages = {174201},
  numpages = {18},
  year = {2021},
  month = {Nov},
  publisher = {American Physical Society},
  doi = {10.1103/PhysRevB.104.174201},
  url = {https://link.aps.org/doi/10.1103/PhysRevB.104.174201}
}

@article{Tikhonov2019-ku,
  title = {Critical behavior at the localization transition on random regular graphs},
  volume = {99},
  ISSN = {2469-9969},
  url = {http://dx.doi.org/10.1103/PhysRevB.99.214202},
  number = {21},
  journal = {Phys. Rev. B},
  publisher = {American Physical Society (APS)},
  author = {Tikhonov,  K. S. and Mirlin,  A. D.},
  year = {2019},
  month = jun,
  pages = {214202}
}

@article{Tikhonov2019-tb,
  title = {Statistics of eigenstates near the localization transition on random regular graphs},
  volume = {99},
  ISSN = {2469-9969},
  url = {http://dx.doi.org/10.1103/PhysRevB.99.024202},
  number = {2},
  journal = {Phys. Rev. B},
  publisher = {American Physical Society (APS)},
  author = {Tikhonov,  K. S. and Mirlin,  A. D.},
  year = {2019},
  month = jan,
  pages = {024202}
}

@article{Roy2020b,
  title = {Localization on Certain Graphs with Strongly Correlated Disorder},
  author = {Roy, Sthitadhi and Logan, David E.},
  journal = {Phys. Rev. Lett.},
  volume = {125},
  issue = {25},
  pages = {250402},
  numpages = {6},
  year = {2020},
  month = {Dec},
  publisher = {American Physical Society},
  doi = {10.1103/PhysRevLett.125.250402},
  url = {https://link.aps.org/doi/10.1103/PhysRevLett.125.250402}
}

@article{sutradhar_2022,
  title = {Scaling of the Fock-space propagator and multifractality across the many-body localization transition},
  author = {Sutradhar, Jagannath and Ghosh, Soumi and Roy, Sthitadhi and Logan, David E. and Mukerjee, Subroto and Banerjee, Sumilan},
  journal = {Phys. Rev. B},
  volume = {106},
  issue = {5},
  pages = {054203},
  numpages = {10},
  year = {2022},
  month = {Aug},
  publisher = {American Physical Society},
  doi = {10.1103/PhysRevB.106.054203},
  url = {https://link.aps.org/doi/10.1103/PhysRevB.106.054203}
}

@article{richter_pal_2022,
  title = {Many-body localization and delocalization dynamics in the thermodynamic limit},
  volume = {105},
  ISSN = {2469-9969},
  url = {http://dx.doi.org/10.1103/PhysRevB.105.L220405},
  number = {22},
  journal = {Phys. Rev. B},
  publisher = {American Physical Society (APS)},
  author = {Richter,  Jonas and Pal,  Arijeet},
  year = {2022},
  month = jun,
  pages = {L220405}
}

@article{derrida_random-energy_1981,
  title = {Random-energy model: {An} exactly solvable model of disordered systems},
  volume = {24},
  copyright = {http://link.aps.org/licenses/aps-default-license},
  issn = {0163-1829},
  shorttitle = {Random-energy model},
  url = {https://link.aps.org/doi/10.1103/PhysRevB.24.2613},
  language = {en},
  number = {5},
  urldate = {2025-03-16},
  journal = {Phys. Rev. B},
  author = {Derrida, Bernard},
  year = {1981},
  pages = {2613-2626},
}

@article{derrida_generalization_1985,
  title = {A generalization of the {Random} {Energy} {Model} which includes correlations between energies},
  volume = {46},
  issn = {0302-072X},
  url = {http://www.edpsciences.org/10.1051/jphyslet:01985004609040100},
  number = {9},
  urldate = {2025-03-16},
  journal = {J. Phys. Lett.},
  author = {Derrida, B.},
  year = {1985},
  pages = {L401--L407}
}

@book{mezard_spin_1987,
  title = {Spin {Glass} {Theory} and {Beyond} : {An} {Introduction} to the {Replica} {Method} and {Its} {Applications}},
  author = {Mezard, M. and Virasoro;, G Parisi;M},
  year = {1987},
  publisher = {World Scientific Publishing Company}
}

@article{b_derrida_solution_1986,
  title = {Solution of the generalised random energy model},
  volume = {19},
  issn = {0022-3719},
  url = {https://iopscience.iop.org/article/10.1088/0022-3719/19/13/015},
  number = {13},
  urldate = {2025-03-16},
  journal = {J. Phys. C: Solid State Phys.},
  author = {{B Derrida} and {E Gardner}},
  year = {1986},
  pages = {2253-2274}
}

@article{carpentier_glass_2001,
  title = {Glass transition of a particle in a random potential, front selection in nonlinear renormalization group, and entropic phenomena in {Liouville} and sinh-{Gordon} models},
  volume = {63},
  copyright = {http://link.aps.org/licenses/aps-default-license},
  issn = {1063-651X, 1095-3787},
  url = {https://link.aps.org/doi/10.1103/PhysRevE.63.026110},
  language = {en},
  number = {2},
  urldate = {2025-03-16},
  journal = {Phys.  Rev. E},
  author = {Carpentier, David and Le Doussal, Pierre},
  year = {2001},
  pages = {026110}
}

@book{Charbonneau2023,
  title = {Spin Glass Theory and Far Beyond: Replica Symmetry Breaking After 40 Years},
  author = {Charbonneau,  Patrick and Marinari,  Enzo and Mézard,  Marc and Parisi,  Giorgio and Ricci-Tersenghi,  Federico and Sicuro,  Gabriele and Zamponi,  Francesco},
  year = {2023},
  publisher = {World Scientific}
}

@BOOK{Datta2013-vq,
  title     = "Cambridge studies in semiconductor physics and microelectronic
               engineering: Electronic transport in mesoscopic systems series
               number 3",
  author    = "Datta, Supriyo",
  publisher = "Cambridge University Press",
  month     =  jun,
  year      =  2013,
  address   = "Cambridge, England"
}

@article{Fisher1981,
  title = {Relation between conductivity and transmission matrix},
  volume = {23},
  ISSN = {0163-1829},
  url = {http://dx.doi.org/10.1103/PhysRevB.23.6851},
  number = {12},
  journal = {Phys. Rev. B},
  publisher = {American Physical Society (APS)},
  author = {Fisher,  Daniel S. and Lee,  Patrick A.},
  year = {1981},
  month = jun,
  pages = {6851-6854},
}

@article{de_luca_anderson_2014,
  title = {Anderson {Localization} on the {Bethe} {Lattice}: {Nonergodicity} of {Extended} {States}},
  volume = {113},
  copyright = {http://link.aps.org/licenses/aps-default-license},
  issn = {0031-9007, 1079-7114},
  shorttitle = {Anderson {Localization} on the {Bethe} {Lattice}},
  url = {https://link.aps.org/doi/10.1103/PhysRevLett.113.046806},
  language = {en},
  number = {4},
  urldate = {2025-03-21},
  journal = {Phys. Rev. Lett.},
  author = {De Luca, A. and Altshuler, B.L. and Kravtsov, V.E. and Scardicchio, A.},
  year = {2014},
  pages = {046806}
}

@article{Herre2023,
  title = {Ergodicity-to-localization transition on random regular graphs with large connectivity and in many-body quantum dots},
  volume = {108},
  issn = {2469-9969},
  url = {http://dx.doi.org/10.1103/PhysRevB.108.014203},
  number = {1},
  journal = {Phys. Rev. B},
  publisher = {American Physical Society (APS)},
  author = {Herre,  Jan-Niklas and Karcher,  Jonas F. and Tikhonov,  Konstantin S. and Mirlin,  Alexander D.},
  year = {2023},
  pages = {014203}
}

@article{Scoquart2024,
  title = {Role of Fock-space correlations in many-body localization},
  volume = {109},
  issn = {2469-9969},
  url = {http://dx.doi.org/10.1103/PhysRevB.109.214203},
  number = {21},
  journal = {Phys. Rev. B},
  publisher = {American Physical Society (APS)},
  author = {Scoquart,  Thibault and Gornyi,  Igor V. and Mirlin,  Alexander D.},
  year = {2024},
  pages = {214203}
}

@article{Aizenman1993-mp,
  title = {Localization at large disorder and at extreme energies: An elementary derivations},
  volume = {157},
  ISSN = {1432-0916},
  url = {http://dx.doi.org/10.1007/BF02099760},
  number = {2},
  journal = {Comm. Math. Phys.},
  publisher = {Springer Science and Business Media LLC},
  author = {Aizenman,  Michael and Molchanov,  Stanislav},
  year = {1993},
  pages = {245-278},
  month = oct
}

@article{Aizenman2013-yd,
  title = {Resonant delocalization for random Schr\"{o}dinger operators on tree graphs},
  volume = {15},
  ISSN = {1435-9863},
  url = {http://dx.doi.org/10.4171/JEMS/389},
  number = {4},
  journal = {J. Eur. Math. Soc.},
  publisher = {European Mathematical Society - EMS - Publishing House GmbH},
  author = {Aizenman,  Michael and Warzel,  Simone},
  year = {2013},
  month = may,
  pages = {1167–1222}
}

@article{derrida_polymers_1988,
  title = {Polymers on disordered trees,  spin glasses,  and traveling waves},
  volume = {51},
  ISSN = {1572-9613},
  url = {http://dx.doi.org/10.1007/BF01014886},
  number = {5–6},
  journal = {J. Stat. Phys.},
  publisher = {Springer Science and Business Media LLC},
  author = {Derrida,  B. and Spohn,  H.},
  year = {1988},
  month = jun
}

@article{Derrida1989,
  title = {Directed Polymers on Disordered Hierarchical Lattices},
  volume = {8},
  issn = {1286-4854},
  url = {http://dx.doi.org/10.1209/0295-5075/8/2/001},
  number = {2},
  journal = {EPL},
  publisher = {IOP Publishing},
  author = {Derrida,  B and Griffiths,  R. B},
  year = {1989},
  pages = {817-840}
}

@article{Derrida1991,
  title = {Mean Field Theory of Directed Polymers in a Random Medium and Beyond},
  volume = {T38},
  issn = {1402-4896},
  url = {http://dx.doi.org/10.1088/0031-8949/1991/T38/002},
  journal = {Physica Scripta},
  publisher = {IOP Publishing},
  author = {Derrida,  B},
  year = {1991},
  pages = {111-116}
}

@book{Mezard1986,
  title = {Spin Glass Theory and Beyond: An Introduction to the Replica Method and Its Applications},
  author = {Mezard,  M and Parisi,  G and Virasoro,  M},
  year = {1986},
  publisher = {World Scientific}
}

@book{comets2017directed,
  title = {Directed Polymers in Random Environments},
  author = {Comets, Francis},
  year = {2017},
  publisher = {Springer}
}

@article{monthus_anderson_2011,
  title = {Anderson localization on the {Cayley} tree: multifractal statistics of the transmission at criticality and off criticality},
  volume = {44},
  issn = {1751-8113, 1751-8121},
  shorttitle = {Anderson localization on the {Cayley} tree},
  url = {https://iopscience.iop.org/article/10.1088/1751-8113/44/14/145001},
  number = {14},
  urldate = {2025-03-21},
  journal = {J. Phys. A: Math. Theor.},
  author = {Monthus, Cécile and Garel, Thomas},
  year = {2011},
  pages = {145001}
}

@article{shiferaw_goldschmidt_2001,
  title = {Localization of a polymer in random media: Relation to the localization of a quantum particle},
  volume = {63},
  ISSN = {1095-3787},
  url = {http://dx.doi.org/10.1103/PhysRevE.63.051803},
  number = {5},
  journal = {Phys. Rev. E},
  publisher = {American Physical Society (APS)},
  author = {Shiferaw,  Yohannes and Goldschmidt,  Yadin Y.},
  year = {2001},
  month = apr,
  pages = {051803}
}

@article{Mu2024-aj,
  title = {Kardar-Parisi-Zhang Physics in the Density Fluctuations of Localized Two-Dimensional Wave Packets},
  volume = {132},
  ISSN = {1079-7114},
  url = {http://dx.doi.org/10.1103/PhysRevLett.132.046301},
  number = {4},
  journal = {Phys. Rev. Lett.},
  publisher = {American Physical Society (APS)},
  author = {Mu,  Sen and Gong,  Jiangbin and Lemarié,  Gabriel},
  year = {2024},
  month = jan,
  pages = {046301}
}

@article{dimitrova_mezard_2011,
  title = {The cavity method for quantum disordered systems: from transverse random field ferromagnets to directed polymers in random media},
  volume = {2011},
  ISSN = {1742-5468},
  url = {http://dx.doi.org/10.1088/1742-5468/2011/01/P01020},
  number = {01},
  journal = {J. Stat. Mech.},
  publisher = {IOP Publishing},
  author = {Dimitrova,  O and Mézard,  M},
  year = {2011},
  month = jan,
  pages = {P01020}
}

@article{somoza_unbinding_2015,
  title = {Unbinding transition in semi-infinite two-dimensional localized systems},
  volume = {91},
  copyright = {http://link.aps.org/licenses/aps-default-license},
  issn = {1098-0121, 1550-235X},
  url = {https://link.aps.org/doi/10.1103/PhysRevB.91.155413},
  language = {en},
  number = {15},
  urldate = {2025-03-21},
  journal = {Phys. Rev.  B},
  author = {Somoza, A. M. and Le Doussal, P. and Ortuño, M.},
  year = {2015},
  pages = {155413}
}

@article{monthus_garel_2012,
  title = {Random transverse field Ising model on the Cayley tree: analysis via boundary strong disorder renormalization},
  volume = {2012},
  ISSN = {1742-5468},
  url = {http://dx.doi.org/10.1088/1742-5468/2012/10/P10010},
  number = {10},
  journal = {J. Stat. Mech.},
  publisher = {IOP Publishing},
  author = {Monthus,  Cécile and Garel,  Thomas},
  year = {2012},
  month = oct,
  pages = {P10010}
}

@article{pietracaprina_forward_2016,
  title = {Forward approximation as a mean-field approximation for the {Anderson} and many-body localization transitions},
  volume = {93},
  copyright = {http://link.aps.org/licenses/aps-default-license},
  issn = {2469-9950, 2469-9969},
  url = {https://link.aps.org/doi/10.1103/PhysRevB.93.054201},
  language = {en},
  number = {5},
  urldate = {2025-03-19},
  journal = {Phys. Rev.  B},
  author = {Pietracaprina, Francesca and Ros, Valentina and Scardicchio, Antonello},
  year = {2016},
  pages = {054201}
}

@article{tarquini_level_2016,
  title = {Level {Statistics} and {Localization} {Transitions} of {Lévy} {Matrices}},
  volume = {116},
  copyright = {http://link.aps.org/licenses/aps-default-license},
  issn = {0031-9007, 1079-7114},
  url = {https://link.aps.org/doi/10.1103/PhysRevLett.116.010601},
  language = {en},
  number = {1},
  urldate = {2025-03-21},
  journal = {Phys. Rev. Lett.},
  author = {Tarquini, E. and Biroli, G. and Tarzia, M.},
  year = {2016},
  pages = {010601}
}

@article{Derrida1990,
  title = {Directed polymers in a random medium},
  volume = {163},
  issn = {0378-4371},
  url = {http://dx.doi.org/10.1016/0378-4371(90)90316-K},
  number = {1},
  journal = {Physica A: Stat. Mech. Appl.},
  publisher = {Elsevier BV},
  author = {Derrida,  B.},
  year = {1990},
   pages = {71-84}
}

@article{Evans1992,
  title = {Improved bounds for the transition temperature of directed polymers in a finite-dimensional random medium},
  volume = {69},
  ISSN = {1572-9613},
  url = {http://dx.doi.org/10.1007/BF01053800},
  DOI = {10.1007/bf01053800},
  number = {1–2},
  journal = {J. Stat. Phys.},
  publisher = {Springer Science and Business Media LLC},
  author = {Evans,  M. R. and Derrida,  B.},
  year = {1992},
  month = oct,
  pages = {427–437}
}

@article{fyodorov_freezing_2008,
  title = {Freezing and extreme-value statistics in a random energy model with logarithmically correlated potential},
  volume = {41},
  issn = {1751-8113, 1751-8121},
  url = {https://iopscience.iop.org/article/10.1088/1751-8113/41/37/372001},
  number = {37},
  urldate = {2025-03-16},
  journal = {J. Phys. A: Math.  Theor.},
  author = {Fyodorov, Yan V and Bouchaud, Jean-Philippe},
  year = {2008},
  pages = {372001}
}

@article{Gardner1989,
  title = {The probability distribution of the partition function of the random energy model},
  volume = {22},
  issn = {1361-6447},
  url = {http://dx.doi.org/10.1088/0305-4470/22/12/003},
  number = {12},
  journal = {J. Phys. A: Math. Gen.},
  publisher = {IOP Publishing},
  author = {Gardner,  E and Derrida,  B},
  year = {1989},
  pages = {1975}
}

@article{Laflorencie2025-mv,
  title = {Cat states carrying long-range correlations in the many-body localized phase},
  author = {Laflorencie, Nicolas and Colbois, Jeanne and Alet, Fabien},
  journal = {Phys. Rev. B},
  volume = {112},
  issue = {22},
  pages = {224207},
  numpages = {25},
  year = {2025},
  month = {Dec},
  publisher = {American Physical Society},
  doi = {10.1103/gtfr-nblw},
  url = {https://link.aps.org/doi/10.1103/gtfr-nblw}
}

@article{biroli2020anomalous,
  title = {Anomalous dynamics on the ergodic side of the many-body localization transition and the glassy phase of directed polymers in random media},
  volume = {102},
  ISSN = {2469-9969},
  url = {http://dx.doi.org/10.1103/PhysRevB.102.064211},
  number = {6},
  journal = {Phys. Rev. B},
  publisher = {American Physical Society (APS)},
  author = {Biroli,  G. and Tarzia,  M.},
  year = {2020},
  month = aug,
  pages = {064211}
}

@article{pieper_high-performance_2016,
  title = {High-performance implementation of {Chebyshev} filter diagonalization for interior eigenvalue computations},
  volume = {325},
  issn = {00219991},
  url = {https://linkinghub.elsevier.com/retrieve/pii/S0021999116303837},
  language = {en},
  urldate = {2025-03-18},
  journal = {J. Comput. Phys.},
  author = {Pieper, Andreas and Kreutzer, Moritz and Alvermann, Andreas and Galgon, Martin and Fehske, Holger and Hager, Georg and Lang, Bruno and Wellein, Gerhard},
  year = {2016},
  pages = {226--243}
}

@article{Andreanov2025,
  title = {From Dyson models to many-body quantum chaos},
  volume = {111},
  ISSN = {2469-9969},
  url = {http://dx.doi.org/10.1103/PhysRevB.111.035147},
  number = {3},
  journal = {Phys. Rev. B},
  publisher = {American Physical Society (APS)},
  author = {Andreanov,  Alexei and Carrega,  Matteo and Murugan,  Jeff and Olle,  Jan and Rosa,  Dario and Shir,  Ruth},
  year = {2025},
  month = jan,
  pages = {035147}
}

@article{Abbout2011,
  title = {Thermal Enhancement of Interference Effects in Quantum Point Contacts},
  volume = {106},
  issn = {1079-7114},
  url = {http://dx.doi.org/10.1103/PhysRevLett.106.156810},
  number = {15},
  journal = {Phys. Rev. Lett.},
  publisher = {American Physical Society (APS)},
  author = {Abbout,  Adel and Lemarié,  Gabriel and Pichard,  Jean-Louis},
  year = {2011},
  pages = {156810}
}

@article{Gorini2013,
  title = {Theory of scanning gate microscopy},
  volume = {88},
  issn = {1550-235X},
  url = {http://dx.doi.org/10.1103/PhysRevB.88.035406},
  number = {3},
  journal = {Phys. Rev. B},
  publisher = {American Physical Society (APS)},
  author = {Gorini,  Cosimo and Jalabert,  Rodolfo A. and Szewc,  Wojciech and Tomsovic,  Steven and Weinmann,  Dietmar},
  year = {2013},
  pages = {035406}
}

@article{Korner2006-lv,
  title = {Probing tails of energy distributions using importance-sampling in the disorder with a guiding function},
  volume = {2006},
  ISSN = {1742-5468},
  url = {http://dx.doi.org/10.1088/1742-5468/2006/04/P04005},
  number = {04},
  journal = {J. Stat. Mech.},
  publisher = {IOP Publishing},
  author = {K\"{o}rner,  Mathias and Katzgraber,  Helmut G and Hartmann,  Alexander K},
  year = {2006},
  month = apr,
  pages = {P04005}
}

@article{Sierant2023-ba,
  title = {Stability of many-body localization in Floquet systems},
  volume = {107},
  ISSN = {2469-9969},
  url = {http://dx.doi.org/10.1103/PhysRevB.107.115132},
  number = {11},
  journal = {Phys. Rev. B},
  publisher = {American Physical Society (APS)},
  author = {Sierant,  Piotr and Lewenstein,  Maciej and Scardicchio,  Antonello and Zakrzewski,  Jakub},
  year = {2023},
  month = mar,
  pages = {115132}
}

@article{Schreiber2015-qi,
  title = {Observation of many-body localization of interacting fermions in a quasirandom optical lattice},
  volume = {349},
  ISSN = {1095-9203},
  url = {http://dx.doi.org/10.1126/science.aaa7432},
  number = {6250},
  journal = {Science},
  publisher = {American Association for the Advancement of Science (AAAS)},
  author = {Schreiber,  Michael and Hodgman,  Sean S. and Bordia,  Pranjal and L\"{u}schen,  Henrik P. and Fischer,  Mark H. and Vosk,  Ronen and Altman,  Ehud and Schneider,  Ulrich and Bloch,  Immanuel},
  year = {2015},
  month = aug,
  pages = {842-845}
}

@article{Bordia2017-qv,
  title = {Probing Slow Relaxation and Many-Body Localization in Two-Dimensional Quasiperiodic Systems},
  volume = {7},
  ISSN = {2160-3308},
  url = {http://dx.doi.org/10.1103/PhysRevX.7.041047},
  number = {4},
  journal = {Phys. Rev. X},
  publisher = {American Physical Society (APS)},
  author = {Bordia,  Pranjal and L\"{u}schen,  Henrik and Scherg,  Sebastian and Gopalakrishnan,  Sarang and Knap,  Michael and Schneider,  Ulrich and Bloch,  Immanuel},
  year = {2017},
  month = nov,
  pages = {041047}
}

@article{Luschen2017-or,
  title = {Observation of Slow Dynamics near the Many-Body Localization Transition in One-Dimensional Quasiperiodic Systems},
  volume = {119},
  ISSN = {1079-7114},
  url = {http://dx.doi.org/10.1103/PhysRevLett.119.260401},
  number = {26},
  journal = {Phys. Rev. Lett.},
  publisher = {American Physical Society (APS)},
  author = {L\"{u}schen,  Henrik P. and Bordia,  Pranjal and Scherg,  Sebastian and Alet,  Fabien and Altman,  Ehud and Schneider,  Ulrich and Bloch,  Immanuel},
  year = {2017},
  month = dec,
  pages = {260401}
}

@article{Amestoy2001-db,
  title = {A Fully Asynchronous Multifrontal Solver Using Distributed Dynamic Scheduling},
  volume = {23},
  ISSN = {1095-7162},
  url = {http://dx.doi.org/10.1137/S0895479899358194},
  number = {1},
  journal = {SIAM J. Matrix Anal. Appl.},
  publisher = {Society for Industrial & Applied Mathematics (SIAM)},
  author = {Amestoy,  Patrick R. and Duff,  Iain S. and L’Excellent,  Jean-Yves and Koster,  Jacko},
  year = {2001},
  month = jan,
  pages = {15-41}
}

@article{Amestoy2006-ni,
  title = {Hybrid scheduling for the parallel solution of linear systems},
  volume = {32},
  ISSN = {0167-8191},
  url = {http://dx.doi.org/10.1016/j.parco.2005.07.004},
  number = {2},
  journal = {Parallel Comput.},
  publisher = {Elsevier BV},
  author = {Amestoy,  Patrick R. and Guermouche,  Abdou and L’Excellent,  Jean-Yves and Pralet,  Stéphane},
  year = {2006},
  month = feb,
  pages = {136-156}
}

@article{Schenk2004-pt,
  title = {Solving unsymmetric sparse systems of linear equations with PARDISO},
  volume = {20},
  ISSN = {0167-739X},
  url = {http://dx.doi.org/10.1016/j.future.2003.07.011},
  number = {3},
  journal = {Future Gener. Comput. Syst.},
  publisher = {Elsevier BV},
  author = {Schenk,  Olaf and G\"{a}rtner,  Klaus},
  year = {2004},
  month = apr,
  pages = {475-487}
}

@article{Thomson2018,
  title = {Time evolution of many-body localized systems with the flow equation approach},
  author = {Thomson, S. J. and Schir\'o, M.},
  journal = {Phys. Rev. B},
  volume = {97},
  issue = {6},
  pages = {060201},
  numpages = {5},
  year = {2018},
  month = {Feb},
  doi = {10.1103/PhysRevB.97.060201},
  url = {https://link.aps.org/doi/10.1103/PhysRevB.97.060201}
}

@article{Prelovsek2018,
  title = {Reduced-basis approach to many-body localization},
  author = {Prelov\ifmmode \check{s}\else \v{s}\fi{}ek, P. and Bari\ifmmode \check{s}\else \v{s}\fi{}i\ifmmode \acute{c}\else \'{c}\fi{}, O. S. and Mierzejewski, M.},
  journal = {Phys. Rev. B},
  volume = {97},
  issue = {3},
  pages = {035104},
  numpages = {11},
  year = {2018},
  month = {Jan},
  doi = {10.1103/PhysRevB.97.035104},
  url = {https://link.aps.org/doi/10.1103/PhysRevB.97.035104}
}

@article{Detomasi2019,
  title = {Efficiently solving the dynamics of many-body localized systems at strong disorder},
  author = {De Tomasi, Giuseppe and Pollmann, Frank and Heyl, Markus},
  journal = {Phys. Rev. B},
  volume = {99},
  issue = {24},
  pages = {241114},
  numpages = {6},
  year = {2019},
  month = {Jun},
  doi = {10.1103/PhysRevB.99.241114},
  url = {https://link.aps.org/doi/10.1103/PhysRevB.99.241114}
}

@article{Colbois2023,
  title = {Breaking the chains: Extreme value statistics and localization in random spin chains},
  author = {Colbois, Jeanne and Laflorencie, Nicolas},
  journal = {Phys. Rev. B},
  volume = {108},
  issue = {14},
  pages = {144206},
  numpages = {20},
  year = {2023},
  month = {Oct},
  doi = {10.1103/PhysRevB.108.144206},
  url = {https://link.aps.org/doi/10.1103/PhysRevB.108.144206}
}

@article{jiang2025quasiconservationlawssuppressedtransport,
  title = {Quasiconservation laws and suppressed transport in weakly interacting localized models},
  author = {Jiang, Jessica K. and Surace, Federica M. and Motrunich, Olexei I.},
  journal = {Phys. Rev. B},
  volume = {112},
  issue = {18},
  pages = {184201},
  numpages = {33},
  year = {2025},
  month = {Nov},
  publisher = {American Physical Society},
  doi = {10.1103/shj5-bvs3},
  url = {https://link.aps.org/doi/10.1103/shj5-bvs3}
}

@article{Evers2008,
  title = {Anderson transitions},
  volume = {80},
  ISSN = {1539-0756},
  url = {http://dx.doi.org/10.1103/RevModPhys.80.1355},
  DOI = {10.1103/revmodphys.80.1355},
  number = {4},
  journal = {Rev. Mod. Phys.},
  publisher = {American Physical Society (APS)},
  author = {Evers,  Ferdinand and Mirlin,  Alexander D.},
  year = {2008},
  month = oct,
  pages = {1355–1417}
}

@article{Sajid2025,
  title = {Thermal avalanches in isolated many-body localized systems},
  author = {Sajid, Muhammad and Yousefjani, Rozhin and Bayat, Abolfazl},
  journal = {Phys. Rev. B},
  volume = {112},
  issue = {15},
  pages = {155140},
  numpages = {11},
  year = {2025},
  month = {Oct},
  publisher = {American Physical Society},
  doi = {10.1103/fvgj-byp7},
  url = {https://link.aps.org/doi/10.1103/fvgj-byp7}
}

@article{RG_XXZ,
  title = {Renormalization group analysis of the many-body localization transition in the random-field XXZ chain},
  author = {Niedda, Jacopo and Testasecca, Giacomo Bracci and Magnifico, Giuseppe and Balducci, Federico and Vanoni, Carlo and Scardicchio, Antonello},
  journal = {Phys. Rev. B},
  volume = {112},
  issue = {14},
  pages = {144201},
  numpages = {14},
  year = {2025},
  month = {Oct},
  publisher = {American Physical Society},
  doi = {10.1103/gcwf-jdlr},
  url = {https://link.aps.org/doi/10.1103/gcwf-jdlr}
}

@article{RG_AM,
author = {Carlo Vanoni  and Boris L. Altshuler  and Vladimir E. Kravtsov  and Antonello Scardicchio },
title = {Renormalization group analysis of the Anderson model on random regular graphs},
journal = {Proceedings of the National Academy of Sciences},
volume = {121},
number = {29},
pages = {e2401955121},
year = {2024},
doi = {10.1073/pnas.2401955121},
URL = {https://www.pnas.org/doi/abs/10.1073/pnas.2401955121}}

@misc{EversBera2025,
      title={Fock space fragmentation in quenches of disordered interacting fermions}, 
      author={Ishita Modak and Rajesh Narayanan and Ferdinand Evers and Soumya Bera},
      year={2025},
      eprint={2510.19510},
      archivePrefix={arXiv},
      primaryClass={cond-mat.dis-nn},
      url={https://arxiv.org/abs/2510.19510}, 
}

@article{Beraetal2017,
  title = {Density Propagator for Many-Body Localization: Finite-Size Effects, Transient Subdiffusion, and Exponential Decay},
  author = {Bera, Soumya and De Tomasi, Giuseppe and Weiner, Felix and Evers, Ferdinand},
  journal = {Phys. Rev. Lett.},
  volume = {118},
  issue = {19},
  pages = {196801},
  numpages = {6},
  year = {2017},
  month = {May},
  publisher = {American Physical Society},
  doi = {10.1103/PhysRevLett.118.196801},
  url = {https://link.aps.org/doi/10.1103/PhysRevLett.118.196801}
}

@article{EversModakBera2023,
  title = {Internal clock of many-body delocalization},
  author = {Evers, Ferdinand and Modak, Ishita and Bera, Soumya},
  journal = {Phys. Rev. B},
  volume = {108},
  issue = {13},
  pages = {134204},
  numpages = {15},
  year = {2023},
  month = {Oct},
  publisher = {American Physical Society},
  doi = {10.1103/PhysRevB.108.134204},
  url = {https://link.aps.org/doi/10.1103/PhysRevB.108.134204}
}

@article{WeinerEversBera2019,
  title = {Slow dynamics and strong finite-size effects in many-body localization with random and quasiperiodic potentials},
  author = {Weiner, Felix and Evers, Ferdinand and Bera, Soumya},
  journal = {Phys. Rev. B},
  volume = {100},
  issue = {10},
  pages = {104204},
  numpages = {11},
  year = {2019},
  month = {Sep},
  publisher = {American Physical Society},
  doi = {10.1103/PhysRevB.100.104204},
  url = {https://link.aps.org/doi/10.1103/PhysRevB.100.104204}
}

@article{ZnidaricProsenPrelovsek2008,
  title = {Many-body localization in the Heisenberg $XXZ$ magnet in a random field},
  author = {{Žnidarič}, Marko and {Prosen}, Tomaž and {Prelovšek}, Peter},
  journal = {Phys. Rev. B},
  volume = {77},
  pages = {064426},
  year = {2008},
  doi = {10.1103/PhysRevB.77.064426}
}

@article{HerbrychPrelovsek2025,
  title = {Spin and energy diffusion versus subdiffusion in disordered spin chains},
  author = {Herbrych, J. and Prelov\ifmmode \check{s}\else \v{s}\fi{}ek, P.},
  journal = {Phys. Rev. B},
  volume = {112},
  issue = {4},
  pages = {045108},
  numpages = {9},
  year = {2025},
  month = {Jul},
  publisher = {American Physical Society},
  doi = {10.1103/kv2f-m8vk},
  url = {https://link.aps.org/doi/10.1103/kv2f-m8vk}
}

@article{HerbrychMierzejewskiPrelovsek2022,
  title = {Relaxation at different length scales in models of many-body localization},
  author = {Herbrych, J. and Mierzejewski, M. and Prelov\ifmmode \check{s}\else \v{s}\fi{}ek, P.},
  journal = {Phys. Rev. B},
  volume = {105},
  issue = {8},
  pages = {L081105},
  numpages = {6},
  year = {2022},
  month = {Feb},
  publisher = {American Physical Society},
  doi = {10.1103/PhysRevB.105.L081105},
  url = {https://link.aps.org/doi/10.1103/PhysRevB.105.L081105}
}

@article{PrelovsekHerbrychMierzejewski2023,
  title = {Slow diffusion and Thouless localization criterion in modulated spin chains},
  author = {Prelov\ifmmode \check{s}\else \v{s}\fi{}ek, P. and Herbrych, J. and Mierzejewski, M.},
  journal = {Phys. Rev. B},
  volume = {108},
  issue = {3},
  pages = {035106},
  numpages = {9},
  year = {2023},
  month = {Jul},
  publisher = {American Physical Society},
  doi = {10.1103/PhysRevB.108.035106},
  url = {https://link.aps.org/doi/10.1103/PhysRevB.108.035106}
}

@dataset{ZenodoData,
  author       = {Alfaro Miranda, Greivin},
  title        = {Large deviations in the many-body localization
                   transition: The case of the random-field XXZ chain
                  },
  month        = mar,
  year         = 2026,
  publisher    = {Zenodo},
  doi          = {10.5281/zenodo.19186108},
  url          = {https://doi.org/10.5281/zenodo.19186108},
}

\end{document}